\documentclass[journal,onecolumn]{IEEEtran}
\usepackage{amsmath, amssymb}
\usepackage{amsthm}

\usepackage{nag}
\usepackage{color}
\usepackage{wasysym}
\usepackage[normalem]{ulem}
\usepackage{todonotes}

\usepackage{graphicx}
\usepackage{caption}
\usepackage{subcaption}

\usepackage{bbold}
\usepackage{rotating}

\usepackage{tikz}
\usetikzlibrary{shapes,arrows,fit}
\usetikzlibrary{plotmarks}
\usetikzlibrary{positioning}
\usetikzlibrary{shapes.geometric, intersections}
\usetikzlibrary{decorations.markings}

\usepackage{pgfplots}
  \pgfplotsset{compat=newest} 
  \pgfplotsset{plot coordinates/math parser=false}

\usepackage{algpseudocode}
\usepackage{mdframed}

\ifCLASSINFOpdf
\else
\fi

\definecolor{TUMpantone540}{RGB}{000,051,089}
\definecolor{TUMpantone300}{RGB}{000,101,189}
\definecolor{TUMpantone301}{RGB}{000,082,147}
\definecolor{TUMpantone285}{RGB}{000,115,207}
\definecolor{TUMpantone542}{RGB}{100,160,200}
\definecolor{TUMpantone283}{RGB}{152,198,234}
\definecolor{TUMdgray}{RGB}{088,088,090}
\definecolor{TUMmgray}{RGB}{156,157,159}
\definecolor{TUMlgray}{RGB}{215,217,218}
\definecolor{TUMyellow}{RGB}{255,180,000}
\definecolor{TUMorange}{RGB}{255,128,000}
\definecolor{TUMblue}{RGB}{000,101,189}
\definecolor{TUMgreen}{RGB}{0,124,48}
\definecolor{TUMred}{RGB}{196,7,27}
\definecolor{TUMgold}{RGB}{255,200,0}
\definecolor{green4}{rgb}{0,0.57,0}
\definecolor{blue4}{rgb}{0,0,0.57}
\definecolor{lavender2}{rgb}{0.9,0.9,0.98}
\definecolor{lightblue}{rgb}{0.74,0.93,1}
\definecolor{white}{rgb}{1,1,1}
\definecolor{lightyellow}{rgb}{0.93,0.93,0.82}
\definecolor{plum}{rgb}{0.86,0.62,0.86}
\definecolor{palegreen}{rgb}{0.6,1,0.6}
\definecolor{markerred}{rgb}{1,0.85,0.85}
\definecolor{hellgrau}{rgb}{0.95,0.95,0.95}

\definecolor{magenta}{rgb}{1.00000,0.00000,1.00000}%

\newcommand{\Defcol}{black}

\newcommand{\CLBcolor}{TUMpantone301}

\newcommand{\UBcolor}{TUMred}
\newcommand{\UBstyle}{dashed}
\newcommand{\UBmarker}{none}

\newcommand{\PTPcolor}{black}
\newcommand{\PTPstyle}{solid}
\newcommand{\PTPmarker}{star}

\newcommand{\CompFcolor}{TUMpantone540}
\newcommand{\CompFstyle}{solid}
\newcommand{\CompFmarker}{star}

\newcommand{\AFcolor}{TUMdgray}
\newcommand{\AFstyle}{solid}
\newcommand{\AFmarker}{asterisk}

\newcommand{\NoSWcolor}{TUMblue}
\newcommand{\NoSWstyle}{solid}
\newcommand{\NoSWmarker}{o}

\newcommand{\CFcolor}{TUMgreen}
\newcommand{\CFstyle}{solid}
\newcommand{\CFmarker}{x}
\newcommand{\CFmarkerBursty}{oplus}

\newcommand{\CFRankovcolor}{palegreen}
\newcommand{\CFRankovstyle}{solid}
\newcommand{\CFRankovmarker}{+}

\newcommand{\NNCcolor}{TUMgold}
\newcommand{\NNCstyle}{solid}
\newcommand{\NNCmarker}{square}

\newcommand{\DFMAcolor}{TUMorange}
\newcommand{\DFMAstyle}{solid}
\newcommand{\DFMAmarker}{diamond}

\newcommand{\DFTDcolor}{plum}
\newcommand{\DFTDstyle}{solid}
\newcommand{\DFTDmarker}{triangle}

\newcommand{\bs}{\bm}
\newcommand{\mc}[1]{\mathcal{#1}}
\newcommand{\Absatz}[0]{\vspace*{5mm}}
\newcommand{\eq}[2]{\begin{equation} #1 \label{#2}\end{equation}}
\newcommand{\tn}[1]{\textnormal{#1}}
\newcommand{\truth}{\mbox{\textbb{1}}}

\newcommand{\G}{\text{G}}
\newcommand{\B}{\text{B}}
\newcommand{\GG}{\text{GG}}
\newcommand{\GB}{\text{GB}}
\newcommand{\BG}{\text{BG}}
\newcommand{\BB}{\text{BB}}

\newcommand{\ve}[1]{\uline{\smash{#1}}}
\newcommand{\mat}[1]{\mathbf{#1}}
\newcommand{\norm}[1]{\left\lVert#1\right\rVert}

\newcommand{\Rxone}{\textnormal{Rx$_1$}}
\newcommand{\Rxtwo}{\textnormal{Rx$_2$}}
\newcommand{\Rxj}{\textnormal{Rx$_j$}}

\newcommand{\epsone}[1]{\ensuremath{\epsilon_1(#1)}}			%
\newcommand{\epstwo}[1]{\ensuremath{\epsilon_2(#1)}} 			%
\newcommand{\epsj}[1]{\ensuremath{\epsilon_j(#1)}}			%
\newcommand{\epsonetwo}[1]{\ensuremath{\epsilon_{12}(#1)}}		%
\newcommand{\epsonenottwo}[1]{\ensuremath{\epsilon_{1\bar{2}}(#1)}}	%
\newcommand{\epsnotonetwo}[1]{\ensuremath{\epsilon_{\bar{1}2}(#1)}}	%
\newcommand{\epsnotonenottwo}[1]{\ensuremath{\epsilon_{\bar{1}\bar{2}}(#1)}}	%

\newcommand{\flowdiv}[3]{\ensuremath{d_{#1}^{#2}#3}}		%

\newcommand{\flowdivRV}[2]{\ensuremath{D_{#1}^{#2}}}		%

\newcommand{\activ}[3]{\ensuremath{e_{#1}^{#2}#3}}		  

\newcommand{\activRV}[2]{\ensuremath{E_{#1}^{#2}}}		

\newcommand{\obs}{\ensuremath{\Omega}}
\newcommand{\obsvar}{\ensuremath{\omega}}
\newcommand{\obsw}{\ensuremath{{\Omega}}}
\newcommand{\obswvar}{\ensuremath{{\omega}}}

\newcommand{\markovRV}{\ensuremath{M}}
\newcommand{\markov}{\ensuremath{m}}

\newcommand{\mixdis}[1]{\ensuremath{\Pi_{#1}}}

\newcommand{\capa}{\ensuremath{\mathfrak C}}

\newcommand{\llr}{\ensuremath{L}}

\newcommand{\tone}{\ensuremath{T_1}}
\newcommand{\ttwo}{\ensuremath{T_2}}
\newcommand{\tr}{\ensuremath{r}}

\newcommand{\w}{\ensuremath{W}}
\newcommand{\wone}{\ensuremath{W_1}}
\newcommand{\wtwo}{\ensuremath{W_2}}
\newcommand{\wonehat}{\ensuremath{\hat{W}_1}}
\newcommand{\wtwohat}{\ensuremath{\hat{W}_2}}

\newcommand{\x}{\ensuremath{X}}
\newcommand{\xone}{\ensuremath{X_1}}
\newcommand{\xtwo}{\ensuremath{X_2}}
\newcommand{\xr}{\ensuremath{X_r}}

\newcommand{\Y}{\ensuremath{Y}}
\newcommand{\yr}{\ensuremath{Y_r}}
\newcommand{\yone}{\ensuremath{Y_1}}
\newcommand{\ytwo}{\ensuremath{Y_2}}

\DeclareMathOperator*{\argmax}{argmax}
\DeclareMathOperator*{\argmin}{argmin}

\theoremstyle{definition}
\newtheorem{theorem}{Theorem}
\newtheorem*{oldtheorem}{Theorem}
\newtheorem{prop}{Proposition}
\newtheorem{lemma}{Lemma}
\newtheorem{cor}{Corollary}

\newtheorem{definition}{Definition}
\newtheorem{remark}{Remark}

\newtheorem{examplex}{Example}
\newenvironment{example}
  {\pushQED{\qed}\renewcommand{\qedsymbol}{$\square$}\examplex}
  {\popQED\endexamplex}

\tikzstyle{block} = [draw,  rectangle, rounded corners,
    minimum height=2em, minimum width=6em]
\tikzstyle{input} = [draw,  rectangle, rounded corners,
    minimum height=2em, minimum width=4em]
\tikzstyle{chan} = [draw, rectangle, rounded corners,
    minimum height=2em, minimum width=4em]
\tikzstyle{output} = [coordinate]
\tikzstyle{pinstyle} = [pin edge={to-,thin,black}]
\tikzstyle{block2} = [draw, rectangle, 
    minimum height=1.5em, minimum width=2em]
\tikzstyle{sum} = [draw, circle, inner sep=0pt, node distance=1cm]

\tikzstyle{factor} = [draw, rectangle, minimum height=5mm, minimum width=5mm]
\tikzstyle{var} = [draw,circle, inner sep=1mm]

\tikzstyle{chan} = [draw, rectangle, rounded corners,minimum height=2em, minimum width=4em]
\tikzstyle{block} = [draw, rectangle, minimum height=4em, minimum width=2em]
\tikzstyle{block2} = [draw, rectangle, minimum height=1.5em, minimum width=2em]
\tikzset{->-/.style={decoration={markings,mark=at position #1 with {\arrow{>}}},postaction={decorate}}}

\begin{document}
\title{Capacity Regions of Two-Receiver Broadcast Erasure Channels with Feedback and  Memory}
\author{Michael Heindlmaier,~\IEEEmembership{Member,~IEEE,}
      and  Shirin Saeedi Bidokhti,~\IEEEmembership{Member,~IEEE,}%
\thanks{Michael Heindlmaier is with the Department for Electrical and Computer Engineering, Technische Universit\"{a}t M\"{u}nchen, Germany (michael.heindlmaier@tum.de).}%
\thanks{Shirin Saeedi Bidokhti was with the Department for Electrical and Computer Engineering, Technische Universit\"{a}t M\"{u}nchen, Germany. She is now with the Department of Electrical and Systems Engineering, University of Pennsylvania, USA (saeedi@seas.upenn.edu).}
\thanks{The authors were supported by the German Federal Ministry of Education and Research in the framework of the Alexander von Humboldt-Professorship. The work of S. Saeedi Bidokhti was also supported by the Swiss National Science Foundation.}%
\thanks{The material in this paper was presented in part at the 2014 Allerton Conf. Comm. Control and Computing, USA \cite{heindlmaier2014oncapacity}, and the 2015 IEEE International Symposium on Information Theory, Honolulu, USA \cite{heindlmaier2015capacity}.}}

\maketitle

\begin{abstract}
The two-receiver broadcast packet erasure channel with feedback and memory is studied. Memory is modeled using a finite-state Markov chain representing a channel state. Two scenarios are considered: (i) when the transmitter has causal knowledge of the channel state (i.e., the state is visible), and (ii) when the channel state is unknown at the transmitter, but observations of it are available at the transmitter through feedback (i.e., the state is hidden). In both scenarios, matching outer and inner bounds on the rates of communication are derived and the capacity region is determined. It is shown that similar results carry over to channels with memory and delayed feedback and memoryless compound channels with feedback.

When the state is visible, the capacity region has a single-letter characterization and is in terms of a linear program. Two optimal coding schemes are devised that use feedback to keep track of the sent/received packets via a network of queues: a probabilistic scheme and a deterministic  backpressure-like algorithm.  The former bases its decisions solely on the past channel state information and the latter follows a max-weight queue-based policy. The performance of the algorithms are analyzed using the frameworks of rate stability in networks of queues, max-flow min-cut duality in networks, and finite-horizon Lyapunov drift analysis.

When the state is hidden, the capacity region does not have a single-letter characterization and is, in this sense, uncomputable. Approximations of the capacity region are provided and two optimal coding algorithms are outlined. The first algorithm is a probabilistic coding scheme that bases its decisions on the past $L$ acknowledgments and its achievable rate region approaches the capacity region exponentially fast in $L$. The second algorithm is a backpressure-like algorithm that performs optimally in the long run.
 
\end{abstract}

\IEEEpeerreviewmaketitle

\section{Introduction}
The capacity of broadcast channels (BCs) remains unresolved both without and with feedback. It was shown in \cite{gamal1978feedback} that feedback does not increase the capacity  of physically degraded BCs. 
Nevertheless, feedback increases the capacity of general BCs \cite{ozarow1984broadcast, kramer2003capacity} and even partial feedback \cite{dueck1980partial,4655434}, noisy feedback \cite{ShayevitzWigger13, VenkataramananPradhan13} and rate-limited feedback \cite{WuWigger16} can help.

The class of memoryless broadcast packet erasure channels (BPECs) is among the few classes of BCs for which the capacity region is known with and without feedback. BPECs are suitable models  for various wireless channels such as channels from satellites to mobile stations (e.g. \cite{lei2009link}).

The capacity region $\mc C$ for the memoryless BPEC without feedback for two receivers consists of all rate pairs
\begin{align}
   (R_1,R_2) \geq 0: \frac{R_1}{1-\epsilon_1} + \frac{R_2}{1-\epsilon_2} \leq 1,
\end{align}
where $\epsilon_1$, $\epsilon_2$ are the erasure probabilities at receiver $1$ and $2$, respectively.
The capacity region $\mc C_\text{fb}$ of a memoryless BPEC with feedback was recently  found in \cite{georgiadis2009broadcast}. It consists of all rate pairs
$(R_1,R_2) \in \mc C_{\text{fb}} = \mc C_1 \cap \mc C_2$, with
\begin{align}
 &\mc C_1 = \left\{ (R_1,R_2) \geq 0: \frac{R_1}{1-\epsilon_1} + \frac{R_2}{1-\epsilon_{12}} \leq 1 \right\} \label{eq:bec+fb_cap1} \\
&\mc C_2 = \left\{ (R_1,R_2) \geq 0: \frac{R_1}{1-\epsilon_{12}} + \frac{R_2}{1-\epsilon_2} \leq 1 \right\}, \label{eq:bec+fb_cap2} 
\end{align}
where $\epsilon_{12}$ is the probability of erasure at both receivers. In particular, feedback increases the capacity. This is of great practical interest  since the required feedback per packet is a low rate (single-bit) ACK/NACK signal. This result has also been extended to several special cases of BPECs  with more receivers %
in \cite{6522177,gatzianas2012feedback,wang2012capacity}. In a similar line of work, the capacity region of two-receiver multiple-input BPECs with feedback has been studied in \cite{6847153} where the capacity region is derived and is shown to be achievable using linear network codes. %

In all the aforementioned works, feedback increases the capacity by providing coding opportunities at the transmitter: Feedback allows the transmitter to track successful packet transmissions for each receiver. Successful packet transmissions can act as receiver side information that can be exploited by index coding \cite{bar2011index, birk2006coding}. Instead of re-transmitting previously lost packets, one can now send linear combinations of packets to simultaneously serve multiple receivers. 
Exploiting receiver side information through index coding has also been applied in the context of information theoretic caching \cite{maddah2014fundamental} recently.

In most previous works, the broadcast channel is assumed to be memoryless. However, many practical communication channels have memory (e.g., see \cite{lutz1991land,lutz1996markov, fontan2001statistical,ibnkahla2004high} that model the channel for satellite links). 
When channels have memory, the role of feedback is dual: (i) to track packet reception, and (ii) to provide information about the channel state.
We model channel memory by a finite-state machine and a set of state-dependent erasure probabilities. 
Finite-state channel models are a common modeling approach for wireless communication channels, see e.g. \cite[Chapter 4.6]{gallager1968information} or \cite{sadeghi2008finite} and the references therein. Using a finite-state Markov channel model, \cite{Viswanathan99} characterizes the capacity of point-to-point time-varying channels with memory and delayed feedback. The feedback-capacity of point-to-point channels with a general form of memory is characterized using infinite letter characterization in \cite{TatikondaMitter09, PermuterWeissmanGoldsmith09, Kim08}. In particular, it is shown that feedback does not increase the capacity if the channel state information is available at both the transmitter and receiver.

The main result of this paper is the  feedback capacity of two-receiver BPECs with hidden and observable memory. More precisely, we study the following two scenarios: 
\begin{enumerate}
\item the transmitter has strictly causal knowledge of the channel state and receives ACK/NACK feedback from both receivers (visible state). 
\item the transmitter receives ACK/NACK feedback from the receivers, but does not receive channel state feedback; i.e., the state evolves according to a hidden Markov model from the transmitter's perspective (hidden state).
\end{enumerate}
In both scenarios, we derive inner and outer bounds on the capacity region, show that the bounds match, and propose optimal achievable schemes that are of low complexity.
We analyze the special case of memoryless finite-state channels separately.

There is prior art deadling with scenarios related to this work:
In the model of \cite{pantelidou2009cross}, the authors deal with correlated broadcast erasure channels but have the \emph{current} channel state (or an estimate of it) available for the current transmission.
Similarly, in \cite{neely2005dynamic, tassiulas1997scheduling} the current channel state is available at the transmitter and coding is not permitted.
In \cite{li2011exploiting, li2013network} the authors focus on obtaining channel state information in a scenario that is related to the case of hidden states, also without permitting coding operations.
Similarly, \cite{ying2011throughput} investigates the case of delayed channel state information for general networks, without permitting coding operations.

In more detail, our contributions are as follows:
\begin{itemize}
\item For both visible and hidden state, we derive tight outer bounds on the capacity region. Our bounds cannot be derived directly using the results of \cite{gamal1978feedback, bergmans1973random, gallager1974capacity} which form the base argument in all previous works. One of the novelties in our work is that  the outer bounds are formulated in terms of linear programs where the solutions guide the design of optimal achievable schemes. The outer bounds, together with the achievable schemes, establish the capacity of two-user BPECs with channel memory and feedback in both scenarios.
\item We devise several optimal and sub-optimal feedback-based coding algorithms and discuss their advantages and disadvantages. Our schemes employ network coding and utilize a virtual network of queues at the transmitter to track the sent packets, as introduced in \cite{georgiadis2009broadcast} and \cite{gatzianas2012feedback}. Analyzing such algorithms and the corresponding flow of packets on these networks is cumbersome even for the memoryless case. In order to simplify the analysis and hence make it applicable to channels with memory, we  develop  a class of probabilistic schemes in both scenarios. In this class, every appropriate coding action is performed with a probability that depends on the previous channel state and feedback. The analysis of probabilistic schemes turns out to be possible using the max-flow min-cut framework. We discuss the optimality of the probabilistic schemes both from an information theoretic perspective (i.e. achieving capacity) and a queueing perspective (i.e. rate stability).
\item While the proposed probabilistic schemes are rate-optimal and easy to analyze, they have several disadvantages from a networking perspective: Their design is dependent on the targeted rate of communication, and they may not perform well for dynamic packet arrivals or varying erasure probabilities. 
We thus propose low complexity and deterministic algorithms that track the packets in the virtual network of queues and choose appropriate coding actions to maximize certain weight function at each time instance. Our proposed achievable schemes extend the max-weight queue-based algorithms of \cite{georgiadis2009broadcast,gatzianas2012feedback, neely2010stochastic}. To analyze our schemes, we use and extend tools from finite-horizon Lyapunov drift analysis to incorporate knowledge about past channel states.
\item We show that simpler coding schemes are sufficient to achieve capacity for finite-state \emph{memoryless} channels, which are a special case of the results above. Further, all above capacity results carry over to delayed feedback.
\item The joint treatment of the topic in both information-theoretic and queueing-theoretic frameworks establish relations between the two fields. 

In particular, we specify the relationship between the outer bound parameters with the injection probabilities in the queueing algorithms (see \eqref{eq:proba_relation_xs}, \eqref{eq:proba_relation_ys} and Remark~\ref{rem:mapping}). This is a step forward towards understanding the interaction of both fields which is currently missing from the literature as discussed in \cite{ephremides1998information}.

\end{itemize}

In a related work that was carried out independently and in parallel by Kuo and Wang \cite{Kuo_Wang2014} (see \cite{kuo2017robust} for the long version), it is shown that some of the coding operations proposed in \cite{6847153} are also useful for BPECs with memory. 
An achievable rate region with opportunistic scheduling is characterized if the \emph{current} channel state information is known at the transmitter. Our results and tools are, however, more general as they distinguish the visible and hidden cases, are complemented by matching outer bounds, and apply to systems with delayed feedback. All our inner and outer bounds apply to channels with memory whereas \cite{Kuo_Wang2014} and \cite{kuo2017robust} assume i.i.d. channels in their analysis and only provide numerical studies of channels with memory.
We note that our achievable coding algorithms (both the probabilistic and deterministic frameworks) use different queueing-theoretic frameworks than \cite{kuo2017robust}. %

Finally, we remark that Dabora and Goldsmith  studied  general broadcast channels with feedback and memory and considered different cooperation scenarios in \cite{dabora2010capacity}. Capacity statements were derived only for partial cooperation among receivers.  The feedback capacity of finite-state point-to-point channels was determined in \cite{yang2005feedback}.

The rest of the paper is organized as follows:
We introduce the notation in Section~\ref{sec:notation} and describe the system models  for channels with observable memory and channels with hidden memory  in Section~\ref{sec:model}. Our main results are summarized in Section~\ref{sec:main}. In Section \ref{sec:outer_bounds}, we derive outer bounds on the capacity. In Section~\ref{sec:new_ach} we present probabilistic coding schemes and deterministic coding algorithms that achieve the capacity. We introduce some simpler, but suboptimal, coding schemes in Section~\ref{sec:ach} and discuss their advantages with respect to optimal schemes. Section~\ref{sec:memoryless} focuses on the special case of finite-state memoryless channels.
The numerical performance of the proposed schemes and  the role of delayed feedback are discussed in Section~\ref{sec:comparison}. We conclude in Section~\ref{sec:concl_becfb}. Proofs of the theorems can be found in the appendices.

\section{Notation}
\label{sec:notation}

Random variables (RVs) are denoted by upper-case letters, e.g. $X$ and their realizations by lower-case letters, e.g. $x$. 
The probability of an RV $X$ taking on the value $x$ given an event $\mc E$ is written as $\Pr[X=x|\mc E]$. Often, the conditional event corresponds to another RV $Y$ taking on some value $y$. This conditional probability is written as $\Pr[X=x|Y=y]$ or equivalently $P_{X|Y}(x|y)$. If the involved RVs are clear from the context, we often write $P(x|y)$ for $P_{X|Y}(x|y)$.
The equivalent expressions $\Pr[X|Y]$ and $P_{X|Y}$ are used to address the conditional probability distribution for any outcome of the RVs. 

The RVs $X,Y,Z$ form a Markov chain if the joint probability mass function (PMF) $P_{XYZ}$ can be written as $P_{XYZ}= P_X P_{Y|X} P_{Z|Y}$. We write this relationship as $X-Y-Z$. 
The conditional expectation of a function $f$ of an RV $X$ given another RV $Z$ is itself an RV and is written as $\mathbb E[f(X)|Z]$. Using the law of total expectation, we have $\mathbb E[f(X)|Z] = \mathbb E\big[\mathbb E[f(X)|YZ] \big| Z \big]$. If $X-Y-Z$ forms a Markov chain, we can write $\mathbb E[f(X)|Z] = \mathbb E\big[\mathbb E[f(X)|Y] \big| Z \big]$.

A finite sequence (or string) of RVs $X_1,X_2,\ldots,X_n$ is denoted by $X^n$. Often, but not always, this refers to a sequence in time. We write $X_t^n$ if the sequence starts at index $t$, i.e. $X_t^n = X_{t}, X_{t+1}, \ldots, X_n$.
Sequences may have subscripts, e.g. $X_j^n$ denotes $X_{j,1},X_{j,2},\ldots,X_{j,n}$. 
This results in one possible ambiguity as $X_j^n$ could also denote $X_j, X_{j+1}, \ldots, X_n$, but in these cases, the meaning will be clear from the context.
Vectors are written with underlined letters and can be used as an alternative to the sequence notation, e.g. $\ve X = X^n$.
These two concepts can be mixed, e.g. $\ve X^n$ denotes the sequence of vectors $\ve X_1, \ldots, \ve X_n$.
Matrices are written boldface, e.g. $\mat A$.

The entropy of a discrete or continuous RV $X$ is written as $H(X)$. Similarly, $H(X|Y)$ denotes the conditional entropy of $X$ given$Y$.
The mutual information between $X$ and $Y$ is denoted by $I(X;Y)$ and their conditional mutual information given $Z$ is written as $I(X;Y|Z)$.

Sets are denoted by calligraphic letters, e.g. $\mc X$. 
The cardinality of the finite set $\mc X$ is denoted by $|\mc X|$. If $\mc X$ is a subset of a larger set $\mc Y$, the complement of $\mc X$ is written as $\mc X^c$ where $\mc X^c = \mc Y \diagdown \mc X$. $\mathbb F_q$ denotes the finite field of order $q$.
The indicator function $\mbox{\textbb{1}}\{\cdot \}$ takes on the value $1$ if the event inside the brackets is true and $0$ otherwise. 
The expression $[a]^+$ is a shorthand notation for $\max(a,0)$.

\section{System Model}
\label{sec:model}
\begin{figure}[t]
\centering
\begin{tikzpicture}[node distance=2mm, scale=0.7, every node/.style={scale=0.7}]

\node at (1.5,0) [block] (Enc) {Encoder};
\node [chan, right = 30mm of Enc.east] (channel) {$P_{Y_{1,t} Y_{2,t}|X_t S_t}$};

\node [block, above right = 10mm of channel.north east] (Dec1) {Decoder 1};
\node [below right=-1mm of Dec1.south east] (Rx1) {Rx$_1$};
\node [right = 4mm of Dec1.east] (W1hat) {$\hat W_1$};

\node [block, below right = 10mm of channel.south east] (Dec2) {Decoder 2};
\node [above right=-1mm of Dec2.north east] (Rx2) {Rx$_2$};
\node [right = 4mm of Dec2.east] (W2hat) {$\hat W_2$};

\node [block2, left = 25mm of Dec1.west] (D1) {$d=1$};
\node [block2, below= 40mm of D1] (D2) {$d=1$};

\node [block2, below= 25mm of D1] (D3) {$d=1$};

\node at (Enc.north west) (Anch1) {};
\node at (Enc.south west) (Anch2) {};
\node at (channel.north east) (Anch3) {};
\node at (channel.south east) (Anch4) {};

\node (W1) [left= 5mm of Anch1.south] {$W_1$};
\node (W2) [left= 5mm of Anch2.north] {$W_2$};

\node at (Enc.south east) (Anch5) {};

\node [chan, below = 15mm of channel] (state_chan) {$P_{S_t|S_{t-1}}$};

\draw[->] (W1) -- (Anch1.south);
\draw[->] (W2) -- (Anch2.north);
\draw[->] (Enc) -- node[above](X){$X_t$} (channel);
\draw[->] (Anch3.south) -| node[left,pos=0.8] (Y1) {$Y_{1,t}$} (Dec1);
\draw[->] (Anch4.north) -| node[left,pos=0.8] (Y2) {$Y_{2,t}$} (Dec2);
\draw[->] (Dec1.west) -- node[above,pos =0.2] (Z1) {$Z_{1,t}$} (D1);
\draw[->] (Dec2.south) |- node[above,pos =0.7] (Z2) {$Z_{2,t}$}  (D2);
\draw[->] (D1) -| node[above,pos =0.2] (Z1del) {$Z_{1,t-1}$} (Enc);
\draw[->] (D2) -| node[above,pos =0.2] (Z2del) {$Z_{2,t-1}$} (Enc);
\draw[->] (Dec1) -- (W1hat);
\draw[->] (Dec2) -- (W2hat);

\draw[->] (state_chan) -- node[right] (S) {$S_{t}$} (channel.south);
\draw[->] (S.west) |- (D3);

\node [draw, circle, inner sep = .5mm, left=10mm of D3] (circle1) {};
\node [draw, circle, inner sep = .5mm, left=15mm of D3] (circle2) {};
\node (circle2north) [above = 3mm of circle2.north] {};
\draw (circle1) -- (circle2north);

\draw (D3) -- node[above,pos =0.5] (Sdel) {$S_{t-1}$} (circle1);
\draw[->] (circle2) -| (Anch5.west);
\draw[->] (Sdel) |- (state_chan);

\end{tikzpicture}
\caption{Block diagram for the BPEC. Depending on whether the switch is closed or not, we call the state visible or hidden. The box marked with $d=1$ represents a delay of one time unit.}
\label{fig:block_bec}
\end{figure}
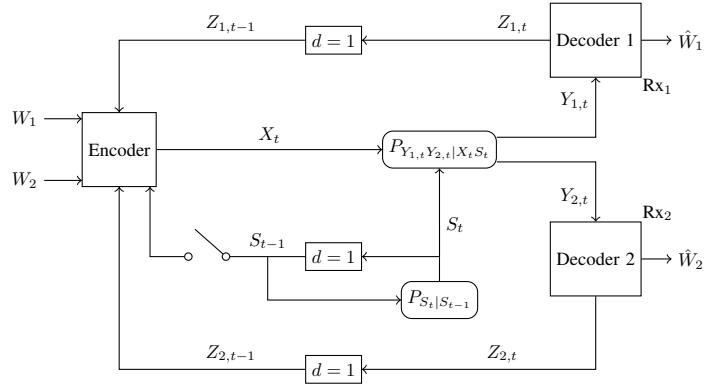

A transmitter communicates two independent messages $W_1$ and $W_2$ (of $nR_1$, $nR_2$ packets, respectively) to two receivers $\text{Rx}_1$ and $\text{Rx}_2$ over $n$ uses of the channel. Communication takes place over a BPEC with memory and feedback as described next.

The input to the broadcast channel  at time $t$, $t=1,\ldots,n$, is denoted by $X_t \in \mc X$.
The channel inputs correspond to packets of $\ell$ bits; we may represent this by choosing $\mc X = \mathbb{F}_{2^\ell}$ with $\ell \gg 1$. The packet length $\ell$ is assumed to be long enough such that the rate loss due to headers is negligible. We further comment on that in Remark~\ref{rm:packet_length}.
Transmission rates are measured in packets per slot, hence all entropies and mutual information terms are considered with logarithms to the base $2^\ell$. 

The channel outputs at time $t$ are written as $Y_{1,t}\in\mathcal{Y}$ and $Y_{2,t}\in\mathcal{Y}$, where $\mathcal{Y}=\mc X \cup \{?\}$.  The output $Y_{j,t}$, $j\in\{1,2\}$, is either $X_t$ (i.e., received perfectly) or $?$ (i.e., erased).

We define binary random variables $Z_{j,t}$, $j\in\{1,2\}$, $t=1,\ldots,n$, to indicate if an erasure occurred at receiver $j$ in time $t$; i.e. $Z_{j,t}=\mbox{\textbb{1}}\{Y_{j,t}=?\}$.
Clearly, $Y_{j,t}$ can be expressed as a function of $X_{t}$ and $Z_{j,t}$, and $Y_{j,t}$ determines $Z_{j,t}$. We denote $(Z_{1,t},Z_{2,t})$ by $\ve Z_t$ and $(Y_{1,t},Y_{2,t})$ by $\ve Y_t$.

The broadcast channel we study has memory that is modeled via a finite-state machine with state $S_t$ at time $t$. The state evolves according to an irreducible aperiodic time-invariant finite-state Markov chain with state space $\mc S$ and steady-state distribution $\pi_s$, $s \in \mc S$. 
We assume that the initial state $S_0$ is distributed according to $\pi$, hence $S^n$ is stationary.
Depending on the current random state of the channel, the erasure probabilities are specified through
the conditional distribution $P_{\ve Z_{t}|S_{t}}$. Note that spatial correlation among the receivers is permitted.
The transition probabilities between channel states are known at the transmitter.
The sequence $\ve Z^n$ is correlated in time in general, hence the channel has memory.
Both $S^n$ and $\ve Z^n$ are ergodic.

After each transmission, an ACK/NACK feedback is available at the encoder from both receivers. 
System-wide knowledge of the feedback facilitates the mathematical treatment of the queueing problem in Section~\ref{sec:new_ach}. 
To realize this, we require a reliable low-rate forward channel from the transmitter to both receivers to inform them about packet losses of the other receiver.
The transmitter uses the feedback information to encode its messages $W_1$, $W_2$. Two setups are considered: 
\begin{itemize}
 \item[(i)] ACK/NACK feedback and previous state information are available at the encoder:
\begin{align}
X_t = f_t(W_1,W_2,\ve Z^{t-1},S^{t-1}).
\end{align}
 \item[(ii)] Only ACK/NACK feedback is available at the encoder:
\begin{align}
X_t = f_t(W_1,W_2,\ve Z^{t-1}).
\end{align}

\end{itemize}

Depending on whether the transmitter knows the previous channel state or not, we call the state \emph{visible} or \emph{hidden} (see also Fig.~\ref{fig:block_bec}). 
When the state is visible, the joint probability mass function of the system factorizes as
\begin{align}
&P_{W_1 W_2 X^n S^n Y_1^n Y_2^n Z_1^n Z_2^n}= 
P_{W_1} P_{W_2} P_{S_0} \left( \prod_{t=1}^n P_{S_t|S_{t-1}} P_{X_t|W_1 W_2 S^{t-1} \ve Z^{t-1} } P_{Z_{1,t}Z_{2,t}|S_t} P_{Y_{1,t}|X_t Z_{1,t}} P_{Y_{2,t}|X_t Z_{2,t}} \right). \nonumber
\end{align}
When the state is hidden, the joint probability mass function of the system factorizes as
\begin{align}
&P_{W_1 W_2 X^n S^n Y_1^n Y_2^n Z_1^n Z_2^n}= 
P_{W_1} P_{W_2} P_{S_0} \left( \prod_{t=1}^n P_{S_t|S_{t-1}} P_{X_t|W_1 W_2 \ve Z^{t-1} }  P_{Z_{1,t} Z_{2,t}|S_t} P_{Y_{1,t}|X_t Z_{1,t}} P_{Y_{2,t}|X_t Z_{2,t}}\right). \nonumber
\end{align}

The state may be visible either because it is explicitly available at the transmitter or because it can be determined from the available feedback. The latter is illustrated via the following example. 
\begin{example}
\label{exampleGilbert}
Consider a Gilbert-Elliot model \cite{gilbert1960capacity,elliott1963estimates} with state space  $\mc S = \{\text{GG},\text{GB}, \text{BG}, \text{BB}\}$ where $\text{G}$ and $\text{B}$ refer to a good and bad state at each user. Suppose that the channel erases the input in state $\text{B}$ and is erasure free in state $\text{G}$; i.e., we have
\begin{align}
&P_{\ve Z_t|S_t}(0,0|\text{GG})=1,\quad P_{\ve Z_t|S_t}(0,1|\text{GB})=1\nonumber\\
&P_{\ve Z_t|S_t}(1,0|\text{BG})=1,\quad P_{\ve Z_t|S_t}(1,1|\text{BB})=1. \label{eq:param_gilbert}
\end{align}
In such a channel, the feedback $\ve Z_t$ determines the channel state $S_t$, hence $H(S_t|\ve Z_t)=0$, and we thus say that the state is visible.
If $\ve Z_t$ does not determine $S_t$, the setup has hidden state, unless $S_t$ is fed back separately. 
We use a Gilbert-Elliot channel model for our simulations in Section~\ref{sec:comparison}.
\end{example}

Given the feedback  and the previous channel state information, the encoder may calculate the statistics of the next channel erasure events.
In the visible case,  the probabilities of erasure events given the \emph{previous} channel state $s$ are given by 
\begin{align}
 \epsonetwo{s}&\ensuremath{r}iangleq P_{\ve Z_t|S_{t-1}}(1,1|s),\quad
 &\epsonenottwo{s}&\ensuremath{r}iangleq P_{\ve Z_t|S_{t-1}}(1,0|s),\nonumber\\
 \epsilon_{\bar 12}(s)&\ensuremath{r}iangleq P_{\ve Z_t|S_{t-1}}(0,1|s),\quad  
  &\epsnotonenottwo{s} &\ensuremath{r}iangleq P_{\ve Z_t|S_{t-1}}(0,0|s),\nonumber\\
   \epsone{s}&\ensuremath{r}iangleq\epsonetwo{s}+\epsonenottwo{s},\quad 
  &\epstwo{s}&\ensuremath{r}iangleq\epsonetwo{s}+\epsnotonetwo{s}. \label{eq:def_epsilon}
\end{align}
Note that these probabilities do not depend on $t$ in our setup.
In the hidden case, the probabilities of erasure events given the past feedback sequence $\ve z^{t-1}$ are given by
\begin{align}
 \epsonetwo{\ve z^{t-1}}&\ensuremath{r}iangleq P_{\ve Z_t|\ve Z^{t-1}}(1,1|\ve z^{t-1}),\quad
 &\epsonenottwo{\ve z^{t-1}}&\ensuremath{r}iangleq P_{\ve Z_t|\ve Z^{t-1}}(1,0|\ve z^{t-1}),\nonumber\\
 \epsilon_{\bar 12}(\ve z^{t-1})&\ensuremath{r}iangleq P_{\ve Z_t|\ve Z^{t-1}}(0,1|\ve z^{t-1}),\quad  
  &\epsnotonenottwo{\ve z^{t-1}} &\ensuremath{r}iangleq P_{\ve Z_t|\ve Z^{t-1}}(0,0|\ve z^{t-1}),\nonumber\\
   \epsone{\ve z^{t-1}}&\ensuremath{r}iangleq\epsonetwo{\ve z^{t-1}}+\epsonenottwo{\ve z^{t-1}},\quad 
  &\epstwo{\ve z^{t-1}}&\ensuremath{r}iangleq\epsonetwo{\ve z^{t-1}}+\epsnotonetwo{\ve z^{t-1}}. \label{eq:def_epsilon_hmm}
\end{align}
In the hidden case, these probabilities depend on $t$.
We use the notations $\epsone{\ve z^L}$, $\epstwo{\ve z^L}$, $\epsonetwo{\ve z^L}$, etc. to refer to the erasure probabilities when they are computed with respect to the distribution $P_{\ve Z_t|\ve Z^{t-1}_{t-L}}$, i.e., based on the past $L$ feedback samples $\ve Z^{t-1}_{t-L} = \ve z^L$ rather than the entire past feedback $\ve z^{t-1}$.

The goal is to have each decoder $\text{Rx}_j$ reliably estimate $\hat W_j = h_j(Y_j^n)$ from its received sequence $Y_j^n$.
\begin{definition}
\label{def:cap_region_becfb}
The packet rate pair $(R_1,R_2)$ is said to be \emph{achievable} if the error probability  $\Pr[\hat W_1 \neq W_1, \hat{W}_2\neq W_2]$ can be made arbitrarily small as $n$ gets large.
The \emph{capacity region} is the closure of the set of achievable rate pairs. For the visible case, we denote the capacity region by $\mc C_{\text{ fb}+s}^\text{mem}$ and for the hidden case, we denote it by  $\mc C_{\text{ fb}}^\text{mem}$.

 \end{definition}

The classic formulation assumes operation in blocks of $n$ slots, where $nR_1$, $nR_2$ packets arrive just before the block being handled. In Section~\ref{sec:new_ach} we consider a dynamic version of the problem, which better models practical packet communication networks:
Packets for \textnormal{Rx$_1$}\, \textnormal{Rx$_2$}\ arrive in each slot with probability $R_1$, $R_2$, respectively, yielding in total $nR_1$, $nR_2$ packets during $n \gg 1$ slots, as is the block-based scheme (see also Remark~\ref{remark:stability}). 
This dynamic version is modeled and analyzed with a network of queues that captures different coding operations and tracks the average rate of delivered packets per slot.

For the dynamic version of the problem, we would like to have all the queues in the network \emph{stable} according to the following two stability definitions (e.g. \cite{neely2010stability}, \cite[Definition 3.1]{georgiadis2006resource}):
\begin{definition}
\label{def:stability}
Let $Q_t$ denote the number of packets in the corresponding queue $Q$ at time $t$.
Queue $Q$ is \emph{rate-stable}~if 
\begin{align}
 \lim_{n\rightarrow \infty} \frac{Q_n}{n} = 0  \qquad \text{ with probability 1.} \label{eq:rate_stability}
\end{align}
Queue $Q$ is \emph{strongly stable} if
\begin{align}
 \limsup_{n\rightarrow \infty} \frac{1}{n} \sum_{t=1}^n \mathbb{E}[Q_t] < \infty. \label{eq:strong_stability}
\end{align}
A network of queues is stable if all queues inside the network are stable \cite[Definition 3.2]{georgiadis2006resource}.
The \emph{network stability region} consists of all rate pairs $(R_1,R_2)$ for which all queues in the network  
are stable. 
 \end{definition}

For the setup in Section~\ref{sec:new_ach}, we will propose coding algorithms that strongly stabilize the queues. We remark that strong stability implies rate stability \cite[Theorem 4]{neely2010stability} in this setup, but not vice versa.

\begin{remark}
\textit{Capacity Regions vs. Stability Regions:}
\label{remark:stability}
Capacity regions are defined via block-coding and decoding error probability as a criterion, 
as in Definition~\ref{def:cap_region_becfb}.
Queue stability deals with dynamic packet arrivals (instead of blocks of packets) and measures if the number of packets in the queues grows sublinearly, as in Definition~\ref{def:stability}. Stability regions can be cast into  corresponding block-based rate regions as follows:
Suppose that a network of queues  is rate-stable for $(R_1,R_2)$; i.e., each queue satisfies \begin{align}
 Q_n = o(n) \qquad \text{ with probability 1}. \label{eq:rate_stability_landau}
\end{align}
In $n$ channel uses, $nR_1 \pm o(n)$ packets arrive for \textnormal{Rx$_1$}\ and $nR_2 \pm o(n)$ packets arrive for \textnormal{Rx$_2$}\ with high probability for $n$ large enough.
Rate stability implies (see \eqref{eq:rate_stability_landau}) that there are only $o(n)$ packets left in the queues after $n$ channel uses. These remaining packets can be delivered in 
$o(n)$ slots. Hence, the total number of slots needed to deliver  $nR_1 \pm o(n)$ packets for user \textnormal{Rx$_1$}\ and $nR_2 \pm o(n)$ packets for \textnormal{Rx$_2$}\ is given by $n+o(n)$. The number of slots of the cleanup phase is negligible compared to $n$, for $n$ getting large, and $(R_1,R_2)$ is achievable in the block-based communication sense.

\end{remark}

 \section{Main Results}
 \label{sec:main}
Our main results are in the form of matching outer and inner bounds on the capacity region of the two-receiver BPEC with memory and ACK/NACK feedback. We also devise low-complexity feedback-based coding algorithms that achieve capacity. We consider the visible and hidden cases and elaborate on the special case of finite-state memoryless channels separately.
\subsection{Visible Case}
Define $\bar{\mc C}_{\text{ fb}+s}^\text{mem}$ as the closure of rate pairs $(R_1,R_2)$ for which there exist variables $x_s$, $y_s$, $s\in \mc S$ such that
 \begin{align}
 &0\leq x_s\leq 1,\quad 0\leq y_s \leq 1, \qquad \forall~s\in \mc S \label{eq:posouter}\\
  &R_1  \leq \sum_{s \in \mc S} \pi_s (1-\epsone{s}) x_s  \label{eq:R1_constr1outer}\\
  &R_1  \leq \sum_{s \in \mc S} \pi_s (1- \epsonetwo{s}) (1-y_s) \label{eq:R1_constr2outer} \\
  &R_2  \leq \sum_{s \in \mc S} \pi_s (1-\epstwo{s}) y_s  \label{eq:R2_constr1outer} \\
  &R_2  \leq \sum_{s \in \mc S} \pi_s (1-\epsonetwo{s}) (1-x_s).  \label{eq:R2_constr2outer}
 \end{align}
We show in Section~\ref{sec:outer_bounds} that $\bar{\mc C}_{\text{ fb}+s}^\text{mem}$ is an outer bound on the capacity region ${\mc C}_{\text{ fb}+s}^\text{mem}$ in the visible case. Moreover, we show in Section~\ref{sec:new_ach}  that the region $\bar{\mc C}_{\text{ fb}+s}^\text{mem}$ is achievable, leading to the following theorem. 
\begin{theorem}
\label{thm:capacity}
The capacity region of the BPEC with feedback, memory, and visible state is 
 \begin{align}
   \mc C_{\text{ fb}+s}^\text{mem} = \bar{\mc C}_{\text{ fb}+s}^\text{mem}.
 \end{align}
\end{theorem}

In our achievable schemes, the transmitter uses the feedback from both receivers to track successful packet transmissions. This is done with the help of a virtual network of queues. Each queue essentially represents a coding opportunity that the transmitter can exploit.
Two capacity achieving schemes  are presented in Section~\ref{sec:new_ach}: 
In the first scheme, the transmitter randomly chooses the coding operation according to a probability distribution that depends on the observed state feedback. The optimization of the corresponding probability distribution results in a single-commodity flow problem in the network of virtual queues. The min-cut version of the max-flow problem can be shown to match the outer bound, proving Theorem~\ref{thm:capacity}.
We show how the parameters $x_s$, $y_s$,  $s\in \mc S$ that appear in the outer bound are related to transmission probabilities of the encoding operations.
The second scheme is a queue-based max-weight backpressure algorithm that operates on the network of queues. This algorithm strongly stabilizes all queues in the network for all rates inside the capacity region.

\subsection{Hidden Case}
Define $\bar{\mc C}_{\text{ fb}}^\text{mem}(L)$, for every integer $L$, as the closure of rate pairs $(R_1,R_2)$ for which there exist parameters $x(\ve z^L)$, $y(\ve z^L)$, $\ve z^L\in\{0,1\}^{2L}$, such that
\begin{align}
0&\leq x(\ve z^L)\leq 1, \quad 0\leq y(\ve z^L)\leq 1 \qquad \forall~\ve z^L\in\{0,1\}^{2L} \label{eq:posouter_hmm_wind_outer_main}\\
R_1&\leq \sum_{\ve z^L} P_{\ve Z^L}(\ve z^L) (1-\epsone{\ve z^L}) x(\ve z^L)  \label{eq:R1_constr1outer_hmm_wind_outer_main}\\
R_1&\leq \sum_{\ve z^L} P_{\ve Z^L}(\ve z^L) (1-\epsonetwo{\ve z^L}) (1-y(\ve z^L))  \label{eq:R1_constr2outer_hmm_wind_outer_main} \\
R_2&\leq \sum_{\ve z^L} P_{\ve Z^L}(\ve z^L) (1-\epstwo{\ve z^L}) y(\ve z^L) \label{eq:R2_constr1outer_hmm_wind_outer_main}\\
R_2&\leq \sum_{\ve z^L} P_{\ve Z^L}(\ve z^L) (1-\epsonetwo{\ve z^L}) (1-x(\ve z^L)). \label{eq:R2_constr2outer_hmm_wind_outer_main}
\end{align}

\begin{theorem}
\label{thm:capacity_hmm}
Given a BPEC with feedback, memory and hidden state\footnote{under the assumption that $P_{\ve Z_t|S_t}$ contains only strictly positive values, as stated in Remark~\ref{rm:exponential_forgetting}.}, the rate region $\bar{\mc C}_{\text{ fb}}^\text{mem}(L)$ is achievable and converges to the capacity region exponentially fast in $L$. More precisely, (i) any rate pair $(R_1,R_2)\in \bar{\mc C}_{\text{ fb}}^\text{mem}(L)$ is achievable and (ii) any rate pair $(R_1,R_2)\in \mc C_{\text{ fb}}^\text{mem}$ is such that 
\begin{align}
(R_1-C(L),R_2-C(L))\in \bar{\mc C}_{\text{ fb}}^\text{mem}(L)
\end{align} where
\begin{align}
-2(1-\sigma)^L\leq C(L)\leq 2(1-\sigma)^L, \qquad 0 < \sigma\leq 1.
\end{align}
\end{theorem}

The region $\bar{\mc C}_{\text{ fb}}^\text{mem}(L)$ is an $L$-th order approximation to the capacity region of the hidden case and  has a similar structure to the capacity region in the visible case: The role of channel states in the right hand side (RHS) of 
\eqref{eq:R1_constr1outer}~-~\eqref{eq:R2_constr2outer} is replaced  in \eqref{eq:R1_constr1outer_hmm_wind_outer_main}~-~\eqref{eq:R2_constr2outer_hmm_wind_outer_main} by a window of $L$ previous feedback samples that determine the approximate system state in the hidden case. As the window gets larger, the approximation gets tight exponentially fast in $L$.%

This similarity allows us to modify the probabilistic and deterministic schemes that we developed for the visible case to be applicable for the hidden case: For the probabilistic scheme, encoding operations are drawn randomly from a distribution that depends on the past window of $L$ feedback sequences, see Section~\ref{sec:probab_scheme_new}. The deterministic scheme allows us to take into account the whole past feedback sequence, as derived in Section~\ref{sec:new_ach_det}.

\subsection{Memoryless Case}
Consider the special case of finite-state \emph{memoryless} broadcast packet erasure channels.
This is an extension of the setup in \cite{georgiadis2009broadcast} to the case of multiple states: The erasure probabilities still depend on the current channel state according to $P_{\ve Z_t|S_t}$, but the state sequence $S^n$ is i.i.d.
Here, strictly causal state feedback does not provide information about future channel statistics, and there is no fundamental difference between the visible and hidden cases.
Theorem~\ref{thm:capacity} includes this setup as a special case.

Define $\underline{\mc C}_{\text{ fb}+s}^\text{mem}$ as the closure of rate pairs $(R_1,R_2)$ for which there exist variables $x_s$, $y_s$, $s\in \mc S$, such that
 \begin{align}
 &0\leq x_s\leq 1,\quad 0\leq y_s \leq 1, \qquad \forall~s\in \mc S \label{eq:pos}\\
 &x_s + y_s \geq 1,~\forall s \in \mc S \label{eq:pa3_pos}\\
  &R_1  \leq \sum_{s \in \mc S} \pi_s (1-\epsone{s}) x_s  \label{eq:R1_constr1}\\
  &R_1  \leq \sum_{s \in \mc S} \pi_s (1- \epsonetwo{s}) (1-y_s) \label{eq:R1_constr2} \\
  &R_2  \leq \sum_{s \in \mc S} \pi_s (1-\epstwo{s}) y_s  \label{eq:R2_constr1} \\
  &R_2  \leq \sum_{s \in \mc S} \pi_s (1-\epsonetwo{s}) (1-x_s).  \label{eq:R2_constr2}
 \end{align}
Note that $\bar{\mc C}_{\text{ fb}+s}^\text{mem}$ and $\underline{\mc C}_{\text{ fb}+s}^\text{mem}$ differ only in the constraints \eqref{eq:pa3_pos}.
We show in Section~\ref{sec:ach} that $\underline{\mc C}_{\text{ fb}+s}^\text{mem}$ is achievable using probabilistic or deterministic schemes that only utilize  ``reactive" coding operations (these coding operations have been used in \cite{georgiadis2009broadcast} for the single-state memoryless case and are a subset of the coding operations that we utilize  to achieve capacity in Theorem \eqref{thm:capacity}).
In general, the additional constraints in \eqref{eq:pa3_pos} make $\underline{\mc C}_{\text{ fb}+s}^\text{mem}$ strictly smaller than ${\mc C}_{\text{ fb}+s}^\text{mem}$. However, for special cases such as the memoryless case, the two regions coincide and reactive coding schemes are capacity achieving.
In Section~\ref{sec:memoryless}, we derive the following theorem:
\begin{theorem}
\label{thm:capacity_memoryless}
The capacity region of the memoryless finite-state BPEC with feedback is given by $\underline{\mc C}_{\text{ fb}+s}^\text{mem}$. 
\end{theorem}

\begin{figure}[t]
\centering

\begin{tikzpicture}
\begin{axis}[scale=0.6,
width=4.82222222222222in,
height=3.80333333333333in,
scale only axis,
xmin=0,
xmax=0.41,
xlabel={$R_1$},
xmajorgrids,
every outer x axis line/.append style={black},
every x tick label/.append style={font=\color{black}},
ymin=0,
ymax=0.51,
ylabel={$R_2$},
ymajorgrids,
every outer y axis line/.append style={black},
every y tick label/.append style={font=\color{black}},
axis x line*=bottom,
axis y line*=left,
legend style={at={(0.0,0.0)},anchor=south west,draw=black,fill=white,legend cell align=left}
]

\addplot [color=TUMred,solid,line width=1.0pt,mark=o,mark options={solid}]
  table[row sep=crcr]{%
-8.81222178561458e-14	0.499999999999775\\
1.26634813746307e-16	0.5\\
1.6826817716975e-16	0.5\\
1.37650134346412e-12	0.499999999986179\\
0.15919999999997	0.435999999999808\\
0.159200000000297	0.435999999999662\\
0.285558025788868	0.347327701200767\\
0.285558025789234	0.347327701200483\\
0.285558025789371	0.347327701200178\\
0.339999999999013	0.223200000001448\\
0.340000000000064	0.223199999998756\\
0.399999999998872	2.03494165962326e-12\\
0.399999999999304	1.59171287261728e-13\\
};
\addlegendentry{$\mc C_{\text{fb}+s}^{\text{mem}}$};

\addplot [
color=TUMpantone301,
solid,
line width=1.0pt,
mark=x,
mark options={solid}
]
  table[row sep=crcr]{%
-4.33177799186168e-13	0.49999999999336\\
-4.68340644044218e-14	0.499999999999881\\
1.45716771982052e-16	0.5\\
1.85268467234323e-15	0.5\\
0.159199999999663	0.435999999999975\\
0.159199999999845	0.435999999999896\\
0.23738610757238	0.381132556089542\\
0.28904977375545	0.339366515837234\\
0.28904977375565	0.339366515837072\\
0.34000000000034	0.223199999998375\\
0.399999999998137	6.18755047199215e-13\\
};
\addlegendentry{$\underline{\mc C}_{\text{fb}+s}^{\text{mem}}$ };

\addplot [color=TUMgold,line width=1.0pt,only marks,mark=+,mark options={solid}]
  table[row sep=crcr]{%
0.0485279276409894	0.477902543689433\\
0.0675775089158853	0.467597066797886\\
0.0868765591877614	0.459788489893051\\
0.105030290097195	0.447122437570101\\
0.123431438486446	0.436984030627993\\
0.141618817378049	0.426932104811651\\
0.160470182307521	0.417330116535105\\
0.179280659056012	0.403927767431969\\
0.198092978441382	0.390567530300344\\
0.218159457538116	0.379416072585456\\
0.237106964297982	0.366068194247707\\
0.256071399173872	0.352751247443674\\
0.268761945242891	0.331692562850989\\
0.288445721987262	0.294549605650576\\
0.308426270377336	0.253246006729796\\
0.326708372027144	0.208346596104029\\
0.354017234630685	0.142372336516328\\
0.378032090832259	0.0721168911702604\\
0.0386228068250366	0.479327955929005\\
0.0579268927814279	0.471505016096308\\
0.0772273503386969	0.463691542112984\\
0.0960074390263397	0.453442169621522\\
0.114570650801451	0.443253601377244\\
0.132918265691995	0.433141315379095\\
0.150213751950921	0.420765986246746\\
0.169875565225925	0.410623679861972\\
0.188686208396767	0.397242398649546\\
0.208690480821263	0.386102352466667\\
0.227631404755946	0.372738105426472\\
0.246586761666256	0.359406028701739\\
0.262517736501606	0.342136602637752\\
0.279436480284157	0.316098546496453\\
0.299303489868723	0.27476060369849\\
0.317572002828929	0.231790200641728\\
0.340114188359386	0.17572566398433\\
0.392564426832327	0.0182542458471402\\
0.370451974685304	0.0906632464341769\\
0.346487941293732	0.161116892691255\\
0.321419956021988	0.22280075069033\\
0.302905379591254	0.266253828660008\\
0.281082574379365	0.306267573043392\\
0.265763361736455	0.339868211395611\\
0.249630657612798	0.357269140055145\\
0.230438572808261	0.370761008563712\\
0.211262044830691	0.384285491076112\\
0.191008724256867	0.395593309813115\\
0.171965585225862	0.409134788182862\\
0.152067254037438	0.420017731539743\\
0.13381532307745	0.43009066172866\\
0.115984028943601	0.442680876800518\\
0.0971919728196658	0.452962212464458\\
0.0781803843941054	0.463305944731502\\
0.0586417828112578	0.471215456414804\\
0.0393028377280732	0.481626404221901\\
0.019652807860898	0.489561072701072\\
0.0491269427127516	0.477660770256913\\
0.0684114744299155	0.467259455871441\\
0.0874785468614662	0.456899571074769\\
0.106325907116662	0.446596739261519\\
0.124953977124004	0.43636686783006\\
0.143365999420958	0.426225675405219\\
0.162444211484524	0.415921723216075\\
0.181486830148404	0.402358648528713\\
0.201677955781159	0.391060632678318\\
0.220848692653468	0.377519024754795\\
0.240032346996015	0.36401115976577\\
0.259234120538482	0.35053455860273\\
0.272056566146852	0.328043931207164\\
0.291895030964922	0.286320777468254\\
0.312149029662796	0.244497821775218\\
0.332729312431061	0.19450404949515\\
0.355999093116929	0.125273194385254\\
};
\addlegendentry{Stable points};

\addplot [color=TUMgold,dashed,line width=1.0pt,mark=triangle*,mark options={solid}]
  table[row sep=crcr]{%
-9.49518241810665e-14	0.499999999999836\\
1.13797860024079e-13	0.499999999999924\\
0.151359999999466	0.421600000000271\\
0.151360000000068	0.421599999999944\\
0.255398422237522	0.348002078590374\\
0.255398422237548	0.34800207859035\\
0.255398422238469	0.348002078588362\\
0.399999999999828	-1.90611415540332e-14\\
0.399999999999968	5.44703171456717e-14\\
};
\addlegendentry{$\bar{\mc C}_{\text{fb}}^{\text{mem}}(L=1)$};

\addplot [color=TUMgold,solid,line width=1.0pt,mark=square,mark options={solid}]
  table[row sep=crcr]{%
1.77429956577796e-14	0.499999999999989\\
1.78407906937378e-14	0.499999999999991\\
1.92467095321176e-14	0.499999999999982\\
0.0504057130299491	0.477305349446207\\
0.0823789928127339	0.461503642303656\\
0.143504181562674	0.427528228237805\\
0.21003992004191	0.385788722711234\\
0.259570585390214	0.351986323190014\\
0.259570585390618	0.351986323189703\\
0.28841382210434	0.298483841288673\\
0.386830695075084	0.0441463592797517\\
0.399999999999743	4.62805629323354e-13\\
};
\addlegendentry{$\bar{\mc C}_{\text{fb}}^{\text{mem}}(L=7)$};

\addplot [color=TUMred,dashed,line width=1.0pt]
  table[row sep=crcr]{%
0	0.5\\
0.193103448275862	0.362068965517241\\
0.4	0\\
};
\addlegendentry{$\mc C_\text{fb}$};

\addplot [color=TUMred,dotted,line width=1.0pt]
  table[row sep=crcr]{%
0	0.5\\
0.4	0\\
};
\addlegendentry{$\mc C$};

\end{axis}
\end{tikzpicture}%
\caption{Capacity regions for visible and hidden cases with $\epsilon_1=0.6$, $\epsilon_2=0.5$.
More details on the channel parameters are given in Section~\ref{sec:results_hidden}.
}
\label{fig:GE_HMM}
\end{figure}
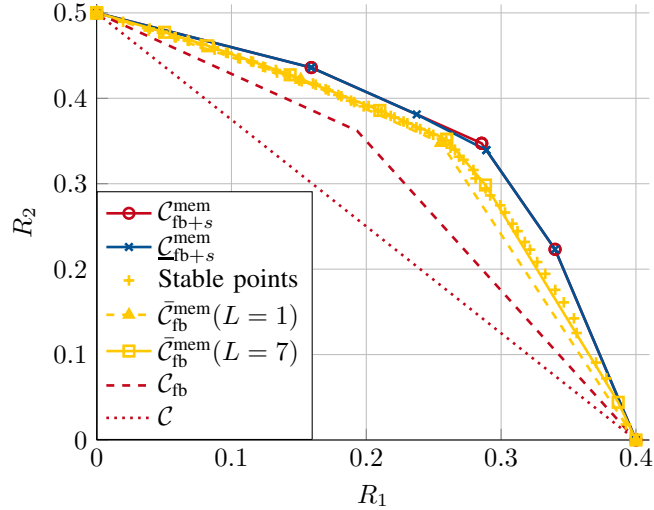

To visualize the main results, Fig.~\ref{fig:GE_HMM} shows the rate regions ${\mc C}_{\text{ fb}+s}^\text{mem}$, $\underline{\mc C}_{\text{ fb}+s}^\text{mem}$, $\mc C_\text{fb}$ and $\mc C$ for an example scenario. Stable rate points are points inside $\bar{\mc C}_{\text{ fb}}^\text{mem}$, and $\bar{\mc C}_{\text{ fb}}^\text{mem}(L)$ show computable approximations of $\bar{\mc C}_{\text{ fb}}^\text{mem}$.
We present further numerical simulations and the role of delayed feedback in Section~\ref{sec:comparison}.

\section{Outer Bounds}
\label{sec:outer_bounds}

\subsection{Visible Case}
\label{sec:proof_bound_visible}
We prove that $\bar{\mc C}_{\text{ fb}+s}^\text{mem}$ is an outer bound on the capacity region. The general idea is to show that for any achievable scheme, there are parameters $x_s,y_s$, $s\in\mathcal{S}$, satisfying  \eqref{eq:posouter} - \eqref{eq:R2_constr2outer}. We find these parameters by relating them to certain mutual information terms that can be found for any given achievable scheme.

In order to bound $R_1$ and $R_2$, for any $\delta>0$, we  start with the following  multi-letter bounds. For $j\in\{1,2\}$, we define $\bar{j} \in \{1,2\}$ such that $\bar{j}\neq j$.
Fano's inequality \cite[Chapter 2.10]{cover2006elements} and the independence of the two messages $W_1$ and $W_2$ lead to the following bounds:
\begin{align}
nR_j&\leq I(W_j;Y_j^n)+n\delta \label{multi1v}\\
nR_j&\leq I(W_j;Y_1^nY_2^n|W_{\bar{j}})+n\delta\label{multi3v}
\end{align}

We next capture the role of channel state feedback in both the bounds above, single-letterize them, and further relate the bounds on $R_j$ and $R_{\bar{j}}$ by establishing a relationship between their corresponding single-letter mutual information terms for each channel state realization.
For $j=1$, the single-letterization is done as follows:
\allowdisplaybreaks
\begin{align}
R_1-\delta&\leq \frac{1}{n}I(W_1;Y_1^n)\nonumber \\
&\leq \frac{1}{n}I(W_1;Y_1^nS^{n-1})\nonumber \\
&=\frac{1}{n}\sum_{t=1}^n I(W_1;Y_{1,t}S_{t-1}|Y_1^{t-1}S^{t-2})\nonumber \\
&=\frac{1}{n}\sum_{t=1}^n \left[I(W_1;S_{t-1}|Y_1^{t-1}Z_1^{t-1}S^{t-2}) + I(W_1;Y_{1,t}|Y_1^{t-1}S^{t-1})\right]\nonumber \\
&\stackrel{(a)}{=}\sum_{t=1}^n \frac{1}{n}I(W_1;Y_{1,t}|Y_1^{t-1}S^{t-1})\nonumber \\
&{\leq}\sum_{t=1}^n \frac{1}{n}I(W_1Y_1^{t-1}S^{t-1};Y_{1,t}|S_{t-1})\nonumber \\
&\stackrel{(b)}{=}\sum_{t=1}^n \frac{1}{n}I(U_{1,t};Y_{1,t}|S_{t-1})\nonumber \\
&\stackrel{(c)}{=} I(U_{1,T};Y_{1,T}|S_{T-1}T)\nonumber \\
&= \sum_{s\in\mathcal{S}}\pi_s I(U_{1,T};Y_{1,T}|T,S_{T-1}=s).\label{no1}
\end{align}
In the above chain of inequalities, $(a)$ follows because $Z_1^{t-1}$ is a function of $Y_1^{t-1}$ and because of the Markov chain
\begin{displaymath}
 W_1-Y_1^{t-1}Z_1^{t-1}S^{t-2}-S_{t-1},
\end{displaymath}
$(b)$ follows by defining $U_{1,t}=(W_1Y_{1}^{t-1}S^{t-1})$, and  $(c)$ follows by a standard random time sharing argument with time sharing random variable $T$. Similarly, one obtains 
\begin{align}
R_1-\delta&\leq \frac{1}{n}I(W_1;Y_1^nY_2^n|W_2)\nonumber\\
&\leq\sum_{s\in\mathcal{S}}\pi_s I(U_{1,T};Y_{1,T}Y_{2,T}|U_{2,T}V_{T}T,S_{T-1}=s),\label{no2}
\end{align}
where $U_{2,T}=(W_2Y_{2}^{T-1}S^{T-1})$ and $V_{T}=(Y_1^{T-1}Y_{2}^{T-1}S^{T-1})$. 
By symmetry, we also have the following bounds:
\begin{align}
R_2-\delta&\leq \sum_{s\in\mathcal{S}}\pi_s I(U_{2,T};Y_{2,T}|T,S_{T-1}=s)\label{no4}\\
R_2-\delta&\leq \sum_{s\in\mathcal{S}}\pi_s I(U_{2,T};Y_{1,T}Y_{2,T}|U_{1,T}V_TT,S_{T-1}=s). \label{no5}
\end{align}
\begin{remark} 
\label{remark1}
Note that by the definitions of $U_{1,T}, U_{2,T}, V_T$,
\begin{itemize}
\item $V_T$ is a  function of $(U_{1,T}U_{2,T})$, and
\item $\underline{Z}_T-TS_{T-1}-U_{1,T}U_{2,T}V_TX_T$ forms a Markov chain.
\end{itemize}
\end{remark}

The following lemma is based on \cite[Lemma 1]{DanaHassibi05}. %
\begin{lemma}
\label{lemma1}
For every $s\in\mathcal{S}$ and $j\in\{1,2\}$, we have:
\begin{alignat}{2}
I(U_{j,T};Y_{j,T}|T,S_{T-1}=s)&= (1-\epsj{s}) I(U_{j,T};X_{T}|T,S_{T-1}=s)\label{eq:lem1eq1}\\
I(U_{j,T};Y_{1,T}Y_{2,T}|U_{\bar{j},T} V_T T,S_{T-1}=s)&= (1-\epsonetwo{s}) I(U_{j,T};X_T|U_{\bar{j},T} V_T T,S_{T-1}=s).\label{eq:lem1eq2}
\end{alignat}
\end{lemma}
Using Lemma~\ref{lemma1}, we now replace the mutual information terms in \eqref{no1}~-~\eqref{no5} to obtain
\begin{align}
&R_j-\delta\leq\sum_{s\in\mathcal{S}}\pi_s (1-\epsj{s}) u^{(j)}_s \quad j=1,2 \label{eq:conv_bound1}\\
&R_j-\delta\leq\sum_{s\in\mathcal{S}}\pi_s (1-\epsonetwo{s}) z^{(j)}_s \quad  j=1,2, \label{eq:conv_bound2}
\end{align}
where $u^{(j)}_s$ and $z^{(j)}_s$ are defined as follows for $j\in\{1,2\}$, $s\in\mathcal{S}$:
\begin{align}
u^{(j)}_s&=I(U_{j,T};X_{T}|T,S_{T-1}=s)\label{defu}\\
z^{(j)}_s&=I(U_{j,T};X_T|U_{\bar{j},T}V_T T,S_{T-1}=s).\label{defz}
\end{align}
The following lemma relates these parameters and is proved in Appendix~\ref{appendix:lemma2}.
\begin{lemma}
\label{lemma2}
For every $j\in\{1,2\}$ and $s\in\mathcal{S}$, we have
\begin{align*}
&u^{(j)}_s+z^{(\bar{j})}_s\leq 1.
\end{align*}
\end{lemma}
Combining the above results and letting $\delta$ go to zero, $(R_1,R_2)$ can be achieved only if there exist variables $u_s^{(1)}$, $u_s^{(2)}$, $z_s^{(1)}$, $z_s^{(2)}$, $s\in\mathcal{S}$, for which the following inequalities hold for all $j\in\{1,2\}$:
\begin{align}
&0\leq u^{(j)}_s, z^{(j)}_s\leq 1 \\
&u^{(j)}_s+z^{(\bar{j})}_s\leq 1 \label{tight}\\
&R_j\leq\sum_{s\in\mathcal{S}}\pi_s (1-\epsj{s}) u^{(j)}_s \\
&R_j\leq\sum_{s\in\mathcal{S}}\pi_s (1-\epsonetwo{s}) z^{(j)}_s .
\end{align}

Finally, we show that the above outer bound matches $ \bar{\mc C}_{\text{ fb+s}}^\text{mem}$ defined in \eqref{eq:posouter} - \eqref{eq:R2_constr2outer}.
This is done by noting that inequality \eqref{tight} can be made tight without changing the rate region. The equivalence of the two regions then becomes clear by setting 
$z^{(1)}_s=1-y_s$, $z^{(2)}_s=1-x_s$, $u^{(1)}_s=x_s$, and $u^{(2)}_s=y_s$.

\subsection{Hidden Case}
\label{sec:outer_bound_hidden}
Following  the same line of arguments as that in  Section \ref{sec:proof_bound_visible}, one can prove an outer bound on the capacity region in the hidden case. Define $\bar{\mc C}_{n,\text{ fb}}^{\text{mem}}$, for every positive integer $n$, as the closure of rate pairs $(R_1,R_2)$ for which there exist variables $x(\ve z^{t-1}), y(\ve z^{t-1})$, $t=1,\ldots,n$, such that
\begin{align}
0&\leq x(\ve z^{t-1}) \leq 1,\quad 0\leq y(\ve z^{t-1})\leq 1 \qquad t=1,\ldots,n, \forall~\ve z^{t-1}\in\{0,1\}^{2(t-1)} \label{eq:posouter_hmm}\\
R_1&\leq\frac{1}{n} \sum_{t=1}^n \sum_{\ve z^{t-1}} P_{\ve Z^{t-1}}(\ve z^{t-1}) (1-\epsone{\ve z^{t-1}}) x(\ve z^{t-1}) \label{eq:R1_constr1outer_hmm}\\
R_1&\leq\frac{1}{n} \sum_{t=1}^n \sum_{\ve z^{t-1}} P_{\ve Z^{t-1}}(\ve z^{t-1}) (1-\epsonetwo{\ve z^{t-1}}) (1-y(\ve z^{t-1})) \label{eq:R1_constr2outer_hmm} \\
R_2&\leq\frac{1}{n} \sum_{t=1}^n \sum_{\ve z^{t-1}} P_{\ve Z^{t-1}}(\ve z^{t-1}) (1-\epstwo{\ve z^{t-1}}) y(\ve z^{t-1}) \label{eq:R2_constr1outer_hmm}\\
R_2&\leq\frac{1}{n} \sum_{t=1}^n \sum_{\ve z^{t-1}} P_{\ve Z^{t-1}}(\ve z^{t-1}) (1-\epsonetwo{\ve z^{t-1}}) (1-x(\ve z^{t-1})) . \label{eq:R2_constr2outer_hmm}
\end{align}
So the capacity region is outer bounded by $\bar{\mc C}_{\text{ fb}}^\text{mem} \ensuremath{r}iangleq \limsup_{n\rightarrow \infty} \bar{\mc C}_{n,\text{ fb}}^{\text{mem}}$. Unfortunately, unlike the outer bound in the visible case, $\bar{\mc C}_{\text{ fb}}^{\text{mem}}$ is not computable. The proof of \eqref{eq:posouter_hmm}~-~\eqref{eq:R2_constr2outer_hmm} is similar to that of the visible case and is omitted for brevity. One can see that compared to the visible case, $\ve z^{t-1}$ plays the role of system's previous state $s$. Here, however, the previous channel state is hidden and the entire past feedback samples (or equivalently $\ve z^{t-1}$) are required for the prediction of future channels. Therefore, the characterization has an averaging over the channel uses, and does not admit a single-letter form.

In order to find computable outer bounds, we establish a sequence of outer and inner approximations for $\bar{\mc C}_{\text{ fb}}^\text{mem}$, indexed by an integer $L$, that have finite-letter characterizations. The parameter $L$ effectively captures how far into past we consider the feedback sequences. To quantify the cost of this truncation, we proceed as follows.

First note that the predicted erasure probabilities in each time slot $t$ can be computed as
\begin{align}
 P_{\ve Z_{t+1}|\ve Z^{t}}(\ve z_{t+1}|\ve z^{t}) = \sum_{s \in \mc S}  P_{S_{t+1}|\ve Z^{t}}(s|\ve z^{t}) P_{\ve Z_{t+1}|S_{t+1}}(\ve z_{t+1}|s).
\label{eq:prediction_erasure_prob}
\end{align}
The distribution $P_{\ve Z_{t+1}|S_{t+1}}$ does not change with $t$ by assumption of time invariance. The distribution $P_{S_{t+1}|\ve Z^{t}}$ does change, however. 
A recursive formula to compute $P_{S_{t+1}|\ve Z^{t}}$ on-the-fly using the previously computed distribution $P_{S_{t}|\ve Z^{t-1}}$ and the new feedback sample $\ve z_{t}$ is 
\begin{align}
 P_{S_{t+1}|\ve Z^{t}}(s|\ve z^{t}) &= \frac{P_{S_{t+1} \ve Z_t | \ve Z^{t-1}}(s, \ve z_t | \ve z^{t-1}) }{ P_{\ve Z_t | \ve Z^{t-1}}(\ve z_t | \ve z^{t-1}) }\nonumber \\
&= \frac{  \sum_{s' \in \mc S} P_{S_{t+1}| S_t}(s|s')  P_{\ve Z_{t}|S_{t}}(\ve z_{t}|s')  P_{S_t|\ve Z^{t-1}}(s'|\ve z^{t-1})
 }{ \sum_{s' \in \mc S} P_{\ve Z_{t}|S_{t}}(\ve z_{t}|s')  P_{S_t|\ve Z^{t-1}}(s'|\ve z^{t-1})   } .
\label{eq:recursion_state}
\end{align} 

Similarly, the distribution of the channel states based only on the past $L$ feedback samples $P_{S_{t+1}|\ve Z^{t}_{t-L+1}}$ can be written as
\begin{align}
 P_{S_{t+1}|\ve Z^{t}_{t-L+1}}(s|\ve z^{t}_{t-L+1}) &= 
 \frac{  \sum_{s' \in \mc S} P_{S_{t+1}| S_t}(s|s')  P_{\ve Z_{t}|S_{t}}(\ve z_{t}|s')  P_{S_t|\ve Z^{t-1}_{t-L+1}}(s'|\ve z^{t-1}_{t-L+1})
 }{ \sum_{s' \in \mc S} P_{\ve Z_{t}|S_{t}}(\ve z_{t}|s')  P_{S_t|\ve Z^{t-1}_{t-L+1}}(s'|\ve z^{t-1}_{t-L+1})   } . \label{eq:recursion_state_L}
\end{align} 
Note that \eqref{eq:recursion_state} and \eqref{eq:recursion_state_L} differ only in whether the distribution of $S_t$ is based on all or only the past $L$ samples.
In practice these two distributions do not differ much for a sufficiently large choice of $L$, where $L$ depends on the mixing time of the Markov chain. This is made precise by the following theorem \cite[Theorem 2.1]{le2000exponential}:
\begin{oldtheorem}[{\cite[Theorem 2.1]{le2000exponential}}]
 Suppose all entries of the state transition matrix $P_{S_t|S_{t-1}}$ and the distribution matrix $P_{\ve Z_t|S_{t}}$ are strictly positive.
For any observed sequence $\ve z^{t}$, we have the following bound on the variational distance:
\begin{align}
 \sum_{s \in \mc S} \left| P_{S_{t+1}|\ve Z^{t}}(s|\ve z^{t}) -  P_{S_{t+1}|\ve Z^{t}_{t-L+1}}(s|\ve z^{t}_{t-L+1})    \right| \leq 2  (1-\sigma)^L,
\end{align}
where $0<\sigma<1$ 
is a characteristic constant of the underlying Markov chain as defined in \cite[Section 2]{le2000exponential}. Roughly speaking, $\sigma$ depends on the smallest entry in the matrix $P_{S_t|S_{t-1}}$ and on the ratio of the largest and smallest values in the matrix $P_{Z_t|S_t}$.

\end{oldtheorem}
\begin{remark}
\label{rm:exponential_forgetting}
 The theorem in its original form \cite[Theorem 2.1]{le2000exponential} is more general and only requires the $r$-th order transition matrix $P_{S_t|S_{t-r}}$ to be strictly positive. We use the weaker form for simplicity.
\end{remark}

\begin{cor}
\label{cor:Pz_decay}
For any observed sequence $\ve z^{t}$, the total variation distance between $P_{\ve Z_{t+1}|\ve Z^{t}}$ and 
$P_{\ve Z_{t+1}|\ve Z^{t}_{t-L+1}}$ is bounded by
 \begin{align}
  \sum_{\ve z_{t+1}} \left| P_{\ve Z_{t+1}|\ve Z^{t}}(\ve z_{t+1}|\ve z^{t}) -  P_{\ve Z_{t+1}|\ve Z^{t}_{t-L+1}}(\ve z_{t+1}|\ve z^{t}_{t-L+1})    \right| \leq 2  (1-\sigma)^L. \label{eq:cor_Pz_decay}
\end{align}
\end{cor}
\begin{IEEEproof}
The proof is deferred to Appendix \ref{ap-cor}.
\end{IEEEproof}

We are now equipped to prove an approximate capacity outer bound using \eqref{eq:posouter_hmm}~-~\eqref{eq:R2_constr2outer_hmm} together with Corollary~\ref{cor:Pz_decay}. We show in Appendix~\ref{sec:proof_bound_hidden_approx} that any achievable rate pair $(R_1,R_2)$ is such that the following inequalities hold for some $x(\ve z^L)$, $y(\ve z^L)$, $\ve z^L\in\{0,1\}^{2L}$:
\begin{align}
0&\leq x(\ve z^L)\leq 1, \quad 0\leq y(\ve z^L)\leq 1 \qquad \forall~\ve z^L\in\{0,1\}^{2L} \label{eq:posouter_hmm_wind_outer}\\
R_1&\leq \sum_{\ve z^L} P_{\ve Z^L}(\ve z^L) (1-\epsone{\ve z^L}) x(\ve z^L) + C(L) \label{eq:R1_constr1outer_hmm_wind_outer}\\
R_1&\leq \sum_{\ve z^L} P_{\ve Z^L}(\ve z^L) (1-\epsonetwo{\ve z^L}) (1-y(\ve z^L)) +  C(L) \label{eq:R1_constr2outer_hmm_wind_outer} \\
R_2&\leq \sum_{\ve z^L} P_{\ve Z^L}(\ve z^L) (1-\epstwo{\ve z^L}) y(\ve z^L) +  C(L)\label{eq:R2_constr1outer_hmm_wind_outer}\\
R_2&\leq \sum_{\ve z^L} P_{\ve Z^L}(\ve z^L) (1-\epsonetwo{\ve z^L}) (1-x(\ve z^L)) +  C(L) \label{eq:R2_constr2outer_hmm_wind_outer}
\end{align}
where
\begin{align}
-2(1-\sigma)^L\leq C(L)\leq 2(1-\sigma)^L.
\end{align}
This proves statement (ii) of Theorem \ref{thm:capacity_hmm}.

\begin{remark}
 For many examples, the parameter $\sigma$ in \eqref{eq:cor_Pz_decay} is close to zero. Thus, $L$ must be large (in the order of $L=10000$) to give meaningful bounds. In numerical examples, we observed that the variational distance decays  faster than predicted by \eqref{eq:cor_Pz_decay}. %
Often, the variational distance is on the order of $10^{-4}$ for values like $L=10$.

\end{remark}

\section{Optimal Coding Schemes}
\label{sec:new_ach}

In this section, we develop codes that achieve the outer bounds stated in Section~\ref{sec:main}.
We show that $\mc C_{\text{ fb}+s}^\text{mem} = \bar{\mc C}_{\text{ fb}+s}^\text{mem}$ and $\mc C_{\text{ fb}}^\text{mem} = \bar{\mc C}_{\text{ fb}}^\text{mem}$.
One of the coding operations that we use to achieve $\mc C_{\text{ fb}+s}^\text{mem} $ was first proposed in \cite{kuo2017robust}  based on a previous result on \cite{6847153}. 
The description in this section does not require knowledge about results and coding schemes in \cite{6847153}.
Our coding schemes build upon well-established results on 
virtual-queue-based algorithms such as \cite{georgiadis2009broadcast,6522177, Neely2009862}.
To analyze the scheme, we track packets through a network of queues as explained next.

\subsection{Queue Model}

\begin{figure}[ht]
\centering
\begin{tikzpicture}[scale=0.5, every node/.style={scale=0.6}]
\draw (0,0) -- ++(0,-1.5cm) -- ++(1.0cm,0) -- ++(0,1.5cm);
\foreach \i in {1,...,3}
  \draw (0,-1.5cm+\i*6pt) -- +(+1.0cm,0);

\draw (1.5,-4.5) -- ++(0,-1.5cm) -- ++(1.0cm,0) -- ++(0,1.5cm);
\foreach \i in {1,...,3}
  \draw (1.5,-6cm+\i*6pt) -- +(+1.0cm,0);

\draw (4,-2) -- ++(0,-1.5cm) -- ++(1.0cm,0) -- ++(0,1.5cm);
\foreach \i in {1,...,3}
  \draw (4,-3.5cm+\i*6pt) -- +(+1.0cm,0);

\draw (2,-8) -- ++(0,-0.9cm) -- ++(1.0cm,0) -- ++(0,0.9cm);

\draw (12,0) -- ++(0,-1.5cm) -- ++(1.0cm,0) -- ++(0,1.5cm);
\foreach \i in {1,...,3}
  \draw (12,-1.5cm+\i*6pt) -- +(+1.0cm,0);

\draw (10.5,-4.5) -- ++(0,-1.5cm) -- ++(1.0cm,0) -- ++(0,1.5cm);
\foreach \i in {1,...,3}
  \draw (10.5,-6cm+\i*6pt) -- +(+1.0cm,0);

\draw (8,-2) -- ++(0,-1.5cm) -- ++(1.0cm,0) -- ++(0,1.5cm);
\foreach \i in {1,...,3}
  \draw (8,-3.5cm+\i*6pt) -- +(+1.0cm,0);

\draw (10,-8) -- ++(0,-0.9cm) -- ++(1.0cm,0) -- ++(0,0.9cm);

\node (Q11) at (0.5,-0.5) {$Q_1^{(1)}$}; 
\node (Q12) at (12.5,-0.5) {$Q_1^{(2)}$}; 

\node (Q21) at (2,-5) {$Q_2^{(1)}$};
\node (Q22) at (11,-5) {$Q_2^{(2)}$};

\node (Q31) at (4.5,-2.5) {$Q_3^{(1)}$};
\node (Q32) at (8.5,-2.5) {$Q_3^{(2)}$};  
  
\node (Q41) at (2.5,-8.5) {$Q_4^{(1)}$};
\node (Q42) at (10.5,-8.5) {$Q_4^{(2)}$};

\draw[<-, thick] (0.5,0) -- +(0,0.5) node[above] {$R_1$};

\draw[->, thick] (0.5,-1.5) to [out=270,in=105] node[left] {$c_{12}^{(1)}$} (Q21.north);
\draw[->, thick] (0.5,-1.5) to [out=225,in=125] node[left]{$c_{14}^{(1)}$} (2.25,-8);
\draw[->,thick] (2,-6) to [out=270,in=75] node[right,pos=0.2]{$c_{24}^{(1)}$} (Q41.north);
\draw[->,thick] (0.5,-1.5) to [out=325,in=75] node[above]{$c_{13}^{(1)}$} (Q31.north);
\draw[->,thick] (4.5,-3.5) to [out=270,in=75] node[right]{$c_{34}^{(1)}$} +(0,-2) to [out=245,in=75] (2.75,-8);
\draw[->,thick] (4.5,-3.5) to [out=270,in=75] node[above]{$c_{32}^{(1)}$} (2.25,-4.5);

\draw[<-, thick] (12.5,0) -- +(0,0.5) node[above] {$R_2$};

\draw[->, thick] (12.5,-1.5) to [out=270,in=75] node[right] {$c_{12}^{(2)}$} (Q22.north);

\draw[->, thick] (12.5,-1.5) to [out=305,in=55] node[right]{$c_{14}^{(2)}$} (10.75,-8);
\draw[->,thick] (11,-6) to [out=270,in=105] node[left,pos=0.2]{$c_{24}^{(2)}$} (Q42.north);
\draw[->,thick] (12.5,-1.5) to [out=215,in=105] node[above]{$c_{13}^{(2)}$} (Q32.north);
\draw[->,thick] (8.5,-3.5) to [out=270,in=105] node[left]{$c_{34}^{(2)}$} +(0,-2) to [out=295,in=105] (10.25,-8);
\draw[->,thick] (8.5,-3.5) to [out=270,in=105] node[above]{$c_{32}^{(2)}$} (10.75,-4.5);

\end{tikzpicture}
\caption{Networked system of queues.}
\label{fig:queues_new}
\end{figure}
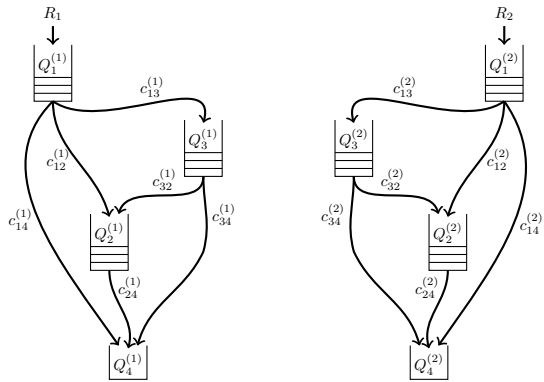

Consider Fig.~\ref{fig:queues_new} that shows a network of queues whose operation we now describe.
The transmitter has two buffers $Q_1^{(1)}$, $Q_1^{(2)}$ to store packets destined for \textnormal{Rx$_1$}, \textnormal{Rx$_2$}, respectively.
We consider dynamic arrivals, where packets for \textnormal{Rx$_1$}, \textnormal{Rx$_2$}~arrive in each slot according to a Bernoulli process with probability $R_1$, $R_2$, respectively. These arrived packets are called \emph{original packets}. An analysis for more general arrival processes is possible. 
The transmitter maintains two additional buffers $Q_2^{(1)}$ (resp. $Q_2^{(2)}$) for packets that have already been sent, but have been received only by \textnormal{Rx$_2$}~(resp. \textnormal{Rx$_1$}). 
Hence buffer $Q_2^{(1)}$ contains packets that are destined for \textnormal{Rx$_1$}~and have been received at \textnormal{Rx$_2$}~but not at \textnormal{Rx$_1$}, and vice versa for $Q_2^{(2)}$. These queues are empty before transmission begins.
If both $Q_2^{(1)}$ and $Q_2^{(2)}$ are nonempty, the transmitter can send the XOR combination of these original packets. 
Such a linear combination of original packets from $Q_2^{(1)}$ and $Q_2^{(2)}$ is called a \emph{coded} packet\footnote{The original packets arriving to $Q_1^{(j)}$ may be coded as well in the sense that error-correcting codes have been employed by lower layers. We use the term \emph{coded} packet to emphasize that information packets for different receivers have been mixed.
Alternative terms would be network-coded or inter-session-coded packet.}.
If both users receive this coded packet, both can decode one desired original packet and two packets per slot are delivered.
Since the coding that is used here is a reaction to previous erasure events, we refer to this coding operation as \emph{reactive} coding\footnote{The term \emph{reactive coding} is also used in \cite{kuo2017robust}, but the meaning is different.}.

Another coding operation turns out to be useful: The transmitter can take one packet from $Q_1^{(1)}$ and $Q_1^{(2)}$ each, compute the XOR combination and send it. 
Note that the packets involved have not been transmitted before (hence have not been received at \textnormal{Rx$_1$}~or \textnormal{Rx$_2$}), so we call this action \emph{proactive} coding or \emph{poisoning}\footnote{We use this term because of its analogy to the poison-remedy approach in \cite{traskov2006network}.}. A poisoned packet is not immediately useful for a receiver upon reception. It may become useful together with a \emph{remedy} packet that enables decoding of original packets that are involved in the linear combination for the poisoned packet:
Assume packet $p_l^{(1)}$ was chosen from $Q_1^{(1)}$ and $p_m^{(2)}$ was chosen from $Q_1^{(2)}$, and the poison packet $p_l^{(1)} + p_m^{(2)}$ was transmitted and received by \textnormal{Rx$_1$}~or \textnormal{Rx$_2$}~or both. 
In that case, $p_l^{(1)}$ is put into an additional queue $Q_3^{(1)}$, likewise $p_m^{(2)}$ is put into $Q_3^{(2)}$.
Assume $p_l^{(1)} + p_m^{(2)}$ was received at \textnormal{Rx$_j$}. If, at a later stage, the corresponding \emph{remedy}
$p_l^{(1)}$ (or $p_m^{(2)}$) is transmitted and received at \textnormal{Rx$_j$}, both $p_l^{(1)}$ and $p_m^{(2)}$ can be decoded at \textnormal{Rx$_j$}. An example why this can be beneficial is provided in Example~\ref{ex:proactive} ahead.

The system exit for \textnormal{Rx$_j$}~is represented by the buffer $Q_4^{(j)}$.
Once a packet reaches the intended receiver, it is moved to this queue and leaves the system.
These buffers are empty by definition.

With slight abuse of notation, let $Q_{l,t}^{(j)}$ denote the number of packets stored in buffer $Q_{l}^{(j)}$ at time $t$. Obviously, $Q_{l,t}^{(j)} \in \mathbb N_0$. 
Define 
\begin{align}
 \ve Q_t = \left( Q_{1,t}^{(1)}, Q_{2,t}^{(1)}, Q_{3,t}^{(1)}, Q_{1,t}^{(2)}, Q_{2,t}^{(2)},Q_{3,t}^{(2)} \right) \in \mathbb N_0^6.
\end{align}
Because $Q_{4,t}^{(1)}= Q_{4,t}^{(2)}=0$ by definition, the vector $\ve Q_t$ determines the queue state at time $t$.

The transmitter selects an action $A_t$ in slot $t$ from the set $\mc A_5 = \{1,2,3,4,5\}$, where
\begin{itemize}%
 \item $A_t=1$ corresponds to sending an uncoded original packet for \textnormal{Rx$_1$}~from $Q_1^{(1)}$ 
 \item $A_t=2$ corresponds to sending an uncoded original packet for \textnormal{Rx$_2$}~from $Q_1^{(2)}$ 
 \item $A_t=3$ corresponds to sending a coded packet from $Q_2^{(1)}$ and $Q_2^{(2)}$
 \item $A_t=4$ corresponds to sending a poisoned packet from $Q_1^{(1)}$ and $Q_1^{(2)}$
 \item $A_t=5$ corresponds to sending a remedy packet either from $Q_3^{(1)}$ or $Q_3^{(2)}$.
\end{itemize}
We will explain in detail how action $A_t=5$ chooses remedy packets in Section~\ref{sec:packet_mov}.

\begin{remark} 
\label{rm:packet_length}
An implementation of the described coding scheme could add a header to each packet. The header should include a flag of $3$ bits to indicate which of the $5$ actions was used. At most two packets will be combined, so one needs two
 fields of at most $\log_2(n R_1+nR_2)$ bits each, to indicate which two out of the $nR_1+nR_2$ packets were involved in the coding operation.
The latter two field sizes grow with the block size $n$, but only logarithmically.
For a typical internet protocol (IP) packet of $\ell=1500 \cdot 8=12000$~bits, this overhead would be less than $1\%$ even if the block size $n$ is as large as $10^9$.
\end{remark}
\begin{remark}
All buffers are physically present at the transmitter.
The packets in $Q_2^{(j)}$ however also have to be stored in \textnormal{Rx$_j$}'s memory. 
Packets in $Q_3^{(j)}$ are involved in a poisoned packet. The corresponding poisoned packet is stored in the memory at \textnormal{Rx$_1$}, \textnormal{Rx$_2$}~or at both.
In order to let both receivers correctly track the packet movement, each receiver needs feedback about the other receiver's packet erasures.
This is realized by the assumed reliable low-rate forward channel from the transmitter to both receivers.
\end{remark}

One can also define a strategy that only uses \emph{reactive} coding, where only the actions $1,2,3$ are permitted. The corresponding set of admissible actions is called $\mc A_3$. In that case, the queues $Q_3^{(j)}$ are not needed and are always empty. These coding operations suffice to achieve capacity for memoryless channels.

A third possibility is to permit only uncoded packet transmissions, i.e. actions $1,2$. The corresponding set of actions is called $\mc A_2$.
In that case, the queues $Q_2^{(j)}$ and $Q_3^{(j)}$ are always empty by definition.

For the visible case, actions at time $t$ are restricted to depend on the \emph{current} queue state $\ve Q_t$ and the \emph{previous} channel state $S_{t-1}$, i.e., the actions are generated by a distribution $P_{A_t|S_{t-1} \ve Q_t}$. We choose distributions that do not depend on $t$.
For the hidden case, actions at time $t$ may depend on the \emph{current} queue state $\ve Q_t$ and all \emph{previous} ACK/NACK messages up to time $t$, $\ve Z^{t-1}$, i.e. the actions are generated by a distribution $P_{A_t|\ve Z^{t-1} \ve Q_t}$.

The following example from \cite{kuo2017robust} demonstrates the necessity of \emph{proactive} coding:
\begin{example}
\label{ex:proactive}
 Consider the Markov chain in Fig.~\ref{fig:example_chain} for $\delta=0$\footnote{A periodic Markov chain is used here for illustration. In general, we consider only aperiodic chains. That would require $\delta>0$.}, i.e. the state sequence is $\ldots,s_1,s_2,s_1,s_2,\ldots$.
The stationary distribution is $\pi_{s_1} = \pi_{s_2} = 0.5$. 
\begin{figure}[t]
\begin{center}
 \begin{tikzpicture}[->,>=stealth',shorten >=1pt,auto,node distance=40mm,
  thick,main node/.style={circle,fill=gray!20,draw},every node/.style={scale=0.6}, every text node part/.style={align=center}]
  \node[main node] (S1) {$s_1$};
  \node[main node] (S2) [right of=S1] {$s_2$};

  \path[every node/.style={font=\sffamily\small}]
    (S1) edge [bend left=20] node {$1-\delta$} (S2)
    edge [loop above] node {$\delta$} (S1)
    (S2) edge [bend left=20] node [above] {$1-\delta$} (S1)
    edge [loop above] node {$\delta$} (S2);

\node (PZ1) [below= 5mm of S1] {$P_{\ve Z_t|S_t}(0,0|s_1)=1$};  

\node (PZ2) [below= 5mm of S2] {$P_{\ve Z_t|S_t}(1,0|s_2)=0.5$\\$P_{\ve Z_t|S_t}(0,1|s_2)=0.5$};  

\draw[->] (S1) -- (PZ1);
\draw[->] (S2) -- (PZ2);
\end{tikzpicture}
\captionof{figure}{Example from \cite{kuo2017robust} explaining the usefulness of poisoning.}
\label{fig:example_chain}
\end{center}
\end{figure}
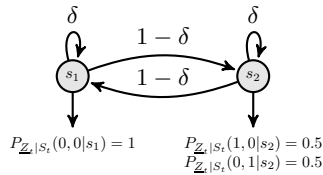

A packet is never erased when we arrive at state $s_1$. In state $s_2$, it is received only at \textnormal{Rx$_1$}~or only at \textnormal{Rx$_2$}, both with a chance of 50\%.
 The setup is visible because the feedback determines the state.
Consider a strategy with \emph{reactive} coding only, with actions $A_t \in \mc A_3$ and corresponding rate region $\underline{\mc C}_{\text{fb}+s}^{\text{mem}}$:
Action $A_t=3$ is fully beneficial only if both \textnormal{Rx$_1$}~and \textnormal{Rx$_2$}~receive the coded packet, so it should be used only after state $s_2$ (when the next state is $s_1$ and no erasure will occur). 
After state $s_1$, one can transmit packets for \textnormal{Rx$_1$}~ and  packets for \textnormal{Rx$_2$}~50\% of the time each. 50\% of these packets are received by the intended receiver and leave the system.
The other half is received by the wrong receiver and goes to the buffer for overheard packets, $Q_2^{(j)}$.
The reactive coding action $A_t=3$ can be used after state $s_2$ only at the pace the overhearing buffers $Q_2^{(1)}$ and $Q_2^{(2)}$ are filled: This will occur at a rate of $\pi_{s_1} P_{A_t|S_{t-1}}(1|s_1) (1-\epsilon_1(s_1)) = 0.5 \cdot 0.5 \cdot 0.5=\tfrac{1}{8}$ for \textnormal{Rx$_1$}~and likewise for \textnormal{Rx$_2$}. Hence $\pi_{s_2} P_{A_t|S_{t-1}}(3|s_2) \leq \tfrac{1}{8}$, leading to $P_{A_t|S_{t-1}}(3|s_2)=0.25$. Reactive coding can be used only 25\% of the time after $s_2$, the remaining time must be distributed among actions $1$ and $2$, leading to a rate of $(R_1,R_2)=(\tfrac{7}{16},\tfrac{7}{16})$. 

Now consider the poison-remedy approach, where one always sends a poisoned packet $p_l^{(1)} + p_m^{(2)}$ after state $s_1$. This packet is received by either \textnormal{Rx$_1$}~or \textnormal{Rx$_2$}~
as state $s_2$ always follows. Through the ACK/NACK feedback the transmitter is informed about where the poison is received. Assume it has been received at \textnormal{Rx$_2$}: The transmitter now chooses $p_l^{(1)}$ as remedy packet. This packet will be received by both receivers because state $s_1$ will follow and packets are never erased. \textnormal{Rx$_1$}~gets $p_l^{(1)}$ and \textnormal{Rx$_2$}~can decode $p_m^{(2)}$ from the poison and remedy, hence one packet is transmitted for each user in two time slots, leading to a rate of $(R_1,R_2)=(\tfrac{1}{2},\tfrac{1}{2})$. This rate point is outside of $\underline{\mc C}_{\text{ fb}}^\text{mem}$ and cannot be achieved using reactive coding only.
The corresponding regions are plotted in Fig.~\ref{fig:rate_example_chain}.
\begin{figure}
\begin{center}

\pgfplotsset{tick label style={font=\scriptsize}}%

\begin{tikzpicture}
\begin{axis}[scale=0.5,
width=4.82222222222222in,
height=3.80333333333333in,
scale only axis,
xmin=0,
xmax=0.8,
xmajorgrids,
every outer x axis line/.append style={black},
every x tick label/.append style={font=\color{black}},
xlabel={$R_1$},
ymin=0,
ymax=0.8,
ymajorgrids,
every outer y axis line/.append style={black},
every y tick label/.append style={font=\color{black}},
ylabel={$R_2$},
axis x line*=bottom,
axis y line*=left,
fill=white,
legend style={at={(0.0,0.0)},anchor=south west,draw=black,fill=white,legend cell align=left, font=\footnotesize}
]

\addplot [color=TUMred,solid,line width=1.0pt,mark=o,mark options={solid}]
  table[row sep=crcr]{%
-3.51062362543375e-13	0.749999999999308\\
-2.01265909418323e-13	0.749999999999662\\
3.38307632042204e-13	0.749999999999773\\
0.499999999999621	0.500000000000176\\
0.500000000000175	0.499999999999621\\
0.749999999999316	-3.47510661730321e-13\\
0.749999999999662	-2.0122546919324e-13\\
0.74999999999983	2.80690642245022e-13\\
};
\addlegendentry{$\bar{\mc C}_{\text{ fb}+s}^\text{mem}$, $\delta=0$};

\addplot [color=TUMpantone301,solid,line width=1.0pt,mark=x,mark options={solid}]
  table[row sep=crcr]{%
-1.26005317291344e-11	0.749999999991943\\
-2.38017938691826e-13	0.749999999999871\\
-1.53668744395929e-13	0.749999999999919\\
-2.35922392732846e-15	0.749999999999998\\
0.249999999999595	0.62500000000015\\
0.62500000000015	0.249999999999595\\
0.749999999990959	-1.60115531944172e-11\\
0.749999999999919	-1.53557722093467e-13\\
0.749999999999963	-2.93653990013354e-14\\
0.749999999999968	-1.37945210809676e-14\\
};
\addlegendentry{$\underline{\mc C}_{\text{ fb}+s}^\text{mem}$, $\delta=0$};

\addplot [color=TUMred,dashed,line width=1.0pt,mark=o,mark options={solid}]
  table[row sep=crcr]{%
-2.35791528474218e-12	0.749999999994791\\
-2.14307092921501e-13	0.749999999999728\\
-1.79786231790332e-13	0.749999999999801\\
1.15115541736926e-14	0.749999999999792\\
1.16965816506395e-13	0.749999999999752\\
3.34524403815473e-13	0.749999999999667\\
0.468749999895809	0.468750000062339\\
0.468749999999729	0.468749999999984\\
0.468749999999984	0.468749999999729\\
0.74999999999427	-3.93342428109061e-12\\
0.749999999999667	3.34741427813845e-13\\
0.74999999999977	-1.687379596929e-14\\
0.7499999999998	-1.80338926299585e-13\\
};
\addlegendentry{$\bar{\mc C}_{\text{ fb}+s}^\text{mem}$, $\delta =0.2$};

\addplot [color=TUMpantone301,dashed,line width=1.0pt,mark=x,mark options={solid}]
  table[row sep=crcr]{%
-6.58752219440117e-12	0.749999999995062\\
-1.79800618838044e-13	0.749999999999898\\
-1.17655885034651e-13	0.749999999999942\\
-3.12250225675825e-15	0.750000000000003\\
0.296052631578945	0.572368421052633\\
0.572368421052632	0.296052631578947\\
0.749999999993958	-9.01081986803831e-12\\
0.749999999999791	1.38930533744031e-13\\
0.749999999999888	-2.19171902848814e-13\\
0.749999999999942	-1.17544862732188e-13\\
};
\addlegendentry{$\underline{\mc C}_{\text{ fb}+s}^\text{mem}$, $\delta=0.2$};

\addplot [color=TUMred,dotted,line width=1.0pt,mark=o,mark options={solid}]
  table[row sep=crcr]{%
-4.37445218007711e-12	0.749999999995742\\
-2.44506568744614e-14	0.749999999999733\\
2.09856111206226e-13	0.749999999999643\\
0.441176470587828	0.441176470588352\\
0.749999999995742	-4.37453635153645e-12\\
0.749999999999761	7.85101452906283e-15\\
0.749999999999799	5.96242115959713e-14\\
};
\addlegendentry{$\bar{\mc C}_{\text{ fb}+s}^\text{mem}$, $\delta=0.4$};

\addplot [color=TUMpantone301,dotted,line width=1.0pt,mark=x,mark options={solid}]
  table[row sep=crcr]{%
-1.08246051011562e-11	0.749999999993652\\
-3.92570698171113e-12	0.749999999998118\\
-3.33066907387547e-16	0.75\\
9.71445146547012e-16	0.749999999999999\\
0.370370370369727	0.49074074074062\\
0.49074074074062	0.370370370369727\\
0.749999999997072	-3.81240872204813e-12\\
0.749999999999895	-2.09873785017578e-13\\
0.749999999999966	-7.19979631469414e-14\\
0.750000000000001	-7.35522753814166e-16\\
};
\addlegendentry{$\underline{\mc C}_{\text{ fb}+s}^\text{mem}$, $\delta=0.4$};

\end{axis}
\end{tikzpicture}%
\captionof{figure}{Rate Regions for the model from Fig.~\ref{fig:example_chain}, for $\delta=0$, $\delta=0.2$ and $\delta=0.4$. $\underline{\mc C}_{\text{fb}+s}^{\text{mem}}$ is strictly smaller than $\bar{\mc C}_{\text{fb}+s}^{\text{mem}}$.}
\label{fig:rate_example_chain}
\end{center}
\end{figure}
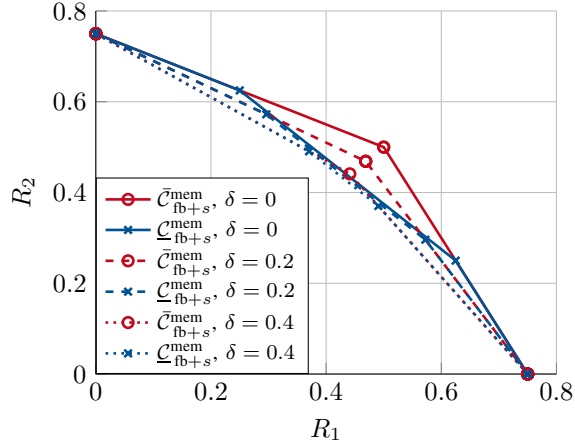
\end{example}

\subsection{Network Flow}
The index set of outgoing neighbors of buffer $Q_{l}^{(j)}$ is written $\mc O_l$, hence $m \in \mc O_l$ if there is a link between $Q_{l}^{(j)}$ and $Q_{m}^{(j)}$.
Define the following three variables related to the link from buffer $Q_l^{(j)}$ to $Q_m^{(j)}$:
Let $C_{lm,t}^{(j)}$ denote the \emph{current link capacity}, i.e. the number of packets that \emph{are allowed} to travel in time slot $t$.
Clearly, $C_{lm,t}^{(j)}$ is in $\{0,1\}$ and is a deterministic function of the action $A_t$ and on the random erasure events $\ve Z_t$. The values of $C_{lm,t}^{(j)}$ are unknown to the transmitter before the transmission because of the erasures.

The \emph{service rate} on the link in slot $t$ is written as $F_{lm,t}^{(j)}$. $F_{lm,t}^{(j)}$ may be different from $C_{lm,t}^{(j)}$ because the transmitter may decide not to move the packet in the time slot. 
We write this as 
\begin{align}
F_{lm,t}^{(j)} = \activRV{lm,t}{(j)} C_{lm,t}^{(j)},
\end{align}
where $\activRV{lm,t}{(j)} \in \{0,1\}$ is a binary random variable indicating if the link will be activated or not,
hence
\begin{align}
 F_{lm,t}^{(j)} \leq C_{lm,t}^{(j)} .
\end{align}
The vector $\ve{\activRV{}{}}_t$ collects all activator variables in slot $t$. $F_{lm,t}^{(j)}$ is a deterministic function of $A_t$, $\ve Z_t$ and $\ve{\activRV{}{}}_t$.
\begin{remark}
 The link activation variables $\ve E_t$ can be necessary to control the service rate of a particular link. The $5$ actions $A_t$ can control from which set of queues we transmit, but not on which of the outgoing links. The outgoing link is chosen by the erasure pattern and the activation variable lets us decide if we want to go over that link.
 With this mechanism the transmitter can control the outgoing link flows independently of each other.
\end{remark}

The \emph{actual} number of packets travelling on the link in time slot $t$ is denoted $\tilde F_{lm,t}^{(j)}$. It can differ from $ F_{lm,t}^{(j)}$ because buffer $Q_l^{(j)}$ can be empty:
\begin{align}
\tilde F_{lm,t}^{(j)} = \mbox{\textbb{1}}\left\{Q_{l,t}^{(j)}>0\right\} \activRV{lm,t}{(j)} C_{lm,t}^{(j)} =  \mbox{\textbb{1}}\left\{Q_{l,t}^{(j)}>0\right\} F_{lm,t}^{(j)} \label{eq:actualFlowRV}
\end{align}

The long-term average link flow $f_{lm}^{(j)}$ and link capacity $c_{lm}^{(j)}$ from $Q_l^{(j)}$ to $Q_m^{(j)}$ are defined by
\begin{align}
 f_{lm}^{(j)}=\lim_{n\rightarrow \infty} \frac{1}{n} \sum_{t=1}^n F_{lm,t}^{(j)}, \qquad c_{lm}^{(j)}=\lim_{n\rightarrow \infty} \frac{1}{n} \sum_{t=1}^n C_{lm,t}^{(j)}  \label{eq:def_link_cap}
\end{align}
where we assume that the limits exist.

To model external packets arrivals,
let $C_{01,t}^{(j)}= F_{01,t}^{(j)}$ denote the indicator random variable if a packet arrived in queue $Q_1^{(j)}$ during time slot $t$. $C_{01,t}^{(j)}$ is independent of all other random variables in the system, and we have
\begin{align}
 \lim_{n\rightarrow \infty} \frac{1}{n} \sum_{t=1}^n C_{01,t}^{(j)}= \mathbb E\left[C_{01,t}^{(j)}\right]=R_j .
\end{align}
This definition of dynamic arrivals is slightly different than the block arrival of $nR_j$ packets as in the outer bounds (see Remark~\ref{remark:stability}).

The \emph{flow divergence} \cite[Chapter 1.1.2]{bertsekas1998network} at buffer $Q_l^{(j)}$ is defined as
\begin{align}
\flowdivRV{l,t}{(j)} = \sum_{m} F_{lm,t}^{(j)} - \sum_{k} F_{kl,t}^{(j)}~. \label{eq:def_flow_div}
\end{align}
The flow divergence is thus the difference of the number of packets that can depart from buffer $Q_l^{(j)}$ minus the number of packets that can arrive at $Q_l^{(j)}$ in slot $t$.
For the long-time average flow divergence, we have
\begin{align}
 \flowdiv{l}{(j)}{} = \sum_{m} f_{lm}^{(j)} - \sum_{k} f_{kl}^{(j)}.
\end{align}
The number of packets in queue $Q_l^{(j)}$ evolve according to 
\begin{align}
 Q_{l,t+1}^{(j)} &= Q_{l,t}^{(j)} - \sum_{m} \tilde F_{lm,t}^{(j)}  +\sum_{k} \tilde F_{kl,t}^{(j)} \nonumber\\
&\leq \left[Q_{l,t}^{(j)} - \sum_{m} F_{lm,t}^{(j)} \right]^+ +\sum_{k} F_{kl,t}^{(j)} \label{eq:queue_dynamics}
\end{align}
where $[x]^+ = \max(x,0)$.
There is an inequality because some queue $Q_k^{(j)}$ might be empty, and $\tilde F_{kl,t}^{(j)} \leq F_{kl,t}^{(j)}$.

Each coding scheme leads to a different evolution of the queue state $\ve Q_t$, as the actions at time $t$ can differ.

The \emph{Rate Stability} Theorem \cite[Theorem 2]{neely2010stability} states that a queue $Q_{l,t}^{(j)}$ evolving according to \eqref{eq:queue_dynamics} is rate-stable if the average arrival rate $\sum_{k} f_{kl}^{(j)}$ is not larger than the average departure rate $\sum_{m} f_{lm}^{(j)}$,
\begin{align}
\sum_{k} f_{kl}^{(j)} &\leq \sum_{m} f_{lm}^{(j)}, \label{eq:rate-stability_suff} \\
\text{ hence if }\qquad  \flowdiv{l}{(j)}{} &\geq 0.
\end{align}
Therefore, satisfying the long-term average capacity constraints is sufficient to ensure rate stability.

\subsection{Packet Movement}
\label{sec:packet_mov}
In this section we summarize when packets can leave buffers and move to another one.
Consider uncoded packet transmission, i.e. actions $A_t=1$ or $A_t=2$:
If an uncoded packet from $Q_1^{(1)}$ is received by \textnormal{Rx$_1$}, it is moved to $Q_4^{(1)}$ and leaves the system.
If such a packet is only received at \textnormal{Rx$_2$}, it is moved to $Q_2^{(1)}$.
If it is erased everywhere, it stays in buffer $Q_1^{(1)}$. Likewise for packets for \textnormal{Rx$_2$}.

For a reactive coding operation, i.e. for $A_t=3$, suppose $p_l^{(1)} \in Q_2^{(1)}$ and $p_m^{(2)} \in Q_2^{(2)}$. The coded packet $p_l^{(1)} + p_m^{(2)}$ is computed and transmitted.
If this packet is received by \textnormal{Rx$_1$}, $p_l^{(1)}$ can be decoded, so it is moved from  $Q_2^{(1)}$ to the system exit $Q_4^{(1)}$. If the coded packet is not received by \textnormal{Rx$_2$}, $p_m^{(2)}$ stays in $Q_2^{(2)}$. This also applies vice versa.

In a degenerate case $A_t=3$ is chosen but either $Q_2^{(1)}$ or $Q_2^{(2)}$ is empty: In this case, only one packet can be delivered per slot, but the queue dynamics as defined in \eqref{eq:queue_dynamics} are satisfied for each of the two queue networks individually (see also \cite{paschos2012scheduling}).

The most complicated packet movement rules apply for the poison/remedy actions and the corresponding buffers:
A packet can move from $Q_1^{(j)}$ to $Q_3^{(j)}$ only if it is involved in a poisoned packet.
This happens if $A_t=4$ and the poisoned packet is not erased at both receivers:
Suppose $p_l^{(1)} \in Q_1^{(1)}$ and $p_m^{(2)} \in Q_1^{(2)}$ and the poisoned packet $p_l^{(1)} + p_m^{(2)}$ is received at \textnormal{Rx$_1$}, at \textnormal{Rx$_2$}~or at both. Then $p_l^{(1)}$ moves to $Q_3^{(1)}$ and $p_m^{(2)}$ moves to $Q_3^{(2)}$.

A packet can leave $Q_3^{(j)}$ only if a remedy packet was transmitted, i.e. if $A_t=5$.
A remedy packet corresponds to one of the packets that was involved in a proactively coded packet and is stored in either $Q_3^{(1)}$ or $Q_3^{(2)}$.
The exact rules describing which packet acts as remedy and how packets leave buffer $Q_3^{(j)}$ are given in 
Table~\ref{tab:packet_movement}. This table appears in similar form in \cite{kuo2017robust}.
\begin{table}
\centering
 \begin{tabular}{|l|c|l|l|l|}
\hline
  \parbox[t]{0.12\textwidth}{poison $p_l^{(1)} + p_m^{(2)}$ received at} & remedy & remedy received at \textnormal{Rx$_1$}~& remedy received at \textnormal{Rx$_2$}~& \parbox[t]{0.23\textwidth}{remedy rec. at \textnormal{Rx$_1$}~ and \textnormal{Rx$_2$}} \\ \hline \hline
  \textnormal{Rx$_1$}~only& $p_m^{(2)}$ 
& \parbox[t]{0.23\textwidth}{\textnormal{Rx$_1$}~decodes both $p_l^{(1)}$ and $p_m^{(2)}$\\$\Rightarrow$ $p_l^{(1)}$ to $Q_4^{(1)}$ (exit) \\$\Rightarrow$ $p_m^{(2)}$ to $Q_2^{(2)}$} 
& \parbox[t]{0.23\textwidth}{\textnormal{Rx$_2$}~gets $p_m^{(2)}$\\$\Rightarrow$ $p_m^{(2)}$ to $Q_4^{(2)}$ (exit)\\$\Rightarrow$ Replace $p_l^{(1)}$ with remedy $p_m^{(2)}$ and move to $Q_2^{(1)}$} 
& \parbox[t]{0.23\textwidth}{both \textnormal{Rx$_1$}~and \textnormal{Rx$_2$}~decode both $p_l^{(1)}$ and $p_m^{(2)}$\\$\Rightarrow$ $p_l^{(1)}$ to $Q_4^{(1)}$ (exit) \\$\Rightarrow$ $p_m^{(2)}$ to $Q_4^{(2)}$ (exit)} \\ \hline
  \textnormal{Rx$_2$}~only& $p_l^{(1)}$ 
& \parbox[t]{0.23\textwidth}{\textnormal{Rx$_1$}~gets $p_l^{(1)}$ \\$\Rightarrow$ $p_l^{(1)}$ to $Q_4^{(1)}$ (exit) \\$\Rightarrow$ Replace $p_m^{(2)}$ with remedy $p_l^{(1)}$ and move to $Q_2^{(2)}$} 
& \parbox[t]{0.23\textwidth}{\textnormal{Rx$_2$}~decodes both $p_l^{(1)}$ and $p_m^{(2)}$\\$\Rightarrow$ $p_m^{(2)}$ to $Q_4^{(2)}$ (exit)\\$\Rightarrow$ $p_l^{(1)}$ to $Q_2^{(1)}$} 
& \parbox[t]{0.23\textwidth}{both \textnormal{Rx$_1$}~and \textnormal{Rx$_2$}~decode both $p_l^{(1)}$ and $p_m^{(2)}$\\$\Rightarrow$ $p_l^{(1)}$ to $Q_4^{(1)}$ (exit) \\$\Rightarrow$ $p_m^{(2)}$ to $Q_4^{(2)}$ (exit)} \\ \hline
  \textnormal{Rx$_1$}~and \textnormal{Rx$_2$}~& $p_l^{(1)}$
& \parbox[t]{0.23\textwidth}{\textnormal{Rx$_1$}~decodes both $p_l^{(1)}$ and $p_m^{(2)}$\\$\Rightarrow$ $p_l^{(1)}$ to $Q_4^{(1)}$ (exit)\\$\Rightarrow$ $p_m^{(2)}$ to $Q_2^{(2)}$} 
& \parbox[t]{0.23\textwidth}{\textnormal{Rx$_2$}~decodes both $p_l^{(1)}$ and $p_m^{(2)}$\\$\Rightarrow$ $p_m^{(2)}$ to $Q_4^{(2)}$ (exit)\\$\Rightarrow$ $p_l^{(1)}$ to $Q_2^{(1)}$} 
& \parbox[t]{0.23\textwidth}{both \textnormal{Rx$_1$}~and \textnormal{Rx$_2$}~decode both $p_l^{(1)}$ and $p_m^{(2)}$\\$\Rightarrow$ $p_l^{(1)}$ to $Q_4^{(1)}$ (exit) \\$\Rightarrow$ $p_m^{(2)}$ to $Q_4^{(2)}$ (exit)} \\ \hline \hline
 \multicolumn{1}{c}{Resulting capacity:} & \multicolumn{1}{c}{} & \multicolumn{1}{c}{$C_{34}^{(1)}=1$, $C_{32}^{(2)}=1$}   &   \multicolumn{1}{c}{$C_{32}^{(1)}=1$, $C_{34}^{(2)}=1$}   & \multicolumn{1}{c}{$C_{34}^{(1)}=1$, $C_{34}^{(2)}=1$}
 \end{tabular}
\caption{Packet movement for poisoned packets in $Q_3^{(1)}$ and $Q_3^{(2)}$.}
\label{tab:packet_movement}
\end{table}
We explain the following two cases in more detail:
\begin{itemize}
 \item Assume poison $p_l^{(1)} + p_m^{(2)}$ was received at \textnormal{Rx$_1$}~only and remedy $p_m^{(2)}$ was again received at \textnormal{Rx$_1$}~only:
\textnormal{Rx$_1$}~can decode both poisoned packets and hence also knows $p_m^{(2)}$, as if this packet was overheard. So $p_m^{(2)}$ should be moved from $Q_3^{(2)}$ to the buffer for overheard packets, $Q_2^{(2)}$.
\item Assume poison $p_l^{(1)} + p_m^{(2)}$ was received at \textnormal{Rx$_1$}~only and remedy $p_m^{(2)}$ was received at \textnormal{Rx$_2$}~only:
Obviously, \textnormal{Rx$_2$}~now knows the desired packet $p_m^{(2)}$. The question is if $p_l^{(1)} + p_m^{(2)}$ is of any use for \textnormal{Rx$_1$}: The key observation made in \cite{kuo2017robust} is that $p_m^{(2)}$ is as useful for \textnormal{Rx$_1$}~as the actually desired packet $p_l^{(1)}$: If \textnormal{Rx$_1$}~obtains $p_m^{(2)}$, it can decode $p_l^{(1)}$. So, in order to deliver $p_l^{(1)}$ to \textnormal{Rx$_1$}, we can also deliver $p_m^{(2)}$. The packet $p_m^{(2)}$ however is already known by \textnormal{Rx$_2$}, so it acts like packets that are in buffer $Q_2^{(1)}$. Hence one can replace $p_l^{(1)}$ with the remedy $p_m^{(2)}$ (because $p_m^{(2)}$ is as useful as $p_l^{(1)}$) and put it in buffer $Q_2^{(1)}$.

At a later stage $p_m^{(2)}$ will be XORed with a packet $p_k^{(2)}$ from $Q_2^{(2)}$ (i.e. a packet that has already been received at \textnormal{Rx$_1$}~but not at \textnormal{Rx$_2$}).
If \textnormal{Rx$_1$}~receives the linear combination $p_m^{(2)} + p_k^{(2)}$, it can decode $p_m^{(2)}$ because $p_k^{(2)}$ is known. With $p_m^{(2)}$, it can decode the desired packet $p_l^{(1)}$. 
This is summarized in the coding example in Table~\ref{tab:coding_ex}.

\end{itemize}

\begin{table}
\centering
 \begin{tabular}{l|c|c|c|c|}
 $t$ & $A_t$ & $X_t$ & $Y_{1,t}$ & $Y_{2,t}$ \\ \hline \hline
 1 & $4$ & $p_l^{(1)} + p_m^{(2)}$ & $p_l^{(1)} + p_m^{(2)}$ & $?$  \\ \hline
 2 & $5$ &$p_m^{(2)}$ & $?$ & $p_m^{(2)}$  \\ \hline
 3 & $2$ &$p_k^{(2)}$ & $p_k^{(2)}$ & $?$ \\ \hline
 4 & $3$ &$p_m^{(2)} + p_k^{(2)}$ & $p_m^{(2)} + p_k^{(2)}$ & $p_m^{(2)} + p_k^{(2)}$ \\\hline
\end{tabular}
\caption{Coding example.}
\label{tab:coding_ex}
\end{table}

Taking into account all events described above, we obtain
\begin{alignat}{3}
  C_{12,t}^{(j)} &= \mbox{\textbb{1}}\{A_t=j\} Z_{j,t}(1-Z_{\bar j,t}), \qquad & F_{12,t}^{(j)} &= C_{12,t}^{(j)} \activRV{12,t}{(j)}, \qquad & \tilde F_{12,t}^{(j)} &= F_{12,t}^{(j)} \mbox{\textbb{1}}\{Q_{1,t}^{(j)}>0\}\label{eq:def_F12t1}\\
  C_{13,t}^{(j)} &= \mbox{\textbb{1}}\{A_t=4\}(1-Z_{1,t}Z_{2,t}), \qquad & F_{13,t}^{(j)} &= C_{13,t}^{(j)} \activRV{13,t}{(j)}, \qquad & \tilde F_{13,t}^{(j)} &= F_{13,t}^{(j)} \mbox{\textbb{1}}\{Q_{1,t}^{(j)}>0\}\label{eq:defF13}\\
  C_{14,t}^{(j)} &= \mbox{\textbb{1}}\{A_t=j\}(1-Z_{j,t}), \qquad & F_{14,t}^{(j)} &= C_{14,t}^{(j)} \activRV{14,t}{(j)}, \qquad & \tilde F_{14,t}^{(j)} &= F_{14,t}^{(j)} \mbox{\textbb{1}}\{Q_{1,t}^{(j)}>0\}\\
  C_{24,t}^{(j)} &= \mbox{\textbb{1}}\{A_t=3\}(1-Z_{j,t}), \qquad & F_{24,t}^{(j)} &= C_{24,t}^{(j)} \activRV{24,t}{(j)}, \qquad & \tilde F_{24,t}^{(j)} &= F_{24,t}^{(j)} \mbox{\textbb{1}}\{Q_{2,t}^{(j)}>0\}\\
  C_{32,t}^{(j)} &= \mbox{\textbb{1}}\{A_t=5\}Z_{j,t}(1-Z_{\bar j,t}), \qquad & F_{32,t}^{(j)} &= C_{32,t}^{(j)} \activRV{32,t}{(j)}, \qquad & \tilde F_{32,t}^{(j)} &= F_{32,t}^{(j)} \mbox{\textbb{1}}\{Q_{3,t}^{(j)}>0\}\label{eq:defF32}\\
  C_{34,t}^{(j)} &= \mbox{\textbb{1}}\{A_t=5\}(1-Z_{j,t}), \qquad & F_{34,t}^{(j)} &= C_{34,t}^{(j)} \activRV{34,t}{(j)}, \qquad & \tilde F_{34,t}^{(j)} &= F_{34,t}^{(j)} \mbox{\textbb{1}}\{Q_{3,t}^{(j)}>0\}.
 \label{eq:def_F34_new}
\end{alignat}
\begin{remark}
 The packet movement for reactive coding ($A_t=4$ and $A_t=5$) in Table~\ref{tab:packet_movement} only considers the non-degenerate case, where $Q_1^{(1)}$ and $Q_1^{(2)}$ are nonempty when $A_t=4$.
The degenerate case when this is not fulfilled is more complicated and studied in detail in Appendix~\ref{sec:deg_packet_movement}.
However, also the degenerate case can be adapted such that the queue and flow dynamics in \eqref{eq:queue_dynamics} and \eqref{eq:def_F12t1} - \eqref{eq:def_F34_new} hold.
\end{remark}

The algorithms developed in the following ensure network stability for rate pairs inside $\bar{\mc C}_{\text{ fb}}^\text{mem}$.

\subsection{Probabilistic Scheme}
\label{sec:probab_scheme_new}
\subsubsection{Visible Case}
\label{sec:probab_scheme_new_visible}
Consider a strategy that bases decisions for actions only on the previous channel state $S_{t-1}$, but not on the queue state $\ve Q_t$.
The decisions for $A_t$ are random and independent from previous decisions, according to a stationary probability distribution $P_{A_t|S_{t-1}}$.
By the sufficient criterion for rate stability in \eqref{eq:rate-stability_suff}, rate pairs $(R_1,R_2)$ can be achieved if there is a distribution $P_{A_t|S_{t-1}}$ such that the following flow optimization problem is feasible:
\begin{align}
 R_j &\leq f_{12}^{(j)} +  f_{13}^{(j)} + f_{14}^{(j)} \label{eq:link_cap_more1}\\
f_{12}^{(j)} + f_{32}^{(j)} &\leq f_{24}^{(j)}  \label{eq:link_cap_more2}\\
f_{13}^{(j)} &\leq f_{32}^{(j)} + f_{34}^{(j)}  \label{eq:link_cap_more3} \\ 
 f_{12}^{(j)} \leq c_{12}^{(j)} &= \sum_{s \in \mc S} \pi_s   (\epsj{s} - \epsonetwo{s}) P_{A_t|S_{t-1}}(j|s) \label{eq:link_cap_more12}\\
 f_{13}^{(j)} \leq c_{13}^{(j)} &= \sum_{s \in \mc S} \pi_s   (1 - \epsonetwo{s}) P_{A_t|S_{t-1}}(4|s) \label{eq:link_cap_more13}\\
 f_{14}^{(j)} \leq c_{14}^{(j)} &= \sum_{s \in \mc S} \pi_s   (1 - \epsj{s}) P_{A_t|S_{t-1}}(j|s) \label{eq:link_cap_more14}\\
 f_{24}^{(j)} \leq c_{24}^{(j)} &= \sum_{s \in \mc S} \pi_s   (1 - \epsj{s}) P_{A_t|S_{t-1}}(3|s) \label{eq:link_cap_more24}\\
 f_{32}^{(j)} \leq c_{32}^{(j)} &= \sum_{s \in \mc S} \pi_s   (\epsj{s} - \epsonetwo{s}) P_{A_t|S_{t-1}}(5|s) \label{eq:link_cap_more32}\\
 f_{34}^{(j)} \leq c_{34}^{(j)} &= \sum_{s \in \mc S} \pi_s   (1 - \epsj{s}) P_{A_t|S_{t-1}}(5|s)\qquad \forall~j\in \{1,2\} \label{eq:link_cap_more34}
\end{align}
The link capacities can be written as in  \eqref{eq:link_cap_more12} - \eqref{eq:link_cap_more34} because of the ergodicity of the influencing random processes $A^n$ and $\ve Z^n$ in this case.
This is a classic flow optimization problem where the individual link capacities on the RHS of \eqref{eq:link_cap_more12} - \eqref{eq:link_cap_more34} can be adjusted by $P_{A_t|S_{t-1}}$. Note that the disconnected flow networks for \textnormal{Rx$_1$}~and \textnormal{Rx$_1$}~are coupled only through the common dependency on $P_{A_t|S_{t-1}}$.
For any feasible rate pair, \eqref{eq:link_cap_more1} - \eqref{eq:link_cap_more3} ensure that $d_l^{(j)}\geq 0$ for every queue $Q_l^{(j)}$.
The exact operation of the probabilistic scheme is as follows:
 \begin{itemize}
  \item Solve the above linear program to find suitable values of $P_{A_t|S_{t-1}}$, $c_{lm}^{(j)}$ and $f_{lm}^{(j)}$.
  \item In each time slot, observe the previous state $s$, sample the next action randomly from $P_{A_t|S_{t-1}}(\cdot|s)$ and transmit from the corresponding queues.
  \item Observe the feedback $\ve Z_t$: $\ve Z_t$ selects at most one outgoing link per queue on which a packet can be moved.
  \item Move packets from queue $Q_l^{(j)}$ to $Q_m^{(j)}$ with probability $\frac{f_{lm}^{(j)}}{c_{lm}^{(j)}}$, i.e. set $E_{lm,t}^{(j)} = 1$ with probability $\frac{f_{lm}^{(j)}}{c_{lm}^{(j)}}$.
  \item Send the information about moved packets over the low-rate reliable forward link.
 \end{itemize}

The strategy of first observing where a transmitted packet is received and deciding later whether this packet should be logically moved or not appears in similar form in \cite{Neely2009862}.
This mechanism allows to asymptotically achieve a rate of $f_{lm}^{(j)}$ on each link, independent of the other other link flow values $f_{lk}^{(j)}$ departing from the same queue.

By the max-flow  min-cut theorem, the flow problem in \eqref{eq:link_cap_more1} - \eqref{eq:link_cap_more34} is feasible if there is a distribution $P_{A_t|S_{t-1}}$ such that
\begin{align}
 R_j &\leq c_{12}^{(j)} +  c_{13}^{(j)} + c_{14}^{(j)} \nonumber\\
     &= \sum_{s \in \mc S} \pi_s (1-\epsonetwo{s}) \left[ P_{A_t|S_{t-1}}(j|s) + P_{A_t|S_{t-1}}(4|s) \right] \label{eq:bec_cut1}\\
 R_j &\leq c_{13}^{(j)} +  c_{14}^{(j)} + c_{24}^{(j)} \nonumber\\
     &=\sum_{s \in \mc S} \pi_s \Big[ (1-\epsonetwo{s})  P_{A_t|S_{t-1}}(4|s)  + (1-\epsj{s}) \left[ P_{A_t|S_{t-1}}(j|s) + P_{A_t|S_{t-1}}(3|s) \right]  \Big] \label{eq:bec_cut2} \\
 R_j &\leq c_{12}^{(j)} +  c_{14}^{(j)} + c_{32}^{(j)} + c_{34}^{(j)} \nonumber\\
     &=\sum_{s \in \mc S} \pi_s (1-\epsonetwo{s}) \left[ P_{A_t|S_{t-1}}(j|s) + P_{A_t|S_{t-1}}(5|s) \right] \label{eq:bec_cut3} \\
 R_j &\leq c_{14}^{(j)} +  c_{24}^{(j)} + c_{34}^{(j)} \nonumber\\
     &=\sum_{s \in \mc S} \pi_s (1-\epsj{s}) \left[ P_{A_t|S_{t-1}}(j|s) + P_{A_t|S_{t-1}}(3|s) + P_{A_t|S_{t-1}}(5|s) \right] \qquad \forall~j \in \{1,2\},\label{eq:bec_cut4}
\end{align}
where each bound in \eqref{eq:bec_cut1}~-~\eqref{eq:bec_cut4} corresponds to a cut in the queue network.

\begin{prop}
\label{prop:cuts}
 Suppose $P_{A_t|S_{t-1}}$ is feasible for a rate pair $(R_1,R_2)$.
 There exists a distribution $P^\star_{A_t|S_{t-1}}$ for which
\begin{itemize}
 \item $(R_1,R_2)$ is feasible and
 \item the bounds in \eqref{eq:bec_cut2} and \eqref{eq:bec_cut3} are redundant.
\end{itemize}
\end{prop}
The proof of Proposition~\ref{prop:cuts} is given in Appendix~\ref{sec:appendix_proof_cuts}.

By Proposition~\ref{prop:cuts}, a rate pair $(R_1,R_2)$ can be achieved if there is a distribution $P_{A_t|S_{t-1}}$ such that
\begin{align}
 R_1 &\leq  \sum_{s \in \mc S} \pi_s (1-\epsone{s}) \left[ P_{A_t|S_{t-1}}(1|s) + P_{A_t|S_{t-1}}(3|s) + P_{A_t|S_{t-1}}(5|s) \right] \label{eq:bec_more_constr1}\\
 R_1 &\leq  \sum_{s \in \mc S} \pi_s (1-\epsonetwo{s}) \left[ P_{A_t|S_{t-1}}(1|s) + P_{A_t|S_{t-1}}(4|s) \right] \label{eq:bec_more_constr2}\\
 R_2 &\leq  \sum_{s \in \mc S} \pi_s (1-\epstwo{s}) \left[ P_{A_t|S_{t-1}}(2|s) + P_{A_t|S_{t-1}}(3|s) + P_{A_t|S_{t-1}}(5|s) \right] \label{eq:bec_more_constr3}\\
 R_2 &\leq  \sum_{s \in \mc S} \pi_s (1-\epsonetwo{s}) \left[ P_{A_t|S_{t-1}}(2|s) + P_{A_t|S_{t-1}}(4|s) \right]. \label{eq:bec_more_constr4}
\end{align}

One can verify that \eqref{eq:bec_more_constr1} - \eqref{eq:bec_more_constr4} is equivalent to the outer bound $\bar{\mc C}_{\text{ fb}+s}^\text{mem}$ in \eqref{eq:posouter} - \eqref{eq:R2_constr2outer}
 by setting 
\begin{align}
P_{A_t|S_{t-1}}(1|s) + P_{A_t|S_{t-1}}(3|s) + P_{A_t|S_{t-1}}(5|s) &= x_s \label{eq:proba_relation_xs}\\
P_{A_t|S_{t-1}}(2|s) + P_{A_t|S_{t-1}}(3|s) + P_{A_t|S_{t-1}}(5|s) &= y_s.\label{eq:proba_relation_ys}
\end{align}
Note that $x_s+y_s = P_{A_t|S_{t-1}}(1|s) + P_{A_t|S_{t-1}}(2|s) + 2 P_{A_t|S_{t-1}}(3|s) + 2 P_{A_t|S_{t-1}}(5|s)$ can be less than $1$, so constraint \eqref{eq:pa3_pos} is not implicitly required.

\begin{remark}\label{rem:mapping}
The mapping from $x_s$, $y_s$ $s \in \mc S$ to $P_{A_t|S_{t-1}}$ defined in \eqref{eq:proba_relation_xs}~-~\eqref{eq:proba_relation_ys} is not unique. In particular, for some valid choices of $P_{A_t|S_{t-1}}$ the constraints in \eqref{eq:bec_cut2} and \eqref{eq:bec_cut3} will \emph{not} be redundant. Proposition~\ref{prop:cuts} however tells us that one can always \emph{find} a distribution $P^\star_{A_t|S_{t-1}}$ 
that satisfies \eqref{eq:proba_relation_xs}~-~\eqref{eq:proba_relation_ys} and
for which \eqref{eq:bec_cut2} and \eqref{eq:bec_cut3} are redundant 
for any values of $x_s$, $y_s$ $s \in \mc S$.
\end{remark}

This shows that $\bar{\mc C}_{\text{ fb}}^\text{mem}$ is achievable with the presented coding scheme and hence $\bar{\mc C}_{\text{ fb}+s}^\text{mem}=\mc C_{\text{ fb}+s}^\text{mem}$. Section~\ref{sec:new_ach_det} shows that $\mc C_{\text{ fb}+s}^\text{mem}$ is achievable using a deterministic algorithm.

\subsubsection{Hidden Case}
\label{sec:probab_scheme_new_hmm}
The relation between $P_{A_t|S_{t-1}(\cdot|s)}$ and parameters $x_s$ and $y_s$ derived for the probabilistic scheme in the visible case can be translated to the hidden case: 
For a given rate pair $(R_1,R_2)$, one must determine a stationary probability distribution $P_{A_t|\ve Z^{t-1}_{t-L}}$, i.e., the window of $L$ past feedback samples takes the role of the previous channel state $S_{t-1}$ in the visible case.
Any value of $x(\ve z^L)$, $y(\ve z^L)$ in the outer bound \eqref{eq:posouter_hmm} - \eqref{eq:R2_constr2outer_hmm} can be translated into appropriate probability values for $P_{A_t|\ve Z^{t-1}}$, as in \eqref{eq:proba_relation_xs} and \eqref{eq:proba_relation_ys}.
The transmitter randomly chooses its next action according to the distribution $P_{A_t|\ve Z^{t-1}}(\cdot|\ve z^L)$.
The scheme can achieve achieve\footnote{in the sense that the achieves rates are at most $2 (1-\sigma)^L$ smaller than the targeted rate.}
$\bar{\mc C}_{\text{ fb}}^\text{mem}(L)$ defined in
\eqref{eq:posouter_hmm_wind_outer_main}~-~\eqref{eq:R2_constr2outer_hmm_wind_outer_main}.

According to Corollary~\ref{cor:Pz_decay}, the region $\bar{\mc C}_{\text{ fb}}^\text{mem}(L)$ converges exponentially fast to $\bar{\mc C}_{\text{ fb}}^\text{mem}$, but also the number of parameters $x(\ve z^{L})$, $y(\ve z^{L})$ grows exponentially with $L$.
We next propose a parameterless deterministic scheme that is optimal in the long-run.

\subsection{Deterministic Scheme}
\label{sec:new_ach_det}

\subsubsection{Visible Case}
In the probabilistic scheme,  
actions are chosen depending on the channel state only. So it might happen that an action is chosen although there is no packet to transmit because the corresponding buffer is empty.
To operate the probabilistic scheme one also needs to compute the corresponding distribution $P_{A_t|S_{t-1}}$ that depends on the arrival rates $R_1$, $R_2$. These rates might be unknown to the transmitter ahead of time.

Both drawbacks can be avoided by a max-weight backpressure-like algorithm \cite{tassiulas1992stability,tassiulas1993dynamic,neely2005dynamic,georgiadis2006resource,neely2010stochastic} that bases its actions on both queue and channel states:
In each slot $t$, %
a weight function is computed for each action. The action with the highest weight is chosen in that slot. In Appendix~\ref{sec:proof_prop_maxweight_joint}, we show that the optimal action is
 \begin{align}
A_t = \argmax_{A_t\in \mc A_5} \sum_{j=1}^2 \sum_{l=1}^3 \sum_{m \in \mc O_l} \left[ Q_{l,t}^{(j)} - Q_{m,t}^{(j)} \right]^+
\mathbb E \left[C_{lm,t}^{(j)} \bigg| \ve Q_t = \ve q, S_{t-1}=s \right].
\label{eq:maxweight_new} 
\end{align}

Table~\ref{tab:new_det_algo} lists the action weights depending on the current queue state $\ve Q_t$ and the previous channel state $S_{t-1}$.

\begin{table}[t]
\centering
 \begin{tabular}{|c|l|}
 \hline
 $A_t$ & Weight depending on $\ve Q_t$ and $S_{t-1}=s$  \\ \hline
 $1$ &  $[1-\epsone{s}] Q_{1,t}^{(1)} + \epsonenottwo{s}  \left[ Q_{1,t}^{(1)}-Q_{2,t}^{(1)} \right]^+ $\\
 $2$ &  $[1-\epstwo{s}] Q_{1,t}^{(2)} + \epsnotonetwo{s}  \left[ Q_{1,t}^{(2)}-Q_{2,t}^{(2)} \right]^+ $\\
 $3$ &  $[1- \epsone{s}]  Q_{2,t}^{(1)} + [1- \epstwo{s}]  Q_{2,t}^{(2)}$\\
 $4$ &  $[1- \epsonetwo{s}]  \left(  \left[Q_{1,t}^{(1)} -  Q_{3,t}^{(1)}\right]^+ + \left[Q_{1,t}^{(2)} -  Q_{3,t}^{(2)} \right]^+ \right) $\\
 $5$ &  $\epsonenottwo{s}  \left[Q_{3,t}^{(1)}-Q_{2,t}^{(1)} \right]^+ + [1-\epsone{s}] Q_{3,t}^{(1)} + \epsnotonetwo{s}  \left[Q_{3,t}^{(2)}-Q_{2,t}^{(2)}\right]^+  + [1-\epstwo{s}] Q_{3,t}^{(2)}   $\\
\hline
 \end{tabular}
\caption{Deterministic scheme for the visible case and $A_t \in \mc A_5$.}
\label{tab:new_det_algo}
\end{table}

\begin{theorem}
The max-weight strategy in \eqref{eq:maxweight_new}  strongly
stabilizes all queues in the network for every rate pair $(R_1+\delta, R_2+\delta) \in \bar{\mc C}_{\text{ fb}+s}^\text{mem}$, $\delta>0$.
\label{prop:maxweight_new}
\end{theorem}

The proof is given in Appendix~\ref{sec:proof_prop_maxweight_joint}.

\subsubsection{Hidden Case}
Similar to the visible case, the following deterministic max-weight backpressure-like algorithm can be defined.
The action is now given by
 \begin{align}
A_t = \argmax_{A_t\in \mc A_5} \sum_{j=1}^2 \sum_{l=1}^3 \sum_{m \in \mc O_l} \left[ Q_{l,t}^{(j)} - Q_{m,t}^{(j)} \right]^+ \mathbb E \left[ C_{lm,t}^{(j)} \bigg| \ve Q_t = \ve q, \ve Z^{t-1}= \ve z^{t-1} \right].
\label{eq:maxweight_new_hmm} 
\end{align}
Table~\ref{tab:new_det_algo_hmm} lists the weights for each action based on the current queue state $\ve Q_t$ and the previous feedback state $\ve Z^{t-1}$.

\begin{table}[t]
\centering
 \begin{tabular}{|c|l|}
 \hline
 $A_t$ & Weight depending on $\ve Q_t$ and $\ve Z^{t-1}=\ve z^{t-1}$  \\ \hline
 $1$ &  $[1-\epsone{\ve z^{t-1}}] Q_{1,t}^{(1)} + \epsonenottwo{\ve z^{t-1}}  \left[ Q_{1,t}^{(1)}-Q_{2,t}^{(1)} \right]^+ $\\
 $2$ &  $[1-\epstwo{\ve z^{t-1}}] Q_{1,t}^{(2)} + \epsnotonetwo{\ve z^{t-1}}  \left[ Q_{1,t}^{(2)}-Q_{2,t}^{(2)} \right]^+ $\\
 $3$ &  $[1- \epsone{\ve z^{t-1}}]  Q_{2,t}^{(1)} + [1- \epstwo{\ve z^{t-1}}]  Q_{2,t}^{(2)}$\\
 $4$ &  $[1- \epsonetwo{\ve z^{t-1}}]  \left(  \left[Q_{1,t}^{(1)} -  Q_{3,t}^{(1)} \right]^+ + \left[ Q_{1,t}^{(2)} -  Q_{3,t}^{(2)} \right]^+ \right) $\\
 $5$ &  $\epsonenottwo{\ve z^{t-1}}  \left[ Q_{3,t}^{(1)}-Q_{2,t}^{(1)} \right]^+  + [1-\epsone{\ve z^{t-1}}] Q_{3,t}^{(1)} + \epsnotonetwo{\ve z^{t-1}}  \left[Q_{3,t}^{(2)}-Q_{2,t}^{(2)} \right]^+  + [1-\epstwo{\ve z^{t-1}}] Q_{3,t}^{(2)}   $\\
\hline
 \end{tabular}
\caption{Deterministic scheme for the hidden case and $A_t \in \mc A_5$.}
\label{tab:new_det_algo_hmm}
\end{table}

\begin{theorem}
The max-weight strategy in \eqref{eq:maxweight_new_hmm}  strongly
stabilizes all queues in the network for every rate pair $(R_1+\delta, R_2+\delta) \in \bar{\mc C}_{\text{ fb}}^\text{mem}$, $\delta>0$.
\label{prop:maxweight_new_hmm}
\end{theorem}

The proof of Theorem~\ref{prop:maxweight_new_hmm} is given in Appendix~\ref{sec:proof_prop_maxweight_joint}.
\begin{remark}
 In each slot, the max-weight strategy in \eqref{eq:maxweight_new_hmm} requires to compute the $5$ weights for each action given in Table~\ref{tab:new_det_algo_hmm}, which are easy to evaluate if $\epsone{\ve z^{t-1}}$, $\epstwo{\ve z^{t-1}}$, $\epsonetwo{\ve z^{t-1}}$ are known.
 The values of $\epsone{\ve z^{t-1}}$, $\epstwo{\ve z^{t-1}}$, $\epsonetwo{\ve z^{t-1}}$ can be recursively computed by
\eqref{eq:prediction_erasure_prob} and \eqref{eq:recursion_state} for every~$t$, which needs $O(|\mc S|)$ multiplications and summations.
The same argumentation applies to the max-weight strategy in \eqref{eq:maxweight_new} for visible states.

To implement the probabilistic scheme we needed to solve a linear program whose number of variables is exponential in the the observation window $L$ to find the injection probabilities. This is only practical for small values of $L$.
\end{remark}

\section{Achievable Rates with Reactive Coding}
\label{sec:ach}
\begin{figure}[h]
\centering
\begin{tikzpicture}[scale=0.5, every node/.style={scale=0.6}]
\draw (0,0) -- ++(0,-1.5cm) -- ++(1.0cm,0) -- ++(0,1.5cm);
\foreach \i in {1,...,3}
  \draw (0,-1.5cm+\i*6pt) -- +(+1.0cm,0);

\draw (3,-2) -- ++(0,-1.5cm) -- ++(1.0cm,0) -- ++(0,1.5cm);
\foreach \i in {1,...,3}
  \draw (3,-3.5cm+\i*6pt) -- +(+1.0cm,0);

\draw (0,-4) -- ++(0,-0.9cm) -- ++(1.0cm,0) -- ++(0,0.9cm);

\draw (8,0) -- ++(0,-1.5cm) -- ++(1.0cm,0) -- ++(0,1.5cm);
\foreach \i in {1,...,3}
  \draw (8,-1.5cm+\i*6pt) -- +(+1.0cm,0);

\draw (5,-2) -- ++(0,-1.5cm) -- ++(1.0cm,0) -- ++(0,1.5cm);
\foreach \i in {1,...,3}
  \draw (5,-3.5cm+\i*6pt) -- +(+1.0cm,0);

\draw (8,-4) -- ++(0,-0.9cm) -- ++(1.0cm,0) -- ++(0,0.9cm);

\node at (0.5,-0.5) {$Q_1^{(1)}$}; 
\node at (8.5,-0.5) {$Q_1^{(2)}$}; 

\node at (0.5,-4.5) {$Q_4^{(1)}$};
\node at (8.5,-4.5) {$Q_4^{(2)}$};  
  
\node at (3.5,-2.5) {$Q_2^{(1)}$};
\node at (5.5,-2.5) {$Q_2^{(2)}$};

\draw[<-, thick] (0.5,0) -- +(0,0.5) node[above] {$R_1$};

\draw[->, thick] (0.5,-1.5) to [out=315,in=75] node[above] {$c_{12}^{(1)}$}(3.5,-2);
\draw[->, thick] (0.5,-1.5) to [out=225,in=100] node[right]{$c_{14}^{(1)}$} +(-0.25,-2.5);
\draw[->,thick] (3.5,-3.5) to [out=225,in=75] node[below]{$c_{24}^{(1)}$} +(-2.75,-0.5);

\draw[<-, thick] (8.5,0) -- +(0,0.5) node[above] {$R_2$};

\draw[->, thick] (8.5,-1.5) to [out=225,in=105] node[above] {$c_{12}^{(2)}$} (5.5,-2);
\draw[->, thick] (8.5,-1.5) to [out=315,in=80] node[left]{$c_{14}^{(2)}$} +(0.25,-2.5);
\draw[->,thick] (5.5,-3.5) to [out=315,in=105] node[below]{$c_{24}^{(2)}$} +(2.75,-0.5);

\end{tikzpicture}
\caption{Networked system of queues for reactive coding, $A_t \in \mc A_3$.}
\label{fig:queues}
\end{figure}
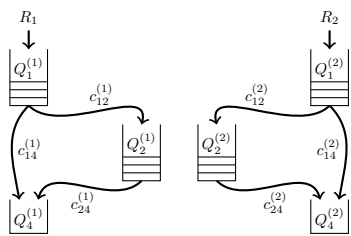

In this section, we investigate the performance of schemes that are reactive; i.e., we are restricted to the set of actions $\mc A_3$. The coding operations in this scheme are conceptually easier as the receivers do not have to store linearly combined packets, but only use them for instantaneous decoding. 
The corresponding simplified queueing network is shown in Fig.~\ref{fig:queues}.
Our focus is on the visible case. All methods can similarly be applied to the hidden case.

\subsection{Probabilistic Scheme}

This scheme adapts the probabilistic scheme in Section~\ref{sec:probab_scheme_new} to the action set $\mc A_3$.
Using this scheme, the rate tuple $(R_1,R_2)$ can be achieved if there is a distribution $P_{A_t|S_{t-1}}$ on $\mc A_3$ such that $\forall~j \in \{1,2\}$: 
\begin{align}
  R_j &\leq f_{14}^{(j)} + f_{12}^{(j)} \label{eq:rate_bound} \\
 f_{12}^{(j)} &\leq f_{24}^{(j)}  \label{eq:flow_cons} \\
  f_{12}^{(j)} \leq c_{12}^{(j)} &= \sum_{s\in \mc S} \pi_s  P_{A_t|S_{t-1}}(j|s) (\epsj{s} - \epsonetwo{s})   \label{eq:bound12} \\
  f_{14}^{(j)} \leq c_{14}^{(j)} &= \sum_{s\in \mc S} \pi_s P_{A_t|S_{t-1}}(j|s) (1-\epsj{s} )  \label{eq:bound13} \\
  f_{24}^{(j)} \leq c_{24}^{(j)} &= \sum_{s\in \mc S} \pi_s P_{A_t|S_{t-1}}(3|s) (1-\epsj{s} ).\label{eq:bound23}
\end{align}
Note that the region described by \eqref{eq:rate_bound}~-~\eqref{eq:bound23} is equivalent to the rate region $\underline{\mc C}_{\text{fb}+s}^\text{mem}$ described in \eqref{eq:R1_constr1}~-~\eqref{eq:R2_constr2}. 
This may be seen by turning the flow problem in \eqref{eq:rate_bound}~-~\eqref{eq:bound23} to the min-cut formulation and
setting 
\begin{align}
P_{A_t|S_{t-1}}(1|s)=1-y_s,\qquad P_{A_t|S_{t-1}}(2|s)=1-x_s,\qquad P_{A_t|S_{t-1}}(3|s)=x_s+y_s-1.
\end{align}
The mapping from $x_s$, $y_s$, $s \in \mc S$ to $P_{A_t|S_{t-1}}$ is unique in this case.
Inequality \eqref{eq:pa3_pos} ensures that $P_{A_t|S_{t-1}}(3|s)\geq0$. This constraint is implicitly required and makes this approach suboptimal in general as it does not appear in the outer bound $\bar{\mc C}_{\text{fb}+s}^\text{mem}$.

\subsection{Deterministic Scheme}
\label{sec:ach_det}
To avoid the drawbacks of the probabilistic scheme, a max-weight algorithm with action set $\mc A_3$ can be defined.
The actions are chosen according to the following criterion:
\begin{align}
A_t =& \argmax_{A_t\in \mc A_3} \sum_{j=1}^2 \sum_{l=1}^2 \sum_{m \in \mc O_l} \left[ Q_{l,t}^{(j)} - Q_{m,t}^{(j)} \right]^+ \mathbb E \left[ C_{lm,t}^{(j)} \bigg| \ve Q_t = \ve q, S_{t-1}=s \right]. \label{eq:maxweight}
\end{align}
The strategy can be shown to strongly stabilize all queues in the network for every rate pair $(R_1+\bar\delta, R_2+\bar\delta) \in \underline{\mc C}_{\text{ fb}+s}^\text{mem}$. The proof is similar to the proof of Theorem~\ref{prop:maxweight_new} and is omitted.

\subsection{Combination of Memoryless Strategies}
\label{sec:minksum}
Looking at the characterization of $\underline{\mc C}_{\text{fb}+s}^{\text{mem}}$ in \eqref{eq:pos} - \eqref{eq:R2_constr2}, one may wonder if this rate region can be attained by simply \emph{combining} memoryless capacity achieving schemes. Let $\mc C_{\text{fb}+s}(s)$, $s\in \mc S$, denote the capacity region of a memoryless BPEC with feedback and erasure probabilities $P_{\ve Z_t|S_{t-1}}(\cdot|s)$. Capacity achieving algorithms  for memoryless BPECs with feedback are derived in \cite{georgiadis2009broadcast}.

A combination of memoryless capacity achieving schemes may be described as follows: %
\begin{itemize}
 \item Choose fractions $\alpha_s\geq 0$ and $\beta_s\geq 0$ such that $\sum_{s\in \mc S} \alpha_s = \sum_{s\in \mc S} \beta_s=1$ and \linebreak $(\alpha_s R_1, \beta_s R_2) \in \pi_s \mc C_{\text{fb}+s}(s)$, for all $s \in \mc S$. 
 \item Take $n \alpha_s R_1$ packets for $\text{Rx}_1$ and $n \beta_s R_2$ packets for $\text{Rx}_2$ to be transmitted only when the previous channel state is equal to $S_{t-1}=s$, $s \in \mc S$. For each previous state $s \in \mc S$, the transmitter chooses an optimal memoryless strategy corresponding to a memoryless BPEC with feedback and erasure probabilities $P_{\ve Z_t|S_{t-1}}(\cdot|s)$.
\end{itemize}

The transmitter needs to maintain a set of queues for each state.
If Algorithm~III of \cite{georgiadis2009broadcast} is chosen as the capacity-achieving algorithm, the coding buffers contain at most one packet per session. Hence, at most $|\mc S|$ packets must be stored at each receiver.

Using the above scheme, for large $n$, one can asymptotically achieve the performance of the memoryless strategy for each state $s$ with the corresponding capacity region $\mc C_{\text{fb}+s}(s)$. The overall rate region achievable by this strategy, called $\mc R_{\oplus}$, is thus a weighted combination of the individual memoryless rate regions (for each state $s$):
\begin{align}
 \mc R_{\oplus} = \bigoplus_{s \in \mc S} \pi_s \mc C_{\text{fb}+s}(s)
\end{align}
where $\oplus$ denotes the set addition operator (Minkowski sum).
Fig.~\ref{fig:Minksum} shows that $\mc R_{\oplus}$ is strictly smaller than $\underline{\mc C}_{\text{fb}+s}^{\text{mem}}$. We outline an explanation why this is in general the case in the following remark.%
\begin{remark}
 The scheme $\mc R_\oplus$ effectively considers the constraints in \eqref{eq:rate_bound}~-~\eqref{eq:bound23} separately for each channel state and adds the corresponding regions. The flow for each state has to satisfy the conservation constraints in \eqref{eq:rate_bound}~-~\eqref{eq:flow_cons}. This is in general more restrictive than having the corresponding constraint for the \emph{sum} of all flows and makes $\mc R_\oplus$ smaller than $\underline{\mc C}_{\text{fb}+s}$.
\end{remark}

\section{Memoryless Case}
\label{sec:memoryless}
This section deals with the case of finite-state \emph{memoryless} broadcast packet erasure channels. For this case, we show that  the region $\underline{\mc C}_{\text{ fb}+s}^\text{mem}$ (described in \eqref{eq:pos}~-~\eqref{eq:R2_constr2}) and the region $\bar{\mc C}_{\text{ fb}+s}^\text{mem}$ (described in \eqref{eq:posouter_hmm_wind_outer_main}~-~\eqref{eq:R2_constr2outer_hmm_wind_outer_main}) match and can be achieved by reactive coding schemes of Section \ref{sec:ach}.
Our argumentation will build on geometric properties of the memoryless rate region $\mc C_\text{fb}$ that we derive next.

\subsection{Memoryless BPEC with Feedback}
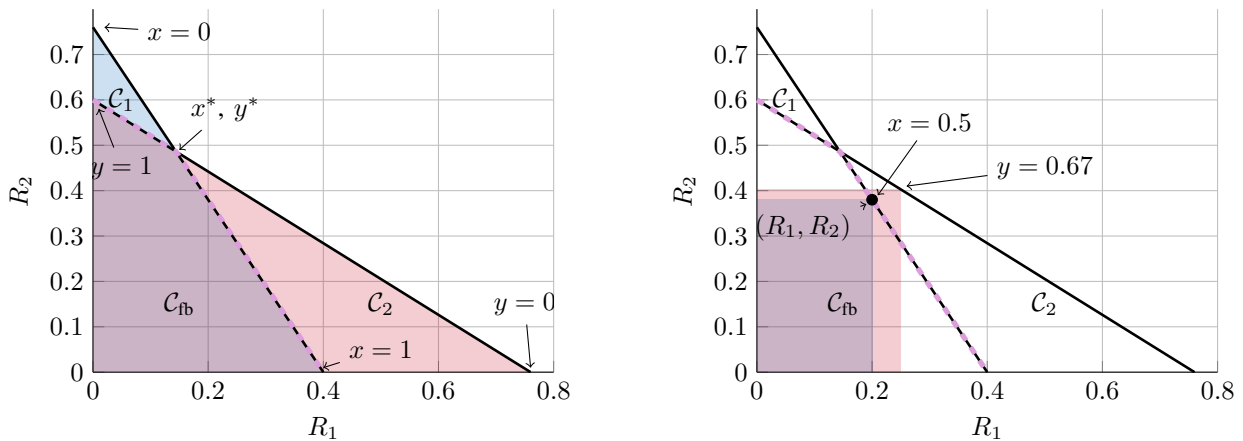
\begin{figure}[h]
 \centering
\begin{subfigure}[t]{0.48\textwidth}
 \centering

\begin{tikzpicture}
\begin{axis}[scale=0.5,
width=4.82222222222222in,
height=3.80333333333333in,
scale only axis,
xmin=0,
xmax=0.8,
xlabel={$R_1$},
xmajorgrids,
ymin=0,
ymax=0.8,
ylabel={$R_2$},
ymajorgrids,
ytick = {0,0.1,0.2,0.3,0.4,0.5,0.6,0.7},
axis x line*=bottom,
axis y line*=left,
legend style={at={(0.0,0.0)},anchor=south west,draw=black,fill=white,legend cell align=left, font=\scriptsize}
]

\addplot[fill=TUMblue,fill opacity=0.2,draw=none] coordinates {(0,0) (0,0.76) (0.4,0)} \closedcycle;
\addplot[fill=TUMred,fill opacity=0.2,draw=none] coordinates {(0,0) (0,0.6) (0.76,0)} \closedcycle;

\addplot [
color=black,
solid,
line width=1.0pt,
]
table[row sep=crcr]{
0 0.6\\
0.76 0\\
};

\addplot [
color=black,
solid,
line width=1.0pt,
]
table[row sep=crcr]{
0 0.76\\
0.4 0\\
};r

\addplot [
color=plum,
dashed,
line width=2.0pt,
]
table[row sep=crcr]{
0 0.6\\
0.1441 0.4863\\
0.4 0\\
};

\node at (axis cs:0.5,0.15) {$\mc C_2$};
\node at (axis cs:0.05,0.6) {$\mc C_1$};
\node at (axis cs:0.15,0.15) {$\mc C_{\text{fb}}$};

 \node (x0) at (axis cs:0.15,0.75) {$x=0$};
 \node (x1) at (axis cs:0.5,0.05) {$x=1$};
 \node (xstar) at (axis cs:0.23,0.58) {$x^*$, $y^*$};
 \draw[->] (x0) -- (axis cs:0.02,0.76);
 \draw[->] (x1) -- (axis cs:0.4,0.01);
 \draw[->] (xstar) -- (axis cs:0.15,0.49);
 
 \node (y0) at (axis cs:0.75,0.15) {$y=0$};
 \node (y1) at (axis cs:0.05,0.45) {$y=1$};
 
 \draw[->] (y1) -- (axis cs:0.01,0.58);
 \draw[->] (y0) -- (axis cs:0.76,0.01);

\end{axis}
\end{tikzpicture}%
\caption{Capacity region $\mc C_\text{fb}$ as intersection of $\mc C_1$ and $\mc C2$.}
\label{fig:fb_geo}
\end{subfigure}
\begin{subfigure}[t]{.48\textwidth}
 \centering

\begin{tikzpicture}
\begin{axis}[scale=0.5,
width=4.82222222222222in,
height=3.80333333333333in,
scale only axis,
xmin=0,
xmax=0.8,
xlabel={$R_1$},
xmajorgrids,
ymin=0,
ymax=0.8,
ylabel={$R_2$},
ymajorgrids,
ytick = {0,0.1,0.2,0.3,0.4,0.5,0.6,0.7},
axis x line*=bottom,
axis y line*=left,
legend style={at={(0.0,0.0)},anchor=south west,draw=black,fill=white,legend cell align=left, font=\scriptsize}
]

\addplot[fill=TUMblue,fill opacity=0.2,draw=none] coordinates {(0,0) (0.2,0) (0.2,0.38) (0,0.38)} \closedcycle;

\addplot[fill=TUMred,fill opacity=0.2,draw=none] coordinates {(0,0) (0.25,0) (0.25,0.4026) (0,0.4026)} \closedcycle;

\addplot [
color=black,
solid,
line width=1.0pt,
]
table[row sep=crcr]{
0 0.6\\
0.76 0\\
};

\addplot [
color=black,
solid,
line width=1.0pt,
]
table[row sep=crcr]{
0 0.76\\
0.4 0\\
};

\addplot [
color=plum,
dashed,
line width=2.0pt,
]
table[row sep=crcr]{
0 0.6\\
0.1441 0.4863\\
0.4 0\\
};

\node at (axis cs:0.5,0.15) {$\mc C_2$};
\node at (axis cs:0.05,0.6) {$\mc C_1$};
\node at (axis cs:0.15,0.15) {$\mc C_{\text{fb}}$};

  \node (x0) at (axis cs:0.3,0.55) {$x=0.5$};
  \draw[->] (x0) -- (axis cs:0.21,0.39);
  \node (y0) at (axis cs:0.5,0.45) {$y=0.67$};
  \draw[->] (y0) -- (axis cs:0.26,0.41);

  \node (R1R2) at (axis cs:0.08,0.32) {$(R_1,R_2)$};
  \draw[->] (R1R2) -- (axis cs:0.19,0.37);

  \filldraw (axis cs:0.2,0.38) circle (2pt);
\end{axis}
\end{tikzpicture}%
\caption{Rate region for a suitable choice of $x$ and $y$ such that only one constraint set is binding.}
\label{fig:fb_geo_illu2}
\end{subfigure}
\caption{Capacity region for $\epsilon_1 = 0.6$, $\epsilon_2=0.4$, $\epsilon_{12}=\epsilon_1 \epsilon_2$.}
\end{figure}
The 2-receiver capacity region of memoryless BPEC with feedback $\mc C_{\text{fb}}$ is characterized in \eqref{eq:bec+fb_cap1}~-~\eqref{eq:bec+fb_cap2} and can be written in the following alternative representation:
A rate pair $(R_1,R_2)$ is in the capacity region $\mc C_\text{fb}$ if there are variables $x$, $y$, such that
\begin{alignat}{3}
 &0&&\leq x &&\leq 1 \label{eq:fb_feas1}\\
 &0&&\leq R_1&&\leq (1-\epsilon_1) x \label{eq:fb_feas_C11}\\
 &0&&\leq R_2&&\leq (1-\epsilon_{12}) (1-x)  \label{eq:fb_feas_C12}\\
 &0&&\leq y &&\leq 1 \label{eq:fb_feas2}\\
 &0&&\leq R_1&&\leq (1-\epsilon_{12}) (1-y)  \label{eq:fb_feas_C21}\\
 &0&&\leq R_2&&\leq (1-\epsilon_2) y. \label{eq:fb_feas_C22}
\end{alignat}
One can see that $\mc C_1$ in \eqref{eq:bec+fb_cap1} is represented by \eqref{eq:fb_feas1}~-~\eqref{eq:fb_feas_C12} and $\mc C_2$ in \eqref{eq:bec+fb_cap2} by \eqref{eq:fb_feas2}~-~\eqref{eq:fb_feas_C22}, as visualized in Fig.~\ref{fig:fb_geo}.
The variables $x$ and $y$ define a point on the boundary of $\mc C_1$ and $\mc C_2$, respectively.
The values $x^*$ and $y^*$ define the intersection of the lines defining $\mc C_1$ and $\mc C_2$.
For $0 \leq x \leq x^*$, the boundary of $\mc C_\text{fb}$ is specified by $\mc C_2$ only; i.e., \eqref{eq:fb_feas_C21} is more restrictive than \eqref{eq:fb_feas_C11} for $R_1$:
\begin{align}
 (1-\epsilon_{12})(1-y)  \leq (1-\epsilon_1)x, \qquad \text{for } 0\leq x \leq x^*. \label{eq_fb_feasb_cond1}
\end{align}
As $\frac{1-\epsilon_{12}}{1-\epsilon_{1}} \geq 1$, the following constraint  implicitly applies also for $0\leq x \leq x^*$:
\begin{align}
 x+y \geq 1 \label{eq:feas_fb_redundant} 
\end{align}
Similarly, for  $x^* \leq x \leq 1$, the boundary of $\mc C_\text{fb}$ is specified by $\mc C_1$; i.e., \eqref{eq:fb_feas_C12} is more restrictive than \eqref{eq:fb_feas_C22} for $R_2$:
\begin{align}
 (1-\epsilon_{12})(1-x)  \leq (1-\epsilon_2)y, \qquad \text{for } x^*\leq x \leq 1. \label{eq_fb_feasb_cond2}
\end{align}
As $\frac{1-\epsilon_{12}}{1-\epsilon_{2}} \geq 1$, the constraint in \eqref{eq:feas_fb_redundant}  implicitly  applies also for $x^*\leq x \leq 1$.
Hence, \eqref{eq:feas_fb_redundant} implicitly holds for all pairs of $x$ and $y$ describing a point on the boundary of $\mc C_{\text{fb}}$ and can be included in the characterization of $\mc C_{\text{fb}}$.

\subsection{Memoryless Compound BPEC with Feedback}
Recall that the capacity region $\mc C_{\text{ fb}+s}^\text{mem}$ differs from the rate region achieved with reactive coding
$\underline{\mc C}_{\text{ fb}+s}^\text{mem} $ only through the constraints 
\begin{align}
 x_s + y_s \geq 1 \qquad \forall~s\in \mc S. \label{eq:sum_memless}
\end{align}
We showed that this constraint is implicitly given for a single channel state.
However, the results in Fig.~\ref{fig:GE_HMM} and Section~\ref{sec:comparison} illustrate that $\underline{\mc C}_{\text{ fb}+s}^\text{mem} $ is strictly smaller than ${\mc C}_{\text{ fb}+s}^\text{mem} $ in general.
For the finite-state memoryless case, note that  $P_{S_t|S_{t-1}}(s|s') = P_{S_t}(s) = \pi_s$.
It follows that $\epsj{s}$ is independent of $s$ and equal to the \emph{average} erasure probability
\begin{align}
 \epsj{s} = \bar \epsilon_j = \sum_{s'\in \mc S} \pi_{s'} P_{Z_j|S_t}(1|s') \qquad \forall~s \in \mc S.
\end{align}
One can define $\epsonetwo{s}=\bar \epsilon_{12}$ correspondingly.
So the capacity region ${\mc C}_{\text{ fb}+s}^\text{mem}$ simplifies to
 \begin{align}
 &0\leq x_s\leq 1,\quad 0\leq y_s \leq 1 \label{eq:posouter_memless}\\
  &R_1  \leq (1-\bar \epsilon_1) \sum_{s \in \mc S} \pi_s  x_s  \label{eq:R1_constr1outer_memless} \\
  &R_1  \leq (1- \bar \epsilon_{12}) \left( 1 - \sum_{s \in \mc S} \pi_s y_s \right) \label{eq:R1_constr2outer_memless} \\
  &R_2  \leq (1- \bar \epsilon_{12})\sum_{s \in \mc S} \pi_s  y_s  \label{eq:R2_constr1outer_memless}  \\
  &R_2  \leq (1- \bar \epsilon_{2}) \left( 1- \sum_{s \in \mc S} \pi_s  x_s \right).  \label{eq:R2_constr2outer_memless}
 \end{align}
By  defining the new variables 
\begin{align}
 \bar x = \sum_{s \in \mc S} \pi_s  x_s, \qquad  \bar y = \sum_{s \in \mc S} \pi_s  y_s
\end{align}
we obtain a characterization that is similar to \eqref{eq:fb_feas1}~-~\eqref{eq:fb_feas_C22} for the single-state memoryless case.
Any feasible choice of $x_s$, $y_s$, $s \in \mc S$ leads to a particular value of $\bar x$ and $\bar y$.
From the previous section we know that we can restrict attention to those pairs $x_s$, $y_s$, $s\in \mc S$ that lead to %
\begin{align}
 \bar x + \bar y \geq 1 \label{eq:sum_memless_star}
\end{align}
without shrinking the rate region.
Moreover, wor every pair of $\bar x$, $\bar y$ satisfying \eqref{eq:sum_memless_star} that resulted from a particular choice of $x_s$, $y_s$, we can find $|\mc S|$ pairs $\hat x_s$, $\hat y_s$ that
\begin{itemize}
 \item yield the same values for $\bar x$, $\bar y$ as the original choice $x_s$, $y_s$, hence
 \item lead to the same bounds in \eqref{eq:posouter_memless}~-~\eqref{eq:R2_constr2outer_memless} and
 \item satisfy \eqref{eq:sum_memless}, as we can set 
\begin{align}
  \hat x_s = \bar x , \qquad \hat y_s = \bar y \qquad \forall~s \in \mc S. \label{eq:ys_star_choice}
\end{align}
\end{itemize}
This shows that $\underline{\mc C}_{\text{ fb}+s}^\text{mem} = {\mc C}_{\text{ fb}+s}^\text{mem} $ for memoryless channels. %
The choice in \eqref{eq:ys_star_choice} means that the probability distributions $P_{A_t|S_{t-1}}$ are independent of $S_{t-1}$, hence the strategy does not need to adapt to the previous state.

\section{Numerical Results}
\label{sec:comparison}

\subsection{Visible Case}
\begin{figure}[h]
 \centering
 \begin{tikzpicture}[->,>=stealth',shorten >=1pt,auto,node distance=30mm,
  thick,main node/.style={circle,inner sep=0.1,fill=gray!20,draw},every node/.style={scale=1.1}, scale=0.6]
  \node[main node] (GG) {$\text{GG}$};
  \node[main node] (GB) [below of=GG] {$\text{GB}$};
  \node[main node] (BG) [right of=GG] {$\text{BG}$};
  \node[main node] (BB) [below of=BG] {$\text{BB}$};

  \path[every node/.style={font=\sffamily\small}]
    (GG) edge [bend right=10] node [below,pos=0.76] {$b_1 b_2$} (BB)
        edge [bend right=20] node[above,rotate=90] {$(1-b_1)b_2$} (GB)
        edge [loop above] node {$(1-b_1)(1-b_2)$} (GG)
        edge [bend left=20] node {$b_1(1-b_2)$} (BG)
    (BG) edge [bend left=20] node [below,rotate=90] {$(1-g_1)b_2$} (BB)
         edge [bend left=10] node [below, pos=0.76] {$g_1 b_2$} (GB)
         edge [bend left=20] node [above] {$g_1 (1-b_2)$} (GG)
         edge [loop above] node {$(1-g_1)(1-b_2) $} (BG)
    
    (GB) edge [bend right=20] node [below] {$b_1(1-g_2)$} (BB)
         edge [bend left=10] node [left, pos=0.88] {$b_1 g_2$} (BG)
         edge [bend right=20] node [above,rotate=90] {$(1-b_1)g_2$} (GG)
         edge [loop below] node {$(1-b_1)(1-g_2)$} (GB)    
         
    (BB) edge [bend right=20] node {$ g_1(1-g_2)$} (GB)
         edge [bend right=10] node[right,pos=0.88] {$g_1 g_2 $} (GG)
         edge [bend left=20] node [below,rotate=90] {$(1-g_1)g_2 $} (BG)
         edge [loop below] node {$(1-g_1)(1-g_2)$} (BB);
\end{tikzpicture}
\caption{Markov Chain of channel state space $\mc S$ with transition probabilities.}
\label{fig:states}
\end{figure}
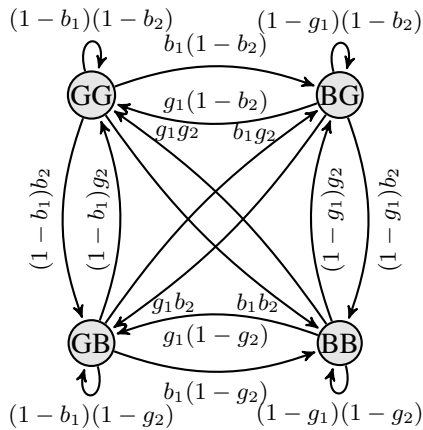

Our main focus will be on the visible Gilbert-Elliot model outlined in Example~\ref{exampleGilbert}:
We assume that the individual channels to \textnormal{Rx$_1$}~and \textnormal{Rx$_2$}~are both Gilbert-Elliot channels with states $\text{G}$ and $\text{B}$. The broadcast channel state space is therefore given by $\mc S = \{\text{GG},\text{GB}, \text{BG}, \text{BB}\}$ where $\text{G}$ and $\text{B}$ respectively refer to a good and bad state at each receiver. 
Transitions from  state $\text{B}$ to state $\text{G}$ occur with probability $g_j$ for \textnormal{Rx$_j$}, $j=1,2$. 
Similarly, a transition from state $\text{G}$ to state $\text{B}$ occurs with probability $b_j$ for \textnormal{Rx$_j$}.
For simplicity, these transitions are independent across the two users.
The corresponding finite-state Markov chain is summarized in Fig.~\ref{fig:states}.

In the visible case, where an erasure occurs always in state $\text{B}$ and never in state $\text{G}$, the long-term average erasure probability at \textnormal{Rx$_j$}~is
\begin{align}
 \epsilon_j = \frac{b_j}{g_j + b_j}\cdot
\end{align}
Given an average erasure probability $\epsilon_j$, $g_j$ determines $b_j$ and specifies the channel to \textnormal{Rx$_j$}.

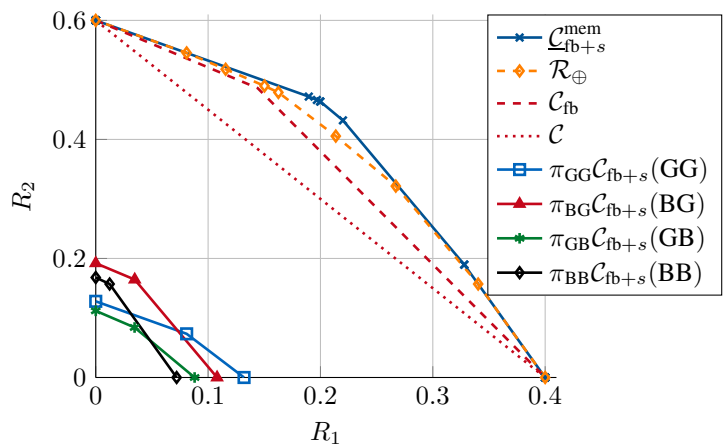
\begin{figure}[t]
 \centering

\begin{tikzpicture}
\begin{axis}[scale=0.5,
width=4.82222222222222in,
height=3.80333333333333in,
scale only axis,
xmin=0,
xmax=0.41,
xlabel={$R_1$},
xmajorgrids,
every outer x axis line/.append style={black},
every x tick label/.append style={font=\color{black}},
ymin=0,
ymax=0.61,
ylabel={$R_2$},
ymajorgrids,
every outer y axis line/.append style={black},
every y tick label/.append style={font=\color{black}},
axis x line*=bottom,
axis y line*=left,
legend style={at={(0.85,1.0)},anchor=north west,draw=black,fill=white,legend cell align=left}
]

\addplot [
color=TUMpantone301,
solid,
line width=1.0pt,
mark=x,
mark options={solid}
]
  table[row sep=crcr]{%
-2.20554111952931e-09	0.599999994315768\\
-2.17065155606877e-09	0.599999994577054\\
-2.12795018064704e-09	0.599999994820831\\
-2.07779714787859e-09	0.599999995048889\\
-1.47167629818701e-09	0.59999999775458\\
-1.723679358967e-11	0.599999999925199\\
7.68199913403489e-12	0.599999999928623\\
0.189600000728751	0.47199999930733\\
0.196832802429257	0.466271047146064\\
0.199830683135434	0.463720834318547\\
0.199830684957968	0.463720831498871\\
0.219999999898067	0.431999999883207\\
0.328000000151881	0.189599999312797\\
0.32800000328372	0.189599991150972\\
0.399999996247596	2.77485687205647e-10\\
0.399999999939633	-1.46718748261776e-12\\
0.399999999957988	1.54624368864376e-11\\
0.399999999958552	1.61638480378201e-11\\
0.399999999960768	1.9636958725755e-11\\
};
\addlegendentry{$\underline{\mc C}_{\text{fb}+s}^{\text{mem}}$ };

\addplot [
color=TUMorange,
dashed,
line width=1.0pt,
mark=diamond,
mark options={solid}
]
table[row sep=crcr]{
0.4 0\\
0.340362231345086 0.157046124124608\\
0.267064246191321 0.321559379691946\\
0.213642813195244 0.405576724313048\\
0.162561486983413 0.478948083780951\\
0.150199255638328 0.489901959656344\\
0.115620688634404 0.517884615035241\\
0.0809186737881689 0.545371359467903\\
0 0.6\\
};
\addlegendentry{$\mc{R}_{\oplus}$};

\addplot [
color=TUMred,
dashed,
line width=1.0pt
]
table[row sep=crcr]{
0 0.6\\
0.144075829383886 0.486255924170616\\
0.4 0\\
};
\addlegendentry{$\mc C_{\text{fb}}$ };

\addplot [
color=TUMred,
dotted,
line width=1.0pt
]
table[row sep=crcr]{
0 0.6\\
0.4 0\\
};
\addlegendentry{$\mc C$ };

\addplot [
color=TUMblue,
solid,
line width=1.0pt,
mark=square,
mark options={solid}
]
table[row sep=crcr]{
0 0.128\\
0.0809186737881689 0.0733713594679029\\
0.132 0\\
};
\addlegendentry{$\pi{}_{\text{GG}}\mc{ C}_{\text{fb}+s}(\text{GG})$};

\addplot [
color=TUMred,
solid,
line width=1.0pt,
mark=triangle*,
mark options={solid}
]
table[row sep=crcr]{
0 0.192\\
0.0347020148462354 0.164513255567338\\
0.108 0\\
};
\addlegendentry{$\pi{}_{\text{BG}}\mc{ C}_{\text{fb}+s}(\text{BG})$};

\addplot [
color=TUMgreen,
solid,
line width=1.0pt,
mark=asterisk,
mark options={solid}
]
table[row sep=crcr]{
0 0.112\\
0.0345785670039232 0.0840173446211026\\
0.088 0\\
};
\addlegendentry{$\pi{}_{\text{GB}}\mc{C}_{\text{fb}+s}(\text{GB})$};

\addplot [
color=black,
solid,
line width=1.0pt,
mark=diamond,
mark options={solid}
]
table[row sep=crcr]{
0 0.168\\
0.0123622313450857 0.157046124124608\\
0.072 0\\
};
\addlegendentry{$\pi{}_{\text{BB}}\mc{C}_{\text{fb}+s}(\text{BB})$};

\end{axis}
\end{tikzpicture}%
\caption{Individual rate regions and Minkowski sum $\mc R_{\oplus}$ for $\epsilon_1=0.6$, $\epsilon_2=0.4$, $g_1=0.3$, $g_2=0.7$. The region $\underline{\mc C}_{\text{fb}+s}^{\text{mem}}$ is strictly larger than $\mc R_{\oplus}$. For comparison, the corresponding capacity regions for memoryless channels with the same average erasure probability are shown for the cases with and without feedback. %
}
\label{fig:Minksum}
\end{figure}

We first compare the regions ${\mc C}_{\text{fb}+s}^{\text{mem}}$, $\underline{\mc C}_{\text{fb}+s}^{\text{mem}}$ and $\mc R_\oplus$ for the strategy combining the memoryless schemes in Section~\ref{sec:minksum}:
Fig.~\ref{fig:Minksum} shows that $\mc R_\oplus$ is smaller than $\underline{\mc C}_{\text{fb}+s}^{\text{mem}}$, despite their similar structure.
$\underline{\mc C}_{\text{fb}+s}^{\text{mem}}$ almost matches the capacity region ${\mc C}_{\text{fb}+s}^{\text{mem}}$ for this example, so ${\mc C}_{\text{fb}+s}^{\text{mem}}$ is omitted for clarity.
For comparison, we also plot the memoryless capacity regions for the same average erasure probabilities with ($\mc C_\text{fb}$) and without feedback ($\mc C$). 
The scaled individual rate regions $\pi_s \mc C_{\text{fb}+s}(s)$ for the $|\mc S|=4$ memoryless states are shown as well. Their Minkowski sum results in $\mc R_\oplus$.

\begin{figure}[t]
 \centering

\begin{tikzpicture}
\begin{axis}[scale=0.5,
width=4.82222222222222in,
height=3.80333333333333in,
scale only axis,
xmin=0,
xmax=0.51,
xlabel={$R_1$},
xmajorgrids,
every outer x axis line/.append style={black},
every x tick label/.append style={font=\color{black}},
ymin=0,
ymax=0.51,
ylabel={$R_2$},
ymajorgrids,
every outer y axis line/.append style={black},
every y tick label/.append style={font=\color{black}},
axis x line*=bottom,
axis y line*=left,
legend style={at={(0,0)},anchor=south west,draw=black,fill=white,legend cell align=left}
]

\addplot [color=TUMred,solid,line width=1.0pt,mark=o,mark options={solid}]
  table[row sep=crcr]{%
-6.3881585785075e-10	0.499999998130779\\
-2.13710755689323e-10	0.499999999398456\\
1.70510008890012e-10	0.499999999336781\\
2.21881249484746e-10	0.499999999328496\\
2.75490883577811e-10	0.499999999319605\\
3.31491344790491e-10	0.499999999310049\\
3.90131420224815e-10	0.499999999299615\\
4.51741009183104e-10	0.499999999288115\\
5.8451112006086e-10	0.499999999262883\\
6.59816385556455e-10	0.499999999245871\\
7.46618900385188e-10	0.499999999221409\\
1.58461724557996e-08	0.499999994231098\\
0.215000000131436	0.424999999812868\\
0.325000010300229	0.349999991654879\\
0.404020468021695	0.291154969886671\\
0.449999999967801	0.189999999846005\\
0.499999994022514	-2.564584586362e-09\\
0.499999999802543	4.94053364191727e-10\\
};
\addlegendentry{$\bar{\mc C}_{\text{fb}+s}^{\text{mem}}$};

\addplot [
color=TUMpantone301,
solid,
line width=1.0pt,
mark=x,
mark options={solid}
]
table[row sep=crcr]{
1.26301822268788e-10 0.499999999894753\\
0.214999997206266 0.425000000464778\\
0.215000002017283 0.424999998333691\\
0.311315164865004 0.359330568917867\\
0.408010680059939 0.282376502247954\\
0.408010682812185 0.282376496593052\\
0.450000000104885 0.189999999091287\\
0.499999999566823 1.33745843283371e-09\\
0.499999999831986 2.51077649096155e-10\\
};
\addlegendentry{$\underline{\mc C}_{\text{fb}+s}^{\text{mem}}$ };

\addplot [
color=TUMorange,
solid,
line width=1.0pt,
mark=diamond,
mark options={solid}
]
table[row sep=crcr]{
0.5 0\\
0.455210237659963 0.170201096892139\\
0.428264129875532 0.229482534017887\\
0.367695526661688 0.300650642794154\\
0.360489762049999 0.30839683975172\\
0.355279524390035 0.313195742859582\\
0.21584812760388 0.417027634083315\\
0.192794235388311 0.432746196957566\\
0 0.5\\
};
\addlegendentry{$\mc{R}_{\oplus}$};

\addplot [
color=TUMred,
dashed,
line width=1.0pt
]
table[row sep=crcr]{
0 0.5\\
0.3 0.3\\
0.5 0\\
};
\addlegendentry{$\mc C_{\text{fb}}$};

\addplot [
color=TUMred,
dotted,
line width=1.0pt
]
table[row sep=crcr]{
0 0.5\\
0.5 0\\
};
\addlegendentry{$\mc C$};

\end{axis}
\end{tikzpicture}%
\caption{Rate regions for $\epsilon_1=0.5$, $\epsilon_2=0.5$, $g_1=0.2$, $g_2=0.3$. 
}
\label{fig:RateReg1}
\end{figure}
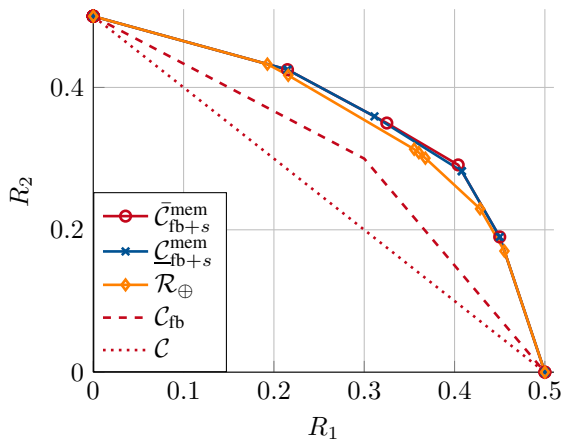

 As a second example, Fig.~\ref{fig:RateReg1} shows the capacity region for a channel with parameters $\epsilon_1=0.5$, $\epsilon_2=0.5$, $g_1=0.2$, $g_2=0.3$.
 Again, the difference between $\underline{\mc C}_{\text{fb}+s}^{\text{mem}}$ and the capacity region ${\mc C}_{\text{fb}+s}^{\text{mem}}$ is small. The strategy for $\mc R_\oplus$ does not achieve capacity, but has a good performance.

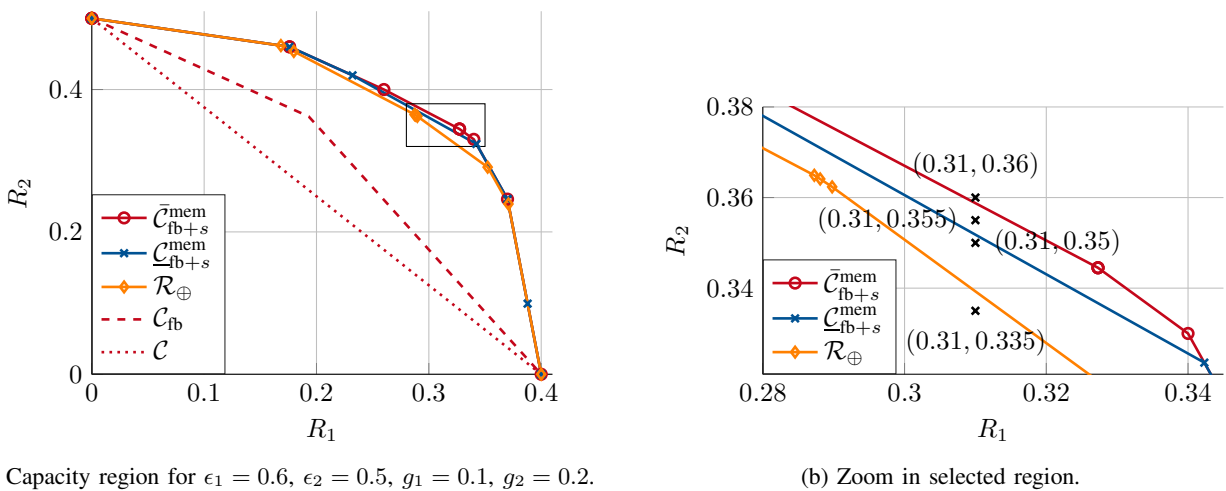
\begin{figure}[t]
 \centering
\begin{subfigure}[t]{0.5\textwidth}
 \centering

\begin{tikzpicture}
\begin{axis}[scale=0.5,
width=4.82222222222222in,
height=3.80333333333333in,
scale only axis,
every outer x axis line/.append style={black},
every x tick label/.append style={font=\color{black}},
xmin=0,
xmax=0.41,
xlabel={$R_1$},
xmajorgrids,
every outer y axis line/.append style={black},
every y tick label/.append style={font=\color{black}},
ymin=0,
ymax=0.51,
ylabel={$R_2$},
ymajorgrids,
axis x line*=bottom,
axis y line*=left,
legend style={at={(0.0,0.0)},anchor=south west,draw=white!15!black,fill=white,legend cell align=left}
]

\addplot [color=TUMred,solid,line width=1.0pt,mark=o,mark options={solid}]
  table[row sep=crcr]{%
2.39391839684799e-16	0.5\\
6.05418493115906e-16	0.5\\
2.48478321251966e-13	0.499999999999199\\
0.176	0.46\\
0.176000000000006	0.459999999999996\\
0.260000000000027	0.399999999999875\\
0.327293982368138	0.34449980835617\\
0.327293982368828	0.344499808355434\\
0.327293982371985	0.344499808351849\\
0.339999999999871	0.330000000000106\\
0.369999999999096	0.246000000001124\\
0.399999999999806	7.97327481816268e-13\\
};
\addlegendentry{$\bar{\mc C}_{\text{fb}+s}^\text{mem}$};

\addplot [
color=TUMpantone301,
solid,
line width=1.0pt,
mark=x,
mark options={solid},
]
table[row sep=crcr]{
7.58856181804537e-10 0.499999999560571\\
3.3139910487206e-09 0.49999999915519\\
0.175999999849385 0.459999999968773\\
0.232069249418006 0.419950535734466\\
0.342301444294153 0.323555955650704\\
0.369999999680172 0.246000000675594\\
0.369999999911246 0.245999999365646\\
0.370000000186229 0.245999997662693\\
0.387896962212648 0.0992449094811143\\
0.399999999169136 4.88041564734677e-09\\
0.399999999246513 5.73071705095629e-09\\
};
\addlegendentry{$\underline{\mc C}_{\text{fb}+s}^\text{mem}$};

\addplot [
color=TUMorange,
solid,
line width=1.0pt,
mark=diamond,
mark options={solid},
]
table[row sep=crcr]{
0.4 0\\
0.370830519918974 0.239189736664416\\
0.352337369234042 0.290970558582224\\
0.289784667049297 0.362354230487169\\
0.288097042959291 0.364101417780352\\
0.287266523040317 0.364911681115936\\
0.179819225225062 0.453528009210992\\
0.168312375909993 0.461747187293183\\
0 0.5\\
};
\addlegendentry{$\mc R_\oplus$};

\addplot [
color=TUMred,
dashed,
line width=1.0pt,
]
table[row sep=crcr]{
0 0.5\\
0.193103448275862 0.362068965517241\\
0.4 0\\
};
\addlegendentry{$\mc C_\text{fb}$};

\addplot [
color=TUMred,
dotted,
line width=1.0pt,
]
table[row sep=crcr]{
0 0.5\\
0.4 0\\
};
\addlegendentry{$\mc C$};

\draw (axis cs:0.28,0.32) -- (axis cs:0.35,0.32) -- (axis cs:0.35,0.38) -- (axis cs:0.28,0.38)-- (axis cs:0.28,0.32);
\end{axis}
\end{tikzpicture}%
\caption{Capacity region for $\epsilon_1 = 0.6$, $\epsilon_2=0.5$, $g_1=0.1$, $g_2=0.2$.}
\label{fig:rate_region_algo_comp}
\end{subfigure}
\begin{subfigure}[t]{0.45\textwidth}
 \centering

\begin{tikzpicture}

\begin{axis}[scale=0.5,
width=4.82222222222222in,
height=2.8in,
scale only axis,
every outer x axis line/.append style={black},
every x tick label/.append style={font=\color{black}},
xmin=0.28,
xmax=0.345,
xlabel={$R_1$},
xmajorgrids,
every outer y axis line/.append style={black},
every y tick label/.append style={font=\color{black}},
ymin=0.321,
ymax=0.38,
ylabel={$R_2$},
ymajorgrids,
axis x line*=bottom,
axis y line*=left,
legend style={at={(0.0,0.0)},anchor=south west,draw=white!15!black,fill=white,legend cell align=left}
]

\addplot [color=TUMred,solid,line width=1.0pt,mark=o,mark options={solid}]
  table[row sep=crcr]{%
2.39391839684799e-16	0.5\\
6.05418493115906e-16	0.5\\
2.48478321251966e-13	0.499999999999199\\
0.176	0.46\\
0.176000000000006	0.459999999999996\\
0.260000000000027	0.399999999999875\\
0.327293982368138	0.34449980835617\\
0.327293982368828	0.344499808355434\\
0.327293982371985	0.344499808351849\\
0.339999999999871	0.330000000000106\\
0.369999999999096	0.246000000001124\\
0.399999999999806	7.97327481816268e-13\\
};
\addlegendentry{$\bar{\mc C}_{\text{fb}+s}^\text{mem}$};

\addplot [color=TUMpantone301,solid,line width=1.0pt,mark=x,mark options={solid}]
  table[row sep=crcr]{%
7.58856181804537e-10	0.499999999560571\\
3.3139910487206e-09	0.49999999915519\\
0.175999999849385	0.459999999968773\\
0.232069249418006	0.419950535734466\\
0.342301444294153	0.323555955650704\\
0.369999999680172	0.246000000675594\\
0.369999999911246	0.245999999365646\\
0.370000000186229	0.245999997662693\\
0.387896962212648	0.0992449094811143\\
0.399999999169136	4.88041564734677e-09\\
0.399999999246513	5.73071705095629e-09\\
};
\addlegendentry{$\underline{\mc C}_{\text{fb}+s}^\text{mem}$};

\addplot [color=TUMorange,solid,line width=1.0pt,mark=diamond,mark options={solid}]
  table[row sep=crcr]{%
0	0\\
0.03	0\\
0.06	0\\
0.17	0\\
0.2	0\\
0.23	0\\
0.34	0\\
0.37	0\\
0.4	0\\
0.370830519918974	0.239189736664416\\
0.352337369234042	0.290970558582224\\
0.289784667049297	0.362354230487169\\
0.288097042959291	0.364101417780352\\
0.287266523040317	0.364911681115936\\
0.179819225225062	0.453528009210992\\
0.168312375909993	0.461747187293183\\
0	0.5\\
0	0.46\\
0	0.44\\
0	0.4\\
0	0.34\\
0	0.3\\
0	0.28\\
0	0.26\\
0	0.24\\
0	0.22\\
0	0.2\\
0	0.16\\
0	0.1\\
0	0.06\\
0	0.04\\
0	0\\
};
\addlegendentry{$\mc R_\oplus$};

\addplot [color=black,only marks,line width=1.0pt,mark=x,mark options={solid}]
  table[row sep=crcr]{%
0.31	0.335\\
};
\draw(axis cs:0.31,0.335) node[label=below:{$(0.31,0.335)$}]{};

\addplot [color=black,only marks,line width=1.0pt,mark=x,mark options={solid}]
  table[row sep=crcr]{%
0.31	0.36\\
};
\draw(axis cs:0.31,0.36) node[label=above:{$(0.31,0.36)$}]{};

\addplot [color=black,only marks,line width=1.0pt,mark=x,mark options={solid}]
  table[row sep=crcr]{%
0.31	0.35\\
};
\draw(axis cs:0.31,0.35) node[label=right:{$(0.31,0.35)$}]{};

\addplot [color=black,only marks,line width=1.0pt,mark=x,mark options={solid}]
  table[row sep=crcr]{%
0.31	0.355\\
};
\draw(axis cs:0.31,0.355) node[label=left:{$(0.31,0.355)$}]{};
\end{axis}
\end{tikzpicture}%
\caption{Zoom in selected region.}
\label{fig:rate_region_algo_comp_zoom}
\end{subfigure}
\caption{Capacity region for $\epsilon_1 = 0.6$, $\epsilon_2=0.5$, $g_1=0.1$, $g_2=0.2$ and zoomed selection. The marked rate points are simulated in Fig.~\ref{fig:fb_algo_comp}.}
\end{figure}

We also compare queue backlog of the deterministic schemes corresponding to ${\mc C}_{\text{fb}+s}^{\text{mem}}$, $\underline{\mc C}_{\text{fb}+s}^{\text{mem}}$ and $\mc R_\oplus$\footnote{For $\mc R_\oplus$, we apply the scheme in \eqref{eq:maxweight} for each state $s$ individually. For single-state channels, the scheme in \eqref{eq:maxweight} is capacity achieving. This explains that the aggregate backlog can be larger than $|\mc S|$, in contrast to the argumentation in Section~\ref{sec:minksum}.}, for the parameter set $\epsilon_1 = 0.6$, $g_1=0.1$, $\epsilon_2=0.5$, $g_2=0.2$:
The corresponding rate regions are shown in Fig.~\ref{fig:rate_region_algo_comp}.
We pick four rate points close to the boundary of all regions:
One point lies inside all three regions, one lies inside $\underline{\mc C}_{\text{fb}+s}^{\text{mem}}$ and $\mc C_{\text{fb}+s}^{\text{mem}}$ but outside of $\mc R_\oplus$, etc. Finally, one rate point lies outside of all regions, as illustrated in Fig.~\ref{fig:rate_region_algo_comp_zoom}.

\begin{figure}[h]
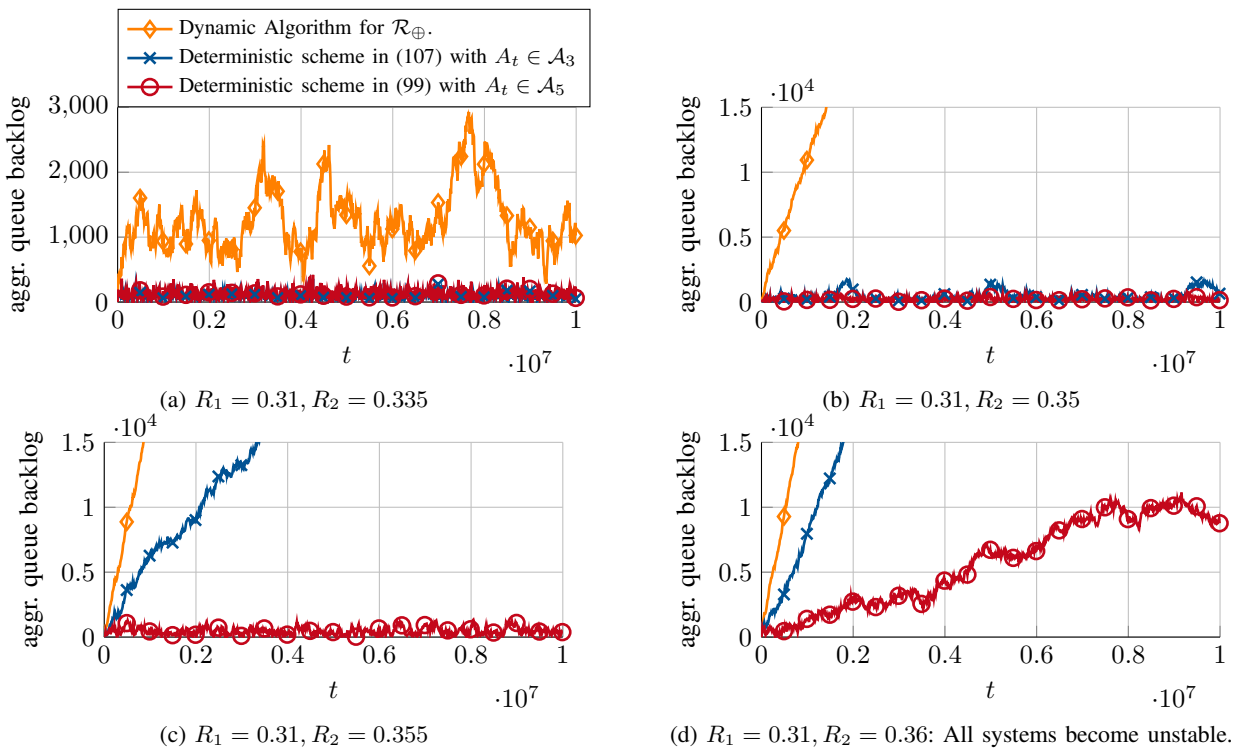

 \centering
\begin{subfigure}[t]{0.5\textwidth}
 \centering

%
\caption{$R_1=0.31,R_2=0.335$}
\label{fig:fb_algo_comp1}
\end{subfigure}
\begin{subfigure}[t]{0.45\textwidth}
 \centering

%
%
\caption{$R_1=0.31,R_2=0.35$}
\label{fig:fb_algo_comp2}
\end{subfigure}
\begin{subfigure}[t]{0.5\textwidth}
 \centering

%
%
\caption{$R_1=0.31,R_2=0.355$}
\label{fig:fb_algo_comp3}
\end{subfigure}
\begin{subfigure}[t]{0.45\textwidth}
 \centering

%
%
\caption{$R_1=0.31,R_2=0.36$: All systems become unstable.}
\label{fig:fb_algo_comp4}
\end{subfigure}
\caption{Aggregate queue backlog for different algorithms and rate points.}
\label{fig:fb_algo_comp}
\end{figure}

For all four points we apply the corresponding algorithms and keep track of the aggregate queue backlog, i.e. the total number of packets stored in all queues in the system.
Results are shown in Fig.~\ref{fig:fb_algo_comp}:
For the point inside all three regions (Fig.~\ref{fig:fb_algo_comp1}), all queues are stable, but the capacity-achieving algorithm in Table~\ref{tab:new_det_algo} has the lowest aggregate backlog on average.
If a point lies outside a rate region (e.g. outside of $\mc R_\oplus$ in Fig.~\ref{fig:fb_algo_comp2}), the corresponding queueing system becomes unstable. 
For the point outside of all regions, all systems are unstable and the average backlog grows slowest for the capacity-achieving scheme.

\begin{figure}[t]
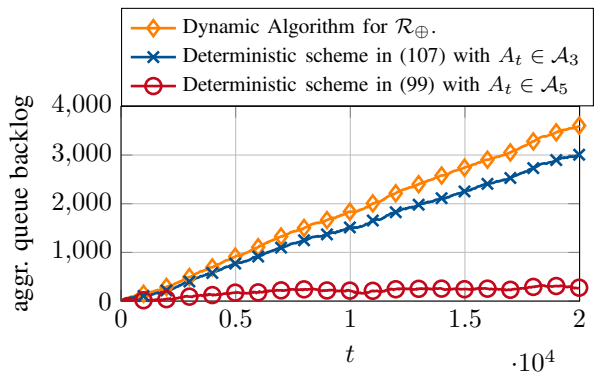

\centering

%
%
\caption{Aggregate queue backlog for the three different strategies, for the state model as in Fig.~\ref{fig:example_chain}, $\delta=0$ and $R_1=R_2=0.499$.}
\label{fig:queue_backlog_example_chain}
\end{figure}

The same behavior can be seen if we apply the algorithms to the $2$-state example in Example~\ref{ex:proactive}, for $\delta=0$ and $R_1=R_2=0.499$. 
Fig.~\ref{fig:queue_backlog_example_chain} shows that the capacity-achieving strategy stabilizes the queueing network whereas the other strategies do not.

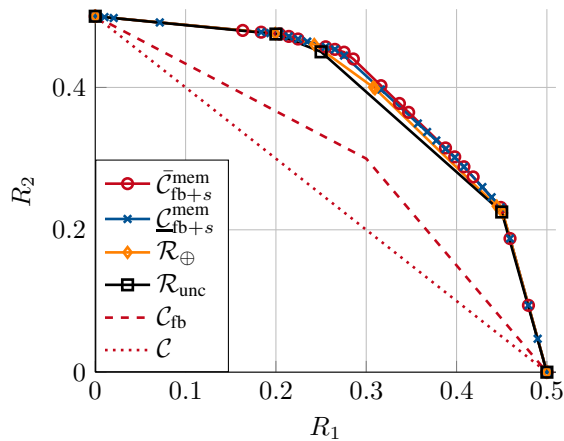
\begin{figure}[t]
 \centering

\begin{tikzpicture}
\begin{axis}[scale=0.5,
width=4.82222222222222in,
height=3.80333333333333in,
scale only axis,
xmin=0,
xmax=0.51,
xlabel={$R_1$},
xmajorgrids,
every outer x axis line/.append style={black},
every x tick label/.append style={font=\color{black}},
ymin=0,
ymax=0.51,
ylabel={$R_2$},
ymajorgrids,
every outer y axis line/.append style={black},
every y tick label/.append style={font=\color{black}},
axis x line*=bottom,
axis y line*=left,
legend style={at={(0.0,0.0)},anchor=south west,draw=black,fill=white,legend cell align=left}
]

\addplot [color=TUMred,solid,line width=1.0pt,mark=o,mark options={solid}]
  table[row sep=crcr]{%
0	0.499999999999932\\
0.163265306122449	0.480089596814315\\
0.183673469387755	0.477600796416103\\
0.204081632653061	0.475111996017783\\
0.214285714285714	0.471683673468783\\
0.224489795918367	0.46803935860037\\
0.255102040816327	0.457106413994151\\
0.26530612244898	0.453462099125346\\
0.275510204081633	0.449531445231133\\
0.285714285714286	0.4399999999996\\
0.316326530612245	0.402499999999943\\
0.336734693877551	0.377499999999999\\
0.346938775510204	0.364999999999998\\
0.387755102040816	0.314999999999764\\
0.397959183673469	0.302499999998257\\
0.408163265306122	0.288571428555139\\
0.418367346938776	0.274285714285535\\
0.448979591836735	0.231428571428432\\
0.459183673469388	0.187755102039516\\
0.479591836734694	0.093877551020059\\
0.5	-1.33330846363577e-13\\
};
\addlegendentry{$\bar{\mc C}_{\text{fb}+s}^\text{mem}$};

\addplot [color=TUMpantone301,solid,line width=1.0pt,mark=x,mark options={solid}]
  table[row sep=crcr]{%
0	0.499999999999567\\
0.0102040816326531	0.498755599800823\\
0.0204081632653061	0.497511199601763\\
0.0714285714285714	0.491289198606256\\
0.183673469387755	0.47760079641611\\
0.193877551020408	0.476356396216967\\
0.204081632653061	0.475111996017666\\
0.214285714285714	0.471683673469352\\
0.224489795918367	0.468039358600564\\
0.23469387755102	0.464395043731735\\
0.255102040816327	0.457106413994066\\
0.26530612244898	0.453462099124704\\
0.275510204081633	0.444803125711156\\
0.316326530612245	0.397082922388266\\
0.357142857142857	0.349362719065287\\
0.36734693877551	0.337432668234525\\
0.377551020408163	0.325502617403748\\
0.387755102040816	0.313572566572809\\
0.397959183673469	0.301642515741179\\
0.408163265306122	0.288571428571261\\
0.428571428571429	0.259999999999835\\
0.438775510204082	0.245714285711868\\
0.448979591836735	0.231428571421912\\
0.459183673469388	0.187755102040288\\
0.479591836734694	0.0938775510200486\\
0.489795918367347	0.0469387755095299\\
0.5	-7.62477234128589e-12\\
};
\addlegendentry{$\underline{\mc C}_{\text{fb}+s}^\text{mem}$};

\addplot [color=TUMorange,solid,line width=1.0pt,mark=diamond,mark options={solid}]
  table[row sep=crcr]{%
0.5	0\\
0.451380552220888	0.223649459783914\\
0.444531237152395	0.233238500879804\\
0.309755862776355	0.398338334490453\\
0.309080643330035	0.399030434422931\\
0.307700091109147	0.400380974639017\\
0.242475465485187	0.460281141028368\\
0.19932478055368	0.475692099932478\\
0	0.5\\
};
\addlegendentry{$\mc R_\oplus$};

\addplot [
color=black,
solid,
line width=1.0pt,
mark=square,
mark options={solid}
]
table[row sep=crcr]{
0 0.5\\
0.2 0.475\\
0.25 0.45\\
0.45 0.225\\
0.5 0\\
};
\addlegendentry{$\mc R_\text{unc}$};

\addplot [
color=TUMred,
dashed,
line width=1.0pt
]
table[row sep=crcr]{
0 0.5\\
0.3 0.3\\
0.5 0\\
};
\addlegendentry{$\mc C_\text{fb}$};

\addplot [
color=TUMred,
dotted,
line width=1.0pt
]
table[row sep=crcr]{
0 0.5\\
0.5 0\\
};
\addlegendentry{$\mc C$};

\end{axis}
\end{tikzpicture}%
\caption{Rate regions for $\epsilon_1=\epsilon_2=0.5$, $g_1=0.8$, $g_2=0.9$. }
\label{fig:curve3}
\vspace*{-3mm}
\end{figure}

Fig.~\ref{fig:curve3} shows an example with oscillatory channel memory.
$\mc R_\text{unc}$ describes the rate region achievable with the action set $\mc A_2$, i.e. only uncoded transmission.
This is useful to distinguish the gains due to channel memory and the gains due to coding. 
This strategy performs well in this example, as the state sequence is highly predictable. The region $\mc R_\oplus$ achieves an even larger region.
Again, $\underline{\mc C}_{\text{fb}+s}^{\text{mem}}$ is only slightly smaller than ${\mc C}_{\text{fb}+s}^{\text{mem}}$.

\subsection{Delayed Feedback}
The result in Theorem~\ref{thm:capacity} extends to scenarios where feedback and channel state become available at the encoder with more than a single symbol-time delay. Consider a  delay of $d\geq1$ time units. In the converse, one can obtain the corresponding bounds by replacing the sequences $S^{T-1}$, $Y_1^{T-1}$ $Z_1^{T-1}$, $Y_2^{T-1}$ and $Z_2^{T-1}$ with $S^{T-d}$, $Y_1^{T-d}$ $Z_1^{T-d}$, $Y_2^{T-d}$ and $Z_2^{T-d}$.

The capacity region $\mc C_{\text{fb}+s}^{\text{mem}}(d)$ and achievable region $\underline{\mc C}_{\text{fb}+s}^{\text{mem}}(d)$ thus have a characterization as in \eqref{eq:posouter}~-~\eqref{eq:R2_constr2outer}, \eqref{eq:pos}~-~\eqref{eq:R2_constr2}, by redefining the erasure probabilities in \eqref{eq:def_epsilon} as
\begin{align}
 \epsonetwo{s}&=P_{\ve Z_t|S_{t-d}}(1,1|s),\quad
 &\epsonenottwo{s}&=P_{\ve Z_t|S_{t-d}}(1,0|s),\nonumber\\
 \epsnotonetwo{s}&=P_{\ve Z_t|S_{t-d}}(0,1|s),\quad  
  &\epsnotonenottwo{s} &=P_{\ve Z_t|S_{t-d}}(0,0|s),\nonumber\\
   \epsone{s}&=\epsonetwo{s}+\epsonenottwo{s},\quad 
  &\epstwo{s}&=\epsonetwo{s}+\epsnotonetwo{s}. \nonumber %
\end{align}
The corresponding deterministic achievable scheme as in Section~\ref{sec:new_ach_det} uses these redefined conditional erasure probabilities to obtain the same description as in Table~\ref{tab:new_det_algo}.

\begin{figure}[t]
\centering

\begin{tikzpicture}
\begin{axis}[scale=0.5,
width=4.82222222222222in,
height=3.80333333333333in,
scale only axis,
xmin=0,
xmax=0.41,
xlabel={$R_1$},
xmajorgrids,
every outer x axis line/.append style={black},
every x tick label/.append style={font=\color{black}},
ymin=0,
ymax=0.51,
ylabel={$R_2$},
ymajorgrids,
every outer y axis line/.append style={black},
every y tick label/.append style={font=\color{black}},
axis x line*=bottom,
axis y line*=left,
legend style={at={(1.0,0.0)},anchor=south west,draw=black,fill=white,legend cell align=left}
]

\addplot [color=TUMred,solid,line width=1.0pt,mark=o,mark options={solid}]
  table[row sep=crcr]{%
3.04262984945902e-09	0.499999999396795\\
3.39388148834496e-09	0.499999999551646\\
0.172999999833664	0.479999999964128\\
0.172999999840085	0.479999999961635\\
0.185170529989744	0.473594457811832\\
0.229999999691038	0.449999999413554\\
0.230000008201643	0.449999992117119\\
0.260477521688112	0.422152517327349\\
0.339999999938638	0.329999999827842\\
0.36230476488635	0.287620946489435\\
0.369999999951003	0.272999999880099\\
0.369999999970275	0.272999999727871\\
0.39999999907244	8.00312011627058e-09\\
0.399999999288013	4.51659832798046e-09\\
};
\addlegendentry{$\bar{\mc C}_{\text{fb}+s}^{\text{mem}}$, delay 1};

\addplot [
color=TUMpantone301,
solid,
line width=1.0pt,
mark=x,
mark options={solid}
]
  table[row sep=crcr]{%
5.42115626001038e-09	0.499999999304043\\
0.172999999900408	0.479999999923169\\
0.172999999974989	0.479999999887034\\
0.215561608619569	0.457599153271377\\
0.215561610481258	0.457599152199471\\
0.341503094666189	0.327144119908171\\
0.369999999966637	0.272999999864625\\
0.370000000027495	0.272999999444877\\
0.396483251465987	0.0320024104856066\\
0.399999997397321	1.65580349119343e-08\\
};
\addlegendentry{$\underline{\mc C}_{\text{fb}+s}^{\text{mem}}$, delay 1};

\addplot [color=TUMred,solid,line width=1.0pt,mark=o,mark options={solid}]
  table[row sep=crcr]{%
4.08231694976058e-11	0.499999999930361\\
5.06030946312297e-11	0.499999999768872\\
4.85336916489221e-10	0.499999995856295\\
6.38944857092594e-10	0.499999995644111\\
0.156949999708745	0.463999999968596\\
0.156949999918903	0.463999999912291\\
0.24783566705575	0.413429922303068\\
0.294999999129591	0.352500000691768\\
0.295000000538491	0.352499998172797\\
0.347500000007089	0.25544999967901\\
0.399999994026922	7.9731871452382e-10\\
0.399999999806744	6.44442089969832e-10\\
};
\addlegendentry{$\bar{\mc C}_{\text{fb}+s}^{\text{mem}}(d=2)$, delay 2};

\addplot [
color=TUMpantone301,
solid,
line width=1.0pt,
mark=asterisk,
mark options={solid}
]
table[row sep=crcr]{
8.88397894701065e-10 0.499999999498033\\
9.72639632900947e-10 0.499999999711467\\
0.156949996778929 0.464000000604486\\
0.15695000070875 0.463999999347534\\
0.221431505722674 0.428121571690184\\
0.299718334955756 0.343777820578036\\
0.347499999349731 0.25545000091145\\
0.347499999771156 0.255449999135915\\
0.347500002477089 0.255449986920886\\
0.399999999878136 2.30874239548158e-10\\
};
\addlegendentry{$\underline{\mc C}_{\text{fb}+s}^{\text{mem}}(d=2)$, delay 2};

\addplot [color=TUMred,solid,line width=1.0pt,mark=o,mark options={solid}]
  table[row sep=crcr]{%
-1.50906845700072e-09	0.499999997511849\\
-5.46040636317802e-10	0.499999999062048\\
4.65750413930488e-10	0.499999999697495\\
0.139243078518205	0.432767999438278\\
0.139243089807093	0.432767992681422\\
0.22330308055164	0.380311162785594\\
0.223303096693208	0.380311152696028\\
0.223303098752961	0.380311150290188\\
0.308476559627929	0.229918522813686\\
0.30847656309711	0.229918515046333\\
0.399999998953125	-2.09943472329055e-10\\
0.399999999851802	4.24361934481254e-12\\
0.399999999926315	7.17781563353004e-11\\
0.39999999992684	6.44289482876204e-11\\
};
\addlegendentry{$\bar{\mc C}_{\text{fb}+s}^{\text{mem}}(d=5)$, delay 5};

\addplot [
color=TUMpantone301,
solid,
line width=1.0pt,
mark=triangle*,
mark options={solid}
]
table[row sep=crcr]{
1.50950512569393e-10 0.499999999859184\\
0.139243083699421 0.432767997283583\\
0.215242765842038 0.385341125219978\\
0.215242766900777 0.385341124175141\\
0.215242767836298 0.385341123230204\\
0.232548136867466 0.363986986464548\\
0.308476562467742 0.229918518641014\\
0.399999998808526 1.24366086078451e-09\\
0.399999999882481 1.72509471699247e-10\\
};
\addlegendentry{$\underline{\mc C}_{\text{fb}+s}^{\text{mem}}(d=5)$, delay 5};

\addplot [color=TUMred,solid,line width=1.0pt,mark=o,mark options={solid}]
  table[row sep=crcr]{%
-8.54477464479197e-10	0.49999999857806\\
-4.87467403248454e-10	0.499999999505802\\
2.99853839674791e-10	0.499999999731377\\
0.137299156056543	0.410737418615813\\
0.199269570285195	0.368531509390085\\
0.199269574063453	0.36853150659245\\
0.199269574714597	0.368531505866853\\
0.199269575388053	0.368531504755233\\
0.199269849586659	0.368531023962609\\
0.399999994552454	1.02901706044989e-08\\
0.399999999005104	-2.6770734240511e-10\\
0.399999999879181	-2.23184387637687e-11\\
0.399999999943554	5.55554213743648e-11\\
};
\addlegendentry{$\bar{\mc C}_{\text{fb}+s}^{\text{mem}}(d=10)$, delay 10};

\addplot [
color=TUMpantone301,
solid,
line width=1.0pt,
mark=diamond*,
mark options={solid}
]
table[row sep=crcr]{
8.97265272955639e-11 0.499999999911609\\
2.04855639298161e-09 0.499999997285639\\
0.137299085813 0.4107374608347\\
0.19914318855396 0.368617583144707\\
0.200209704805009 0.366899387297235\\
0.399999971905306 5.34943439534591e-08\\
0.399999999012641 9.56804233676889e-10\\
0.399999999874006 1.71880204557762e-10\\
};
\addlegendentry{$\underline{\mc C}_{\text{fb}+s}^{\text{mem}}(d=10)$, delay 10};

\addplot [
color=TUMred,
dashed,
line width=1.0pt
]
table[row sep=crcr]{
0 0.5\\
0.193103448275862 0.362068965517241\\
0.4 0\\
};
\addlegendentry{$\mc C_{\text{fb}}$};

\addplot [
color=TUMred,
dotted,
line width=1.0pt
]
table[row sep=crcr]{
0 0.5\\
0.4 0\\
};
\addlegendentry{$\mc C$};

\end{axis}
\end{tikzpicture}%
\caption{Capacity regions with delayed feedback, for $\epsilon_1=0.6$, $g_1=0.1$, $\epsilon_2=0.5$, $g_2=0.1$.}
\label{fig:delay}
\vspace*{-5mm}
\end{figure}
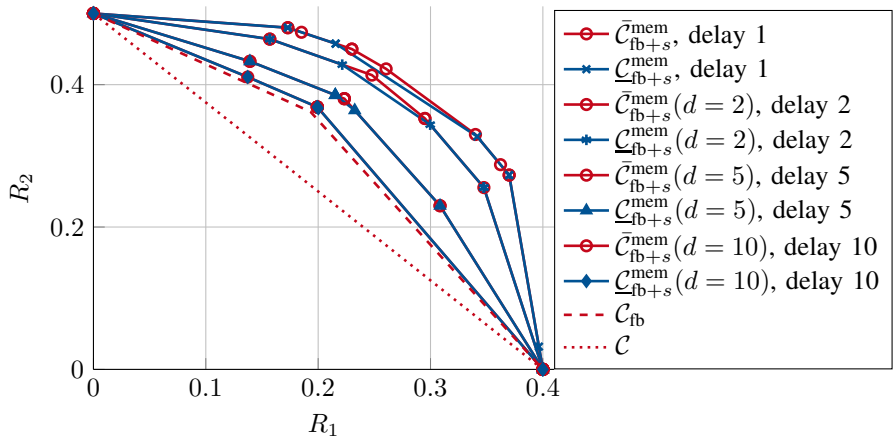

Fig.~\ref{fig:delay} shows the effect of feedback delay for a Gilbert-Elliot channel with parameters  $\epsilon_1=0.6$, $g_1=0.1$, $\epsilon_2=0.5$, $g_2=0.1$. Observe that delayed feedback shrinks both ${\mc C}_{\text{fb}+s}^{\text{mem}}(d)$ and $\underline{\mc C}_{\text{fb}+s}^{\text{mem}}(d)$, as the state information becomes less useful. After a feedback delay of $d=10$ time units, the region $\underline{\mc C}_{\text{fb}+s}^{\text{mem}}(d=10)$ is almost the same as for the memoryless case. In general this depends on the convergence speed of the state Markov chain towards its stationary distribution.
It is interesting to see that, as $d$ increases, the difference between $\underline{\mc C}_{\text{fb}+s}^{\text{mem}}$ and ${\mc C}_{\text{fb}+s}^{\text{mem}}$ becomes smaller.
$\underline{\mc C}_{\text{fb}+s}^{\text{mem}}$ and ${\mc C}_{\text{fb}+s}^{\text{mem}}$ match for the memoryless single-state BPEC, a result further generalized in Section~\ref{sec:memoryless}.
~
~
\subsection{Hidden Case}
\label{sec:results_hidden}
For the hidden case, we consider a variation on the Gilbert-Elliot model used before.
Suppose there is a nonzero erasure probability in both states $\text{G}$ and $\text{B}$, where $\epsilon_{j\text{G}}$ is the erasure probability at \textnormal{Rx$_j$}~when in state $\text{G}$ and $\epsilon_{j\text{B}}$ when in state $\text{B}$.
Typically one chooses $\epsilon_{j\text{G}} < \epsilon_{j\text{B}}$.

The long-term average erasure probability at \textnormal{Rx$_j$}\ is 
\begin{align}
 \epsilon_j = \pi_\text{G} \epsilon_{j\text{G}}   + \pi_\text{B} \epsilon_{j\text{B}} 
= \frac{g_j \epsilon_{j\text{G}} + b_j \epsilon_{j\text{B}}}{g_j + b_j}\cdot
\end{align}
Because all erasure events can happen in every state, the corresponding setup is hidden if only ACK/NACK feedback is available.
In Fig.~\ref{fig:GE_HMM} in Section~\ref{sec:main} we plot the $L$-th order approximations $\bar{\mc C}_{\text{fb}}^{\text{mem}}(L)$ of the outer bounds, as defined in Section~\ref{sec:outer_bounds}, for $L=1$ and $L=7$. We observe that $\bar{\mc C}_{\text{fb}}^{\text{mem}}(L_1) \subset \bar{\mc C}_{\text{fb}}^{\text{mem}}(L_2)$ for $L_1<L_2$.
We also plot rate pairs that lead to a stable queueing network when the deterministic scheme in Table~\ref{tab:new_det_algo_hmm} is used. The decision if the queueing system is stable or not was made by inspection after a simulation of $n=10^7$ time slots.
Both curves provide an almost equivalent characterization of the capacity region ${\mc C}_{\text{fb}}^{\text{mem}}$.
The exact channel parameters for Fig.~\ref{fig:GE_HMM} are $\epsilon_1=0.6$, $g_1=0.1$, $b_1=0.15$, $\epsilon_{1\text{G}}=0.2$, $\epsilon_{1\text{B}}=0.866$ and $\epsilon_2=0.5$, $g_2=0.2$, $b_2=0.2$, $\epsilon_{2\text{G}}=0.2$, $\epsilon_{2\text{B}}=0.8$.

As a second example, define the following $3$-state channel  with transition matrix $\mat P$, $\mat P_{ij} = P_{S_t|S_{t-1}}(j|i)$, and erasure distribution matrix $\mat E$. Each row of $\mat E$ represents $[P_{\ve Z_t|S_t}(0,0|s)$, $P_{\ve Z_t|S_t}(0,1|s)$, $P_{\ve Z_t|S_t}(1,0|s)$, $P_{\ve Z_t|S_t}(1,1|s)]$:
\begin{align}
 \mat P = \left[\begin{array}{ccc}
           0.7 & 0.2 & 0.1 \\
	   0.2 & 0.4 & 0.4 \\
	   0.3 & 0.01 & 0.69 
          \end{array}\right],\quad
\mat E = \left[\begin{array}{cccc}
           0.75 & 0.1 & 0.1 & 0.05 \\
	   0.2  & 0.2 & 0.3 & 0.3 \\
	   0    & 0.1 & 0.2 & 0.7 
          \end{array}\right]. \label{eq:3state_eras}
\end{align}
Note that the channels to \textnormal{Rx$_1$} and \textnormal{Rx$_2$}~are correlated in this case.

\begin{figure}[t]
\centering

\begin{tikzpicture}
\begin{axis}[scale=0.5,
width=4.82222222222222in,
height=3.80333333333333in,
scale only axis,
xmin=0,
xmax=0.51,
xmajorgrids,
every outer x axis line/.append style={black},
every x tick label/.append style={font=\color{black}},
xlabel={$R_1$},
ymin=0,
ymax=0.6,
ymajorgrids,
every outer y axis line/.append style={black},
every y tick label/.append style={font=\color{black}},
ylabel={$R_2$},
axis x line*=bottom,
axis y line*=left,
legend style={at={(0.0,0.0)},anchor=south west,draw=black,fill=white,legend cell align=left}
]

\addplot [color=TUMred,solid,line width=1.0pt,mark=o,mark options={solid}]
  table[row sep=crcr]{%
-5.58092808078392e-12	0.555263157876719\\
-8.10594646960539e-13	0.555263157894103\\
-3.08780778723872e-16	0.555263157894737\\
5.20417042793042e-17	0.555263157894737\\
0.093157894705711	0.484210526338276\\
0.267231449902559	0.345303345930997\\
0.267231449903516	0.34530334593007\\
0.324473684208545	0.275526315791607\\
0.324473684211066	0.275526315788461\\
0.502631578947032	-3.63868657426991e-13\\
0.502631578947177	4.16819356807707e-14\\
};
\addlegendentry{$\mc C_{\text{fb}+s}^{\text{mem}}$};

\addplot [color=TUMgold,only marks,line width=1.0pt,mark=+,mark options={solid}]
  table[row sep=crcr]{%
0.469118582739949	0.0492140760681642\\
0.441880086233247	0.0880774496183674\\
0.417174604809814	0.125602160321713\\
0.392523031716515	0.162931540274615\\
0.365732273488589	0.199067649305405\\
0.341330955347996	0.236190279022649\\
0.315097697591857	0.271193074784636\\
0.292641740911934	0.30258188068003\\
0.269893803918854	0.334339075616503\\
0.24537750772094	0.35636142095887\\
0.220859290965153	0.375975303695941\\
0.19634766200394	0.395614670244555\\
0.171837453824113	0.415280208060107\\
0.147323890427864	0.434971468604454\\
0.122802863820865	0.45468706890349\\
0.0987781710565633	0.476872986403819\\
0.0740945727444993	0.495996065836188\\
0.0494038558499961	0.514911567034977\\
0.0247057367869251	0.533834321309752\\
0.0124106562703054	0.545797403111735\\
0.0370557410869422	0.5243721125246\\
0.0617501155754624	0.505452833959691\\
0.0860000054393555	0.484079959116311\\
0.111117199227872	0.466985470044647\\
0.135064535537004	0.444826342255179\\
0.159581366381497	0.425122690428664\\
0.185126267918948	0.40772051033859\\
0.208602981114786	0.385791730189527\\
0.233117247381302	0.366165267611426\\
0.257640717716879	0.346563485499091\\
0.282134667235022	0.319354170346977\\
0.304812187525055	0.287696715856281\\
0.327179769137508	0.253239513134846\\
0.35352217334459	0.217623884333806\\
0.380218529645509	0.181595902585498\\
0.404841774038517	0.144267625839706\\
0.429521071358265	0.106933324734184\\
0.45672224658631	0.0688919970190702\\
0.481520283116404	0.0295319092552103\\
};
\addlegendentry{Stable points};

\addplot [color=TUMgold,solid,line width=1.0pt,mark=square,mark options={solid}]
  table[row sep=crcr]{%
6.21669667241927e-15	0.555263157894728\\
6.40996417999428e-15	0.555263157894819\\
8.97495273900154e-15	0.555263157894684\\
2.89769378332633e-14	0.555263157894624\\
0.26481515446957	0.342812620179489\\
0.343807745739912	0.234564493611037\\
0.502631578947307	1.1290063190437e-13\\
0.502631578947313	1.13754505489706e-13\\
};
\addlegendentry{$\bar{\mc C}_{\text{fb}}^{\text{mem}}(L=7)$};

\addplot [color=TUMgold,dashed,line width=1.0pt,mark=triangle*,mark options={solid}]
  table[row sep=crcr]{%
-1.86498386178791e-13	0.5552631578945\\
-4.15743828252602e-14	0.555263157894628\\
-3.70467545529607e-14	0.555263157894631\\
5.78287417951628e-14	0.55526315789468\\
0.264485455911835	0.342909930991009\\
0.264485455912795	0.342909930989939\\
0.385065789456216	0.1753815789712\\
0.502631578947183	-1.18620391287294e-13\\
0.502631578947311	6.62248034188906e-14\\
};
\addlegendentry{$\bar{\mc C}_{\text{fb}}^{\text{mem}}(L=1)$};

\addplot [color=TUMred,dashed,line width=1.0pt]
  table[row sep=crcr]{%
0	0.555263157894737\\
0.228099268556442	0.366522194500975\\
0.502631578947369	0\\
};
\addlegendentry{$\mc C_\text{fb}$};

\addplot [color=TUMred,dotted,line width=1.0pt]
  table[row sep=crcr]{%
0	0.555263157894737\\
0.502631578947369	0\\
};
\addlegendentry{$\mc C$};

\end{axis}
\end{tikzpicture}%
\caption{Capacity region for hidden case, $3$-state model as described in \eqref{eq:3state_eras}. }
\label{fig:3states_HMM}
\vspace*{-5mm}
\end{figure}
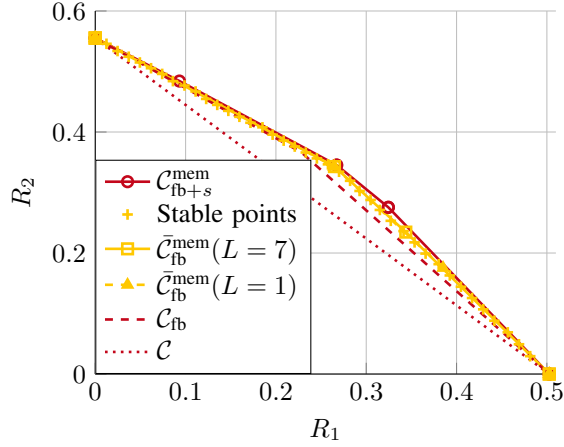

The long-term average erasure probabilities are given by $\epsilon_1 = 0.497$, $\epsilon_2=0.445$, $\epsilon_{12}=0.329$.
The packet loss probability is very low in the first state, moderate in the second state, and very high in the third state. Results are shown in Fig.~\ref{fig:3states_HMM}.
For this example, the gain of ${\mc C}_{\text{fb}}^{\text{mem}}$ compared to the equivalent memoryless region ${\mc C}_{\text{fb}}$ is only moderate. We can also see that the first-oder approximation $\bar{\mc C}_{\text{fb}}^{\text{mem}}(L=1)$ already provides  an accurate description of the hidden capacity region ${\mc C}_{\text{fb}}^{\text{mem}}$.

\section{Conclusion}
\label{sec:concl_becfb}
This paper studied two-receiver finite-state broadcast packet erasure channels (BPECs) with feedback and memory, represented by a finite-state model. 
Two different cases were investigated: In the case of a visible channel state, the transmitter knows the channel state strictly causally, in addition to the channel output feedback. In the case of a hidden channel state, the transmitter only has channel output feedback. For both situation, we derived novel outer bounds on the capacity region.
We formulate coding schemes as a queueing problem that can be analyzed with approaches from network control.
The coding schemes can be formulated as linear network flow problems, where we showed that the dual min-cut representation matches the outer bounds. Hence the coding schemes are optimal and achieve any point inside the capacity region.
We complemented these results with suboptimal algorithms, delayed feedback and derivations about finite-state memoryless BPECs.

There are multiple directions to extend the work in this paper: The $K$-receiver setup is well studied for the memoryless case, and we think some of the results could carry over to the case with channel memory using the techniques presented in this paper.
Moreover, thinking towards practical application, schemes dealing with lossy feedback, unknown channel models, and time-varying channels are highly desirable.

\appendices

\section{Proof of Lemma \ref{lemma2}}
\label{appendix:lemma2}
Using \eqref{defu} and \eqref{defz} for $j=1,2$, we have
\begin{align}
u_s^{(1)} + v_s^{(2)}&\leq u^{(j)}_s+I(U_{\bar{j},T}V_T;X_T|U_{{j},T}T,S_{T-1}=s)\\
&=I(U_{\bar{j},T}U_{{j},T}V_T;X_T|T,S_{T-1}=s) \leq 1,
\end{align}
where the last inequality holds because $\log_{2^\ell} |\mc X| = 1$.

\section{Proof of Corollary \ref{cor:Pz_decay}}
\label{ap-cor}
\begin{align}
  &\sum_{\ve z_{t+1}} \left| P_{\ve Z_{t+1}|\ve Z^{t}}(\ve z_{t+1}|\ve z^{t}) -  P_{\ve Z_{t+1}|\ve Z^{t}_{t-L+1}}(\ve z_{t+1}|\ve z^{t}_{t-L+1})    \right| \nonumber\\
  =&\sum_{\ve z_{t+1}} \left| \sum_{s \in \mc S} \left( P_{\ve Z_{t+1} S_{t+1}|\ve Z^{t}}(\ve z_{t+1}, s|\ve z^{t}) -  P_{\ve Z_{t+1} S_{t+1}|\ve Z^{t}_{t-L+1}}(\ve z_{t+1}, s|\ve z^{t}_{t-L+1}) \right)   \right| \nonumber \\
  =&\sum_{\ve z_{t+1}} \left| \sum_{s \in \mc S} \left( P_{S_{t+1}|\ve Z^{t}}(s|\ve z^{t}) -  P_{ S_{t+1}|\ve Z^{t}_{t-L+1}}(s|\ve z^{t}_{t-L+1}) \right)  P_{\ve Z_{t+1} |S_{t+1}}(\ve z_{t+1}|s) \right| \nonumber \\
  \leq&  \sum_{s \in \mc S} \left| P_{S_{t+1}|\ve Z^{t}}(s|\ve z^{t}) -  P_{ S_{t+1}|\ve Z^{t}_{t-L+1}}(s|\ve z^{t}_{t-L+1}) \right| \sum_{\ve z_{t+1}}P_{\ve Z_{t+1} |S_{t+1}}(\ve z_{t+1}|s) \nonumber \\
  =&\sum_{s_t} \left| P_{S_{t+1}|\ve Z^{t}}(s|\ve z^{t}) -  P_{ S_{t+1}|\ve Z^{t}_{t-L+1}}(s|\ve z^{t}_{t-L+1}) \right|,
\end{align}
where the inequality is due to the triangle inequality.

\section{Approximation of Outer Bounds for the Hidden Case}
\label{sec:proof_bound_hidden_approx}
We prove inequality  \eqref{eq:R1_constr1outer_hmm_wind_outer}. The other inequalities follow similarly. 
\begin{align}
R_1\leq& \frac{1}{n} \sum_{t=1}^n \sum_{\ve z^{t-1}} P_{\ve Z^{t-1}}(\ve z^{t-1}) (1-\epsone{\ve z^{t-1}}) x(\ve z^{t-1})\nonumber \\
\stackrel{(a)}{\leq} &\frac{1}{n} \sum_{t=1}^n \sum_{\ve z^{t-1}} P_{\ve Z^{t-1}}(\ve z^{t-1}) \left( (1-\epsone{\ve z^{t-1}_{t-L}} + 2 (1-\sigma)^L \right) x(\ve z^{t-1}) \nonumber \\
= &\frac{1}{n} \sum_{t=1}^n \left[ \left( \sum_{\ve z^{t-1}} P_{\ve Z^{t-1}}(\ve z^{t-1}) (1-\epsone{\ve z^{t-1}_{t-L}} )x(\ve z^{t-1}) \right) \right. \nonumber\\
&\qquad  + 2  (1-\sigma)^L  \left(\sum_{\ve z^{t-1}} P_{\ve Z^{t-1}}(\ve z^{t-1})x(\ve z^{t-1})\right)  \Bigg]\nonumber \\
\stackrel{(b)}{\leq}&  \frac{1}{n} \sum_{t=1}^n \sum_{\ve z^{t-1}_{t-L}}\hspace*{-1mm}\left(P_{\ve Z^{t-1}_{t-L}}(\ve z^{t-1}_{t-L}) \left(1 - \epsone{\ve z^{t-1}_{t-L}}\right)   \left(\sum_{\ve z^{t-L-1}}P_{\ve Z^{t-L-1}|\ve Z_{t-L}^{t-1}}(\ve z^{t-L-1}|\ve z_{t-L}^{t-1})x(\ve z^{t-1})\right)\right)+2  (1-\sigma)^L \label{eq:appxx}
\end{align}
where $(a)$ follows from Corollary~\ref{cor:Pz_decay} and $(b)$ follows from $x(\ve z^{t-1})\leq 1$ for all $t$ and $\ve z^{t-1}$.
If the sequence $\ve Z^n$ is stationary, i.e. if $P_{\ve Z^{t-1}_{t-L}}(\ve z^L)$ does not depend on $t$ but only on $\ve z^L$, then we can write the right hand side of \eqref{eq:appxx} as
\begin{align}
  \sum_{\ve z^L} P_{\ve Z^{t-1}_{t-L}}(\ve z^L) \left(1 - \epsone{\ve z^L}\right)  \left(\frac{1}{n}\sum_{t=1}^n\sum_{\ve z^{t-L-1}}P_{\ve Z^{t-L-1}|\ve Z_{t-L}^{t-1}}(\ve z^{t-L-1}|\ve z^L)x(\ve z^{t-1})\right) + 2 (1-\sigma)^L  .
 \end{align}
Abbreviating $ \left(\frac{1}{n}\sum_{t=1}^n\sum_{\ve z^{t-L-1}}P_{\ve Z^{t-L-1}|\ve Z_{t-L}^{t-1}}(\ve z^{t-L-1}|\ve z^L)x(\ve z^{t-1})\right)$ by $x(\ve z^L)$, we obtain
\begin{align}
R_1\leq  \sum_{\ve z^L} P_{\ve Z^L}(\ve z^L) (1-\epsone{\ve z^L}) x(\ve z^L)+ 2  (1-\sigma)^L.
\end{align}
With similar steps and adaption of Lemma~\ref{lemma1}, we obtain \eqref{eq:R1_constr1outer_hmm_wind_outer} - \eqref{eq:R2_constr2outer_hmm_wind_outer}.
Similar steps apply for the inner approximation.

\section{Packet Movement and Network Flow in the Degenerate Case}
\label{sec:deg_packet_movement}
\newcommand{\packetA}{\ensuremath{p_l^{(1)} } }
\newcommand{\packetAstar}{\ensuremath{ p_{\star}^{(1)} } }
\newcommand{\packetB}{\ensuremath{p_m^{(2)}} }
\newcommand{\packetBstar}{\ensuremath{ p_{\star}^{(2)} } }
One problem arises if we want to use proactive coding, i.e. $A_t=4$, but $Q_1^{(1)}$ or $Q_1^{(2)}$ or both are empty.
The flow-based models and backpressure schemes in Section~\ref{sec:new_ach_det}
require \eqref{eq:queue_dynamics} and \eqref{eq:def_F12t1}~-~\eqref{eq:def_F34_new} to hold.
This section shows that these conditions can be satisfied in the degenerate case.

Section~\ref{sec:packet_mov} describes the packet movement in the non-degenerate case, i.e. both $Q_1^{(1)}$ and $Q_1^{(2)}$ are nonempty when $A_t=4$.
The original packets involved in the poisoned packet are moved to $Q_3^{(1)}$, $Q_3^{(2)}$, respectively, if the poisoned packet is not erased at both \textnormal{Rx$_1$}\ and \textnormal{Rx$_2$}.
That is, either both original packets move to $Q_3^{(1)}$, $Q_3^{(2)}$, or none of them.
The flow dynamics in \eqref{eq:def_F12t1}~-~\eqref{eq:def_F34_new} show that a packet leaves $Q_3^{(1)}$ if and only if a packet leaves $Q_3^{(2)}$, provided that both $Q_3^{(1)}$ and $Q_3^{(2)}$ are nonempty.
This implies that $Q_3^{(1)}$ and $Q_3^{(2)}$ always contain the same number of packets, provided that both queues are empty in the beginning and only non-degenerate cases occur.

Suppose $Q_{1,t}^{(1)}>0$, $Q_{1,t}^{(2)}=0$ and $A_t=4$:
In this case, the poisoned packet only consists of a single original packet \ensuremath{p_l^{(1)} } , because there is no packet in $Q_1^{(2)}$. Hence, this packet is uncoded in principle and could follow the same packet movement rules as if $A_t=1$ was chosen.
However, that would violate the flow dynamics in \eqref{eq:defF13}, which require that a packet moves from $Q_1^{(1)}$ to $Q_3^{(1)}$ for $A_t=4$ unless erased at both \textnormal{Rx$_1$}\ and \textnormal{Rx$_2$}.

To deal with this case,
we split the queues $Q_3^{(j)}$ into two subqueues $Q_3^{(j)\text{nondeg}}$ and $Q_3^{(j)\text{deg}}$, with 
\begin{equation}
 Q_{3,t}^{(j)}= Q_{3,t}^{(j)\text{nondeg}} + Q_{3,t}^{(j)\text{deg}}.
\end{equation}
$Q_3^{(j)\text{nondeg}}$ contains only original packets that were involved in a non-degenerate poisoned packet. Each packet in $Q_3^{(1)\text{nondeg}}$ is linked to a packet in $Q_3^{(2)\text{nondeg}}$ with which it was combined in a poisoned packet. 
$Q_3^{(j)\text{deg}}$ contains original packets for degenerate poisoned packets. 
$Q_3^{(1)\text{nondeg}}$ and $Q_3^{(2)\text{nondeg}}$ always contain the same number of packets, but the number of packets in $Q_3^{(1)\text{deg}}$ and $Q_3^{(2)\text{deg}}$ may be different.

In the degenerate case we move the poisoned packet \ensuremath{p_l^{(1)} }  from $Q_1^{(1)}$ to $Q_3^{(1)\text{deg}}$ if it is not erased at both \textnormal{Rx$_1$}\ and \textnormal{Rx$_2$}, as suggested by \eqref{eq:defF13}. 
We perform this movement even if \ensuremath{p_l^{(1)} }  was received by \textnormal{Rx$_1$}\ and hence already reached its destination\footnote{The header information can be used to distinguish whether an uncoded packet was transmitted due to $A_t=1$ or $A_t=4$.}. Hence, this packet movement is suboptimal in principle, but ensures that \eqref{eq:defF13} is satisfied.
This applies similarly for $Q_{1,t}^{(1)}=0$, $Q_{1,t}^{(2)}>0$ and $A_t=4$.
\begin{remark}
Note that the packet movement from $Q_1^{(1)}$ to $Q_3^{(1)}$ is independent of the number of packets in $Q_1^{(2)}$, as required by \eqref{eq:defF13}, and vice versa. 
\end{remark}

For $A_t=5$, i.e. when moving packets out of $Q_3^{(j)}$, we want ensure that also \eqref{eq:defF32} and \eqref{eq:def_F34_new} hold, regardless if we move the packet out of $Q_3^{(j)\text{nondeg}}$ or $Q_3^{(j)\text{deg}}$.
Consider the following policy:
\begin{itemize}
 \item Whenever $Q_3^{(1)\text{nondeg}}$ and $Q_3^{(2)\text{nondeg}}$ are nonempty and $A_t=5$, we pick the pair of original packets in $Q_3^{(1)\text{nondeg}}$ and $Q_3^{(2)\text{nondeg}}$ that arrived first and apply the remedy packet transmission as described in Table~\ref{tab:packet_movement}.
  \item When both $Q_3^{(1)\text{nondeg}}$ and $Q_3^{(2)\text{nondeg}}$ are empty, $Q_3^{(1)\text{deg}}$ is empty, and $Q_3^{(2)\text{deg}}$ is nonempty, one can retransmit the packet \ensuremath{p_m^{(2)}} \ as a remedy packet. This remedy transmission then follows the same rules as an uncoded transmission ($A_t=2$) and hence satisfies \eqref{eq:defF32} and \eqref{eq:def_F34_new}.
 Similarly for $Q_3^{(1)\text{deg}}$ nonempty and $Q_3^{(2)\text{deg}}$ empty.
 \item Suppose $Q_3^{(1)\text{nondeg}}$ and $Q_3^{(2)\text{nondeg}}$ are empty and both $Q_3^{(1)\text{deg}}$ and $Q_3^{(2)\text{deg}}$ are nonempty: Hence $Q_3^{(1)\text{deg}}$ contains \ensuremath{p_l^{(1)} } \ and $Q_3^{(2)\text{deg}}$ contains \ensuremath{p_m^{(2)}} , but these packets arrived in different time slots at $Q_3^{(1)\text{deg}}$, $Q_3^{(2)\text{deg}}$, respectively. They were not combined to a poisoned packet, but they were both degenerate poisoned packets.
 In this case, it matters at which receivers \ensuremath{p_l^{(1)} } \ and \ensuremath{p_m^{(2)}} \ were received, respectively, and how the corresponding remedy packet looks like. All different possibilities are described in Table~\ref{tab:packet_movement_degenerate}.
\end{itemize}
\begin{example}
For further illustration, we pick an example:
Suppose poison \ensuremath{p_l^{(1)} } \ is received at \textnormal{Rx$_1$}\ only and \ensuremath{p_m^{(2)}} \ is also received at \textnormal{Rx$_1$}\ only.
This corresponds to the first line in Table~\ref{tab:packet_movement_degenerate}.
Note that \ensuremath{p_l^{(1)} } \ is - in principle - already known at its destination \textnormal{Rx$_1$}. However, we do not account for this, i.e. we do not move it to $Q_4^{(1)}$.

We choose \ensuremath{p_m^{(2)}} \ as the remedy packet and consider the following cases:

\begin{itemize}
 \item If the remedy packet is received at \textnormal{Rx$_1$}\ only, we \emph{release} \ensuremath{p_l^{(1)} } \ at \textnormal{Rx$_1$}. That is, \ensuremath{p_l^{(1)} } \ has been received at \textnormal{Rx$_1$}\ before and was not counted, but we count it now and move it to $Q_4^{(1)}$.
Because the remedy \ensuremath{p_m^{(2)}} \ was received at \textnormal{Rx$_1$}\, we move it from $Q_3^{(2)}$ to $Q_2^{(2)}$. Note that we could have done that already before, as \ensuremath{p_m^{(2)}} \ was received as a poisoned packet at \textnormal{Rx$_1$}\ before.
\item If the remedy packet is received at \textnormal{Rx$_2$}\ only, we move \ensuremath{p_m^{(2)}} \ to $Q_4^{(2)}$ as it was received at \textnormal{Rx$_2$}.
The flow dynamics in \eqref{eq:defF32} require that \ensuremath{p_l^{(1)} } \ should move from $Q_3^{(1)}$ to $Q_2^{(1)}$: We can achieve this by replacing \ensuremath{p_l^{(1)} } \ with a dummy packet \ensuremath{ p_{\star}^{(1)} } \ that is known to \textnormal{Rx$_2$}\ and identified via a flag. Once this dummy packet \ensuremath{ p_{\star}^{(1)} } \ is combined in a reactive coding operation ($A_t=3$), and received at \textnormal{Rx$_1$}\, we release (the previously already received) \ensuremath{p_l^{(1)} } \ at \textnormal{Rx$_1$}, satisfying the flow dynamics.
\item If the remedy packet \ensuremath{p_m^{(2)}} \ is received at \textnormal{Rx$_1$}\ and \textnormal{Rx$_2$}, we release the (previously received) packet \ensuremath{p_l^{(1)} } \ at \textnormal{Rx$_1$} and move it to $Q_4^{(1)}$. Packet \ensuremath{p_m^{(2)}} \ is received at \textnormal{Rx$_2$}, so we can also move it to $Q_4^{(2)}$.
\end{itemize}
The other cases require slightly different operations for the remedy packet, but the procedure is similar.
\end{example}

\begin{remark}
In all cases, the packet movement from $Q_3^{(1)}$ to $Q_2^{(1)}$ or $Q_4^{(1)}$ is independent of the number of packets contained in $Q_3^{(2)}$, as required by \eqref{eq:defF32} and \eqref{eq:def_F34_new}.
 \end{remark}

\begin{table}
\centering
 \begin{tabular}{|l|l|c|l|l|l|}
\hline
 \parbox[t]{0.09\textwidth}{poison \ensuremath{p_l^{(1)} } \ received at} &  \parbox[t]{0.09\textwidth}{poison \ensuremath{p_m^{(2)}} \ received at} & remedy & remedy received at \textnormal{Rx$_1$}~& remedy received at \textnormal{Rx$_2$}~& \parbox[t]{0.21\textwidth}{remedy rec. at \textnormal{Rx$_1$}~ and \textnormal{Rx$_2$}} \\ \hline \hline
 \textnormal{Rx$_1$}\ only & \textnormal{Rx$_1$}~only& \ensuremath{p_m^{(2)}}  
& \parbox[t]{0.15\textwidth}{Release \ensuremath{p_l^{(1)} }  at \textnormal{Rx$_1$}\ 
\\$\Rightarrow$ \ensuremath{p_l^{(1)} }  to $Q_4^{(1)}$ (exit) 
\\$\Rightarrow$ \ensuremath{p_m^{(2)}}  to $Q_2^{(2)}$} 
& \parbox[t]{0.18\textwidth}{
$\Rightarrow$ \ensuremath{p_m^{(2)}}  to $Q_4^{(2)}$ (exit)
\\$\Rightarrow$ Replace \ensuremath{p_l^{(1)} }  with \ensuremath{ p_{\star}^{(1)} }  and move to $Q_2^{(1)}$} 
& \parbox[t]{0.20\textwidth}{
Release \ensuremath{p_l^{(1)} }  at \textnormal{Rx$_1$} 
\\$\Rightarrow$ \ensuremath{p_l^{(1)} }  to $Q_4^{(1)}$ (exit) \\$\Rightarrow$ \ensuremath{p_m^{(2)}}  to $Q_4^{(2)}$ (exit)} \\ \hline
  \textnormal{Rx$_1$}\ only & \textnormal{Rx$_2$}~only& \ensuremath{p_m^{(2)}}  
& \parbox[t]{0.15\textwidth}{Release \ensuremath{p_l^{(1)} }  at \textnormal{Rx$_1$}\ 
\\$\Rightarrow$ \ensuremath{p_l^{(1)} }  to $Q_4^{(1)}$ (exit) 
\\$\Rightarrow$ Put \ensuremath{p_m^{(2)}}  to $Q_2^{(2)}$} 
& \parbox[t]{0.18\textwidth}{
$\Rightarrow$ \ensuremath{p_m^{(2)}}  to $Q_4^{(2)}$ (exit)
\\$\Rightarrow$ Replace \ensuremath{p_l^{(1)} }  with \ensuremath{ p_{\star}^{(1)} }  and move to $Q_2^{(1)}$}
& \parbox[t]{0.20\textwidth}{Release both \ensuremath{p_l^{(1)} }  and \ensuremath{p_m^{(2)}} \\$\Rightarrow$ \ensuremath{p_l^{(1)} }  to $Q_4^{(1)}$ (exit) \\$\Rightarrow$ \ensuremath{p_m^{(2)}}  to $Q_4^{(2)}$ (exit)} \\ \hline
\textnormal{Rx$_1$}\ only & \textnormal{Rx$_1$}~and \textnormal{Rx$_2$}~& \ensuremath{p_l^{(1)} } 
& \parbox[t]{0.15\textwidth}{
Release \ensuremath{p_l^{(1)} }  at \textnormal{Rx$_1$} 
\\$\Rightarrow$ \ensuremath{p_l^{(1)} }  to $Q_4^{(1)}$ (exit)\\$\Rightarrow$ Put \ensuremath{p_m^{(2)}}  to $Q_2^{(2)}$} 
& \parbox[t]{0.18\textwidth}{
Release \ensuremath{p_m^{(2)}}  at \textnormal{Rx$_2$} 
\\$\Rightarrow$ \ensuremath{p_m^{(2)}}  to $Q_4^{(2)}$ (exit)\\$\Rightarrow$ Put \ensuremath{p_l^{(1)} }  to $Q_2^{(1)}$} 
& \parbox[t]{0.20\textwidth}{Release both \ensuremath{p_l^{(1)} }  and \ensuremath{p_m^{(2)}} \\$\Rightarrow$ \ensuremath{p_l^{(1)} }  to $Q_4^{(1)}$ (exit) \\$\Rightarrow$ \ensuremath{p_m^{(2)}}  to $Q_4^{(2)}$ (exit)} \\ \hline \hline
\textnormal{Rx$_2$}\ only & \textnormal{Rx$_1$}~only& $\ensuremath{p_l^{(1)} }  \oplus \ensuremath{p_m^{(2)}} $
& \parbox[t]{0.15\textwidth}{
\textnormal{Rx$_1$}\ decodes \ensuremath{p_l^{(1)} }  
\\$\Rightarrow$ \ensuremath{p_l^{(1)} }  to $Q_4^{(1)}$ (exit)\\$\Rightarrow$ Put \ensuremath{p_m^{(2)}}  to $Q_2^{(2)}$} 
& \parbox[t]{0.18\textwidth}{
\textnormal{Rx$_2$}\ decodes \ensuremath{p_m^{(2)}}  
\\$\Rightarrow$ \ensuremath{p_m^{(2)}}  to $Q_4^{(2)}$ (exit)\\$\Rightarrow$ Put \ensuremath{p_l^{(1)} }  to $Q_2^{(1)}$} 
& \parbox[t]{0.20\textwidth}{\textnormal{Rx$_1$}\ decodes \ensuremath{p_l^{(1)} } , \textnormal{Rx$_2$}\ decodes \ensuremath{p_m^{(2)}} \\$\Rightarrow$ \ensuremath{p_l^{(1)} }  to $Q_4^{(1)}$ (exit) \\$\Rightarrow$ \ensuremath{p_m^{(2)}}  to $Q_4^{(2)}$ (exit)} \\ \hline
\textnormal{Rx$_2$}\ only & \textnormal{Rx$_2$}~only& \ensuremath{p_l^{(1)} } 
& \parbox[t]{0.18\textwidth}{
\textnormal{Rx$_1$}\ gets \ensuremath{p_l^{(1)} }  
\\$\Rightarrow$ \ensuremath{p_l^{(1)} }  to $Q_4^{(1)}$ (exit)\\$\Rightarrow$ Replace \ensuremath{p_m^{(2)}}  with \ensuremath{ p_{\star}^{(2)} }  and move to $Q_2^{(2)}$} 
& \parbox[t]{0.18\textwidth}{
Release  \ensuremath{p_m^{(2)}}  at \textnormal{Rx$_2$}\ 
\\$\Rightarrow$ \ensuremath{p_m^{(2)}}  to $Q_4^{(2)}$ (exit)\\$\Rightarrow$ Put \ensuremath{p_l^{(1)} }  to $Q_2^{(1)}$} 
& \parbox[t]{0.20\textwidth}{
\textnormal{Rx$_1$}\ receives \ensuremath{p_l^{(1)} } , release  \ensuremath{p_m^{(2)}}  at \textnormal{Rx$_2$}\ 
\\$\Rightarrow$ \ensuremath{p_l^{(1)} }  to $Q_4^{(1)}$ (exit) \\$\Rightarrow$ \ensuremath{p_m^{(2)}}  to $Q_4^{(2)}$ (exit)} \\ \hline
\textnormal{Rx$_2$}\ only & \textnormal{Rx$_1$}\ and \textnormal{Rx$_2$}\ & \ensuremath{p_l^{(1)} } 
& \parbox[t]{0.15\textwidth}{
\textnormal{Rx$_1$}\ gets \ensuremath{p_l^{(1)} }  
\\$\Rightarrow$ \ensuremath{p_l^{(1)} }  to $Q_4^{(1)}$ (exit)\\$\Rightarrow$ Put \ensuremath{p_m^{(2)}}  to $Q_2^{(2)}$} 
& \parbox[t]{0.18\textwidth}{
Release  \ensuremath{p_m^{(2)}}  at \textnormal{Rx$_2$}\ 
\\$\Rightarrow$ \ensuremath{p_m^{(2)}}  to $Q_4^{(2)}$ (exit)\\$\Rightarrow$ Put \ensuremath{p_l^{(1)} }  to $Q_2^{(1)}$} 
& \parbox[t]{0.20\textwidth}{
\textnormal{Rx$_1$}\ receives \ensuremath{p_l^{(1)} } , release  \ensuremath{p_m^{(2)}}  at \textnormal{Rx$_2$}\ 
\\$\Rightarrow$ \ensuremath{p_l^{(1)} }  to $Q_4^{(1)}$ (exit) \\$\Rightarrow$ \ensuremath{p_m^{(2)}}  to $Q_4^{(2)}$ (exit)} \\ \hline \hline
\textnormal{Rx$_1$}\ and \textnormal{Rx$_2$}\ & \textnormal{Rx$_1$}\ & \ensuremath{p_m^{(2)}} 
& \parbox[t]{0.15\textwidth}{
Release \ensuremath{p_l^{(1)} }  at \textnormal{Rx$_1$}\ 
\\$\Rightarrow$ \ensuremath{p_l^{(1)} }  to $Q_4^{(1)}$ (exit)\\$\Rightarrow$ Put \ensuremath{p_m^{(2)}}  to $Q_2^{(2)}$} 
& \parbox[t]{0.18\textwidth}{
\textnormal{Rx$_2$}\ receives  \ensuremath{p_m^{(2)}}   
\\$\Rightarrow$ \ensuremath{p_m^{(2)}}  to $Q_4^{(2)}$ (exit)\\$\Rightarrow$ Put \ensuremath{p_l^{(1)} }  to $Q_2^{(1)}$} 
& \parbox[t]{0.20\textwidth}{
Release  \ensuremath{p_l^{(1)} }  at \textnormal{Rx$_1$}\ 
\\$\Rightarrow$ \ensuremath{p_l^{(1)} }  to $Q_4^{(1)}$ (exit) \\$\Rightarrow$ \ensuremath{p_m^{(2)}}  to $Q_4^{(2)}$ (exit)} \\ \hline
\textnormal{Rx$_1$}\ and \textnormal{Rx$_2$}\ & \textnormal{Rx$_2$}\ & \ensuremath{p_m^{(2)}} 
& \parbox[t]{0.15\textwidth}{
Release \ensuremath{p_l^{(1)} }  at \textnormal{Rx$_1$}\ 
\\$\Rightarrow$ \ensuremath{p_l^{(1)} }  to $Q_4^{(1)}$ (exit)\\$\Rightarrow$ Put \ensuremath{p_m^{(2)}}  to $Q_2^{(2)}$} 
& \parbox[t]{0.18\textwidth}{
Release \ensuremath{p_m^{(2)}}  at \textnormal{Rx$_2$}\ 
\\$\Rightarrow$ \ensuremath{p_m^{(2)}}  to $Q_4^{(2)}$ (exit)\\$\Rightarrow$ Put \ensuremath{p_l^{(1)} }  to $Q_2^{(1)}$} 
& \parbox[t]{0.20\textwidth}{
Release  \ensuremath{p_l^{(1)} }  at \textnormal{Rx$_1$}\ and \ensuremath{p_m^{(2)}}  at \textnormal{Rx$_2$}\ 
\\$\Rightarrow$ \ensuremath{p_l^{(1)} }  to $Q_4^{(1)}$ (exit) \\$\Rightarrow$ \ensuremath{p_m^{(2)}}  to $Q_4^{(2)}$ (exit)} \\ \hline
\textnormal{Rx$_1$}\ and \textnormal{Rx$_2$}\ & \textnormal{Rx$_1$}\ and \textnormal{Rx$_2$}\ & \ensuremath{p_l^{(1)} } 
& \parbox[t]{0.15\textwidth}{
Release \ensuremath{p_l^{(1)} }  at \textnormal{Rx$_1$}\ 
\\$\Rightarrow$ \ensuremath{p_l^{(1)} }  to $Q_4^{(1)}$ (exit)\\$\Rightarrow$ Put \ensuremath{p_m^{(2)}}  to $Q_2^{(2)}$} 
& \parbox[t]{0.18\textwidth}{
Release \ensuremath{p_m^{(2)}}  at \textnormal{Rx$_2$}\ 
\\$\Rightarrow$ \ensuremath{p_m^{(2)}}  to $Q_4^{(2)}$ (exit)\\$\Rightarrow$ Put \ensuremath{p_l^{(1)} }  to $Q_2^{(1)}$} 
& \parbox[t]{0.20\textwidth}{
Release  \ensuremath{p_l^{(1)} }  at \textnormal{Rx$_1$}\ and \ensuremath{p_m^{(2)}}  at \textnormal{Rx$_2$}\ 
\\$\Rightarrow$ \ensuremath{p_l^{(1)} }  to $Q_4^{(1)}$ (exit) \\$\Rightarrow$ \ensuremath{p_m^{(2)}}  to $Q_4^{(2)}$ (exit)} \\ \hline \hline
\multicolumn{1}{c}{Resulting capacity:} & \multicolumn{1}{c}{} & \multicolumn{1}{c}{} & \multicolumn{1}{c}{$C_{34}^{(1)}=1$, $C_{32}^{(2)}=1$}   &   \multicolumn{1}{c}{$C_{32}^{(1)}=1$, $C_{34}^{(2)}=1$}   & \multicolumn{1}{c}{$C_{34}^{(1)}=1$, $C_{34}^{(2)}=1$}
 \end{tabular}
\caption{Packet movement for poisoned packets in $Q_3^{(1)}$ and $Q_3^{(2)}$ - degenerate cases.}
\label{tab:packet_movement_degenerate}
\end{table}

\clearpage
\section{Proof of Proposition~\ref{prop:cuts}}
\label{sec:appendix_proof_cuts}

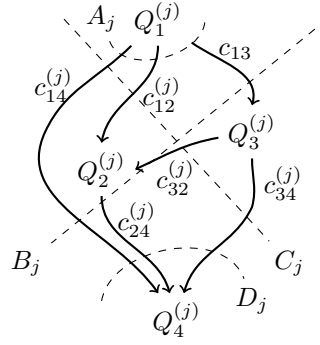
\begin{figure}[ht]
\centering
\begin{tikzpicture}[scale=0.5]
  
  \node (Q11) at (0.5,-0.5) {$Q_1^{(j)}$};

  \node (Q21) at (-1,-4.5) {$Q_2^{(j)}$};

  \node (Q31) at (3,-3.5) {$Q_3^{(j)}$};

\node (Q41) at (1,-8.5) {$Q_4^{(j)}$};

\draw[->, thick] (Q11) to [out=270,in=80] node[right] {$c_{12}^{(j)}$} (Q21.north);
\draw[->, thick] (Q11) to [out=225,in=65] node[left]{$c_{14}^{(j)}$} +(-3,-3) to [out=245,in=125] (Q41);

\draw[->,thick] (Q21) to [out=270,in=105] node[right,pos=0.2]{$c_{24}^{(j)}$} (Q41);
\draw[->,thick] (Q11) to [out=325,in=75] node[above]{$c_{13}$} (Q31.north);
\draw[->,thick] (Q31) to [out=270,in=75] node[right]{$c_{34}^{(j)}$} +(0,-2) to [out=245,in=75] (Q41);
\draw[->,thick] (Q31) to [out=190,in=25] node[below]{$c_{32}^{(j)}$} (Q21.east);

\node (A) at (-1,-0.5) {$A_j$};
\draw[dashed] (A) to [out=290,in=270] (1.75,-0.5);

\node (C) at (4,-7) {$C_j$};
\draw[dashed] (-2,-0.5) to (C);

\node (B) at (-3,-7) {$B_j$};
\draw[dashed] (5,-0.5) to (B);

\node (D) at (3,-8) {$D_j$};
\draw[dashed] (-1,-8) to [out=80,in=110] (D);

\end{tikzpicture}
\caption{Queue network and cuts.}
\label{fig:queues_cuts}
\end{figure}

We define the variables $A_j$, $B_j$, $C_j$, $D_j$, $j \in \{1,2\}$ as the cut values in \eqref{eq:bec_cut1}~-~\eqref{eq:bec_cut4} (see Fig.~\ref{fig:queues_cuts}):
\begin{align}
 A_j &= c_{12}^{(j)} +  c_{13} + c_{14}^{(j)} \nonumber\\
     &= \sum_{s \in \mc S} \pi_s (1-\epsonetwo{s}) \left[ P_{A_t|S_{t-1}}(j|s) + P_{A_t|S_{t-1}}(4|s) \right] \\
 B_j &= c_{13} +  c_{14}^{(j)} + c_{24}^{(j)} \nonumber\\
     &=\sum_{s \in \mc S} \pi_s \Big[ (1-\epsonetwo{s})  P_{A_t|S_{t-1}}(4|s)  + (1-\epsj{s}) \left[ P_{A_t|S_{t-1}}(j|s) + P_{A_t|S_{t-1}}(3|s) \right]  \Big]  \\
 C_j &= c_{12}^{(j)} +  c_{14}^{(j)} + c_{32}^{(j)} + c_{34}^{(j)} \nonumber\\
     &=\sum_{s \in \mc S} \pi_s (1-\epsonetwo{s}) \left[ P_{A_t|S_{t-1}}(j|s) + P_{A_t|S_{t-1}}(5|s) \right]  \\
 D_j &= c_{14}^{(j)} +  c_{24}^{(j)} + c_{34}^{(j)} \nonumber\\
     &=\sum_{s \in \mc S} \pi_s (1-\epsj{s}) \left[ P_{A_t|S_{t-1}}(j|s) + P_{A_t|S_{t-1}}(3|s) + P_{A_t|S_{t-1}}(5|s) \right]
\end{align}
We omit the superscript $(j)$ for $c_{13}$ to emphasize that $c_{13}^{(1)}= c_{13}^{(2)}= c_{13} $.

Our goal is to show that for any link capacities $c_{rl}^{(j)}$ and associated cut values $A_j$, $B_j$, $C_j$, $D_j$ induced by a distribution $P_{A_t|S_{t-1}}$, there is another distribution $P^\star_{A_t|S_{t-1}}$ with associated link capacities $c_{rl}^{(j)\star}$ and cut values $A^\star_j$, $B^\star_j$, $C^\star_j$, $D^\star_j$ such that 
\begin{align}
 \min\left\{ A^\star_j, B^\star_j, C^\star_j, D^\star_j \right\} &= \min\left\{ A_j, B_j, C_j, D_j \right\} \label{eq:cuts_criterion1} \\
 \min\left\{ A^\star_j, B^\star_j, C^\star_j, D^\star_j \right\} &= \min\left\{ A^\star_j, D^\star_j \right\}. \label{eq:cuts_criterion2}
\end{align}
That is, the minimal cut value does not change under $P^\star_{A_t|S_{t-1}}$, but the cuts $B^\star_j$ and $C^\star_j$ are redundant.
The next lemma states one sufficient condition:

\begin{lemma}
\label{lem:flow_div_equality}
 The cut values $B_j$ and $C_j$ are redundant for both $j=1,2$ if $P_{A_t|S_{t-1}}$ leads to
 \begin{align}
c_{13} &= c_{32}^{(j)} + c_{34}^{(j)}\quad \forall~j\in \{1,2\}. \label{eq:flow_div_equality}
\end{align}
\end{lemma}
\begin{IEEEproof}
The condition in \eqref{eq:flow_div_equality} is equivalent to 
\begin{align}
\sum_{s \in \mc S} \pi_s   (1 - \epsonetwo{s}) P_{A_t|S_{t-1}}(4|s) & = \sum_{s \in \mc S} \pi_s   (1 - \epsonetwo{s}) P_{A_t|S_{t-1}}(5|s). 
\end{align}
 One can verify that $C_j = A_j$ in this case, so $C_j$ can be omitted.
Additionally, $D_j \leq B_j$ because
\begin{align}
 B_j - D_j =&c_{13} +  c_{14}^{(j)} + c_{24}^{(j)} - \left( c_{14}^{(j)} +  c_{24}^{(j)} + c_{34}^{(j)} \right) \nonumber \\
=& \left(\sum_{s \in \mc S} \pi_s   (1 - \epsonetwo{s}) P_{A_t|S_{t-1}}(4|s)\right) - \left(\sum_{s \in \mc S} \pi_s   (1 - \epsj{s}) P_{A_t|S_{t-1}}(5|s)\right) \nonumber \\
=& \left(\sum_{s \in \mc S} \pi_s   (1 - \epsonetwo{s}) P_{A_t|S_{t-1}}(5|s)\right) - \left(\sum_{s \in \mc S} \pi_s   (1 - \epsj{s}) P_{A_t|S_{t-1}}(5|s)\right) \nonumber \\
\geq & 0,
\end{align}
where the last inequality is due to $1 - \epsonetwo{s} \geq 1 - \epsj{s}$ for all $s \in \mc S$ and $j\in \{1,2\}$.
\end{IEEEproof}

\begin{remark}
 Note that one \emph{cannot} simply set $P_{A_t|S_{t-1}}(4|s) = P_{A_t|S_{t-1}}(5|s)$, $\forall s \in \mc S$, to ensure \eqref{eq:flow_div_equality}.
 By the mapping in \eqref{eq:proba_relation_xs}~-~\eqref{eq:proba_relation_ys}, this would require $x_s + y_s \geq 1$, $\forall s \in \mc S$ and hence permit only a limited set of values for $x_s$ and $y_s$.
\end{remark}

For any given distribution $P_{A_t|S_{t-1}}$, we choose $P^\star_{A_t|S_{t-1}}$ such that $\forall s \in \mc S$:
\begin{align}
P^\star_{A_t|S_{t-1}}(0|s) &= P_{A_t|S_{t-1}}(0|s) \label{eq:pstar1} \\
 P^\star_{A_t|S_{t-1}}(1|s) &= P_{A_t|S_{t-1}}(1|s)  \\
 P^\star_{A_t|S_{t-1}}(2|s) &= P_{A_t|S_{t-1}}(2|s) \\ 
 P^\star_{A_t|S_{t-1}}(4|s) &= P_{A_t|S_{t-1}}(4|s) \\
 P^\star_{A_t|S_{t-1}}(3|s)+ P^\star_{A_t|S_{t-1}}(5|s) & = P_{A_t|S_{t-1}}(3|s)+ P_{A_t|S_{t-1}}(5|s) \label{eq:sum_constr}
\end{align}
By choosing $P^\star_{A_t|S_{t-1}}(3|s) \neq P_{A_t|S_{t-1}}(3|s)$, $P^\star_{A_t|S_{t-1}}(5|s) \neq P_{A_t|S_{t-1}}(5|s)$, the link capacities $c_{24}^{(j)\star}$, $c_{34}^{(j)\star}$ and $c_{32}^{(j)\star}$, $j \in \{1,2\}$ are varied. 
The other link capacities are not affected, hence 
\begin{align}
 c_{12}^{(j)\star} = c_{12}^{(j)}, \quad
 c_{13}^{(j)\star} = c_{13}^{(j)}, \quad
 c_{14}^{(j)\star} = c_{14}^{(j)}, ~ j \in \{1,2\}. 
\end{align}
In the following we therefore omit the superscript $\star$ for those capacities that stay constant with $P_{A_t|S_{t-1}}$ and $P^\star_{A_t|S_{t-1}}$.

Note that $A^\star_j=A_j$, $D^\star_j=D_j$, $j \in \{1,2\}$ because 
\begin{itemize}
 \item $A^\star_j$ is not affected by changing $P^\star_{A_t|S_{t-1}}(3|s)$ and $P^\star_{A_t|S_{t-1}}(5|s)$ and
 \item $D^\star_j$ depends only on the sum $P^\star_{A_t|S_{t-1}}(3|s)+P^\star_{A_t|S_{t-1}}(5|s)$ that is kept constant \eqref{eq:sum_constr}, hence
 \begin{align}
  c_{24}^{(j)} + c_{34}^{(j)} &= c_{24}^{(j)\star} + c_{34}^{(j)\star}. \label{eq:sum_constr_equal}
   \end{align}
\end{itemize}

By changing $P^\star_{A_t|S_{t-1}}(3|s)$ and $P^\star_{A_t|S_{t-1}}(5|s)$ under the sum-constraint \eqref{eq:sum_constr} one can obtain the maximal and minimal link capacities for $c_{24}^{(j)\star}$ and $c_{34}^{(j)\star}$, as follows:
\begin{enumerate}
 \item $c_{24}^{(j)\star} = D_j - c_{14}^{(j)}$, $c_{34}^{(j)\star}=0$, $j\in\{1,2\}$, \\by setting $P^\star_{A_t|S_{t-1}}(3|s)=P_{A_t|S_{t-1}}(3|s)+P_{A_t|S_{t-1}}(5|s)$ and $P^\star_{A_t|S_{t-1}}(5|s)=0$, $\forall s \in \mc S$.
 \item $c_{24}^{(j)\star}=0$, $c_{34}^{(j)\star} = D_j - c_{14}^{(j)}$, $j\in\{1,2\}$, \\by setting $P^\star_{A_t|S_{t-1}}(3|s)=0$ and $P^\star_{A_t|S_{t-1}}(5|s)=P_{A_t|S_{t-1}}(3|s)+P_{A_t|S_{t-1}}(5|s)$,  $\forall s \in \mc S$.
\end{enumerate}
Due to the continuity of the link capacities with respect to $P^{\star}_{A_t|S_{t-1}}$, any convex combination of $\left( D_j - c_{14}^{(j)}, 0 \right)$ and $\left(0, D_j - c_{14}^{(j)}\right)$ can be obtained for $\left( c_{24}^{(j)\star},c_{34}^{(j)\star}\right)$. 

We now show that choosing $P^\star_{A_t|S_{t-1}}$ as in \eqref{eq:pstar1}~-~\eqref{eq:sum_constr} suffices to ensure the desired criteria \eqref{eq:cuts_criterion1}~-~\eqref{eq:cuts_criterion2}.
We distinguish two cases. For one case, we use Lemma~\ref{lem:flow_div_equality}.

\begin{itemize}%
 \item Case I: $A_j \leq D_j$ for at least one $j \in \{1,2\}$. 
 This happens if we have
 \begin{align}
  c_{12}^{(j)} + c_{13} \leq c_{24}^{(j)} + c_{34}^{(j)} \quad \text{ for some } j \in \{1,2\}. \label{eq:case1crit}
 \end{align}
 In this case we choose $P^\star_{A_t|S_{t-1}}(5|s)$ such that $c_{34}^{(j)\star} = c_{13} - c_{32}^{(j)\star}$.
 By Lemma~\ref{lem:flow_div_equality}, this suffices to ensure the criteria \eqref{eq:cuts_criterion1}~-~\eqref{eq:cuts_criterion2}.\\
 We can always find such values for $P^\star_{A_t|S_{t-1}}(5|s)$ because, by definition of Case I in \eqref{eq:case1crit},
 there is a $j\in\{1,2\}$ for which 
 \begin{align}
  D_j - c_{14}^{(j)} \geq c_{12}^{(j)} + c_{13}. \label{eq:case1_labeleq}
 \end{align}
 The LHS of \eqref{eq:case1_labeleq} is the maximal possible link capacity for $c_{34}^{(j)\star}$. The RHS of \eqref{eq:case1_labeleq} is larger than (or equal to) $c_{13} - c_{32}^{(j)\star}$, which is the desired value for $c_{34}^{(j)\star}$.
 We can adjust $c_{34}^{(j)\star}$ between $0$ and $D_j - c_{14}^{(j)}$. As $c_{13} - c_{32}^{(j)\star}$ lies in this interval, we can choose values for $P_{A_t|S_{t-1}}^\star(5|s)$ such that $c_{34}^{(j)\star} = c_{13} - c_{32}^{(j)\star}$.
 \item Case II: $ D_j < A_j $ for both $j \in \{1,2\}$. \\
 In this case the sufficient criterion in Lemma~\ref{lem:flow_div_equality} cannot be guaranteed, since it is possible that  
there is no distribution $P^\star_{A_t|S_{t-1}}(5|s)$ such that $c_{13} = c_{32}^{(j)\star} + c_{34}^{(j)\star}$.
To satisfy the criteria \eqref{eq:cuts_criterion1}~-~\eqref{eq:cuts_criterion2}, we need for both $j \in \{1,2\}$:
\begin{alignat}{2}
 D_j &\leq B_j^\star \qquad& \Leftrightarrow \qquad  c_{34}^{(j)\star} &\leq c_{13} \label{eq:CaseIIcrit1} \\
  D_j &\leq C_j^\star & \Leftrightarrow \qquad c_{24}^{(j)\star} &\leq c_{32}^{(j)\star} + c_{12}^{(j)} \label{eq:CaseIIcrit2}
\end{alignat}
In this case we choose $P^\star_{A_t|S_{t-1}}(5|s)$ such that $\max \left\{c_{34}^{(1)\star}, c_{34}^{(2)\star} \right\}$ is as large as possible, but at most equal to $c_{13}$, in order not to violate \eqref{eq:CaseIIcrit1}. Two sub-cases must be distinguished:
\begin{itemize}[leftmargin=+17mm]
 \item Case IIa: One can choose $P^\star_{A_t|S_{t-1}}(5|s)$ such that $\max \left\{c_{34}^{(1)\star}, c_{34}^{(2)\star} \right\} = c_{13}$.\\
 The condition in \eqref{eq:CaseIIcrit1} is satisfied by construction, so we have to check only \eqref{eq:CaseIIcrit2}.
 Note that the following inequalities always hold for Case IIa:
 \begin{align}
  c_{24}^{(j)} + c_{34}^{(j)} &< c_{12}^{(j)} + c_{13}, \qquad \text{because } D_j < A_j, \text{ and} \nonumber \\
  c_{24}^{(j)\star} + c_{34}^{(j)\star} &< c_{12}^{(j)} + c_{13}, \qquad \text{because of \eqref{eq:sum_constr_equal}. Hence, }\nonumber \\
  c_{24}^{(j)\star}  &< c_{12}^{(j)} + \left( c_{13} - c_{34}^{(j)\star} \right).
 \end{align}
  For condition \eqref{eq:CaseIIcrit2} to hold, the term $\left( c_{13} - c_{34}^{(j)\star} \right)$ should be smaller than (or equal to) $c_{32}^{(j)\star}$ for both $j=\in \{1,2\}$. \\
 For $j_{\max} = \argmax_{j \in \{1,2\}} c_{34}^{(j)\star} $, we have $ c_{13} - c_{34}^{(j_{\max})\star}  = 0$, satisfying the condition. 
 For $j_{\min}= \argmin_{j \in \{1,2\}} c_{34}^{(j)\star} $, we have $c_{13} - c_{34}^{(j_{\min})\star}  \geq 0$. We next show that  $c_{13} - c_{34}^{(j_{\min})\star}  \leq c_{32}^{(j_{\min})\star}$ holds for this case as well.
 As $c_{13} = c_{34}^{(j_{\max})\star}$, we have
 \begin{align}
  c_{13} - c_{34}^{(j_{\min})\star} = & c_{34}^{(j_{\max})\star} - c_{34}^{(j_{\min})\star} 
  \nonumber \\
  =& \sum_{s \in \mc S} \pi_s \left( \epsilon_{j_{\min}}(s) - \epsilon_{j_{\max}}(s)\right) P^\star_{A_t|S_{t-1}}(5|s) 
  \geq  0 .\nonumber
 \end{align}
The following statement shows that $c_{13} - c_{34}^{(j_{\min})\star}  \leq c_{32}^{(j_{\min})\star}$:
\begin{align}
 &c_{13} - c_{34}^{(j_{\min})\star} - c_{32}^{(j_{\min})\star} \nonumber \\
 = &\sum_{s \in \mc S} \pi_s \left[\left( \epsilon_{j_{\min}}(s) - \epsilon_{j_{\max}}(s) \right) - \left( \epsilon_{j_{\min}}(s) - \epsonetwo{s} \right) \right]P^\star_{A_t|S_{t-1}}(5|s) \nonumber \\
 = & \sum_{s \in \mc S} \pi_s \left[ \epsonetwo{s} - \epsilon_{j_{\max}}(s)\right]P^\star_{A_t|S_{t-1}}(5|s) \leq 0,
\end{align}
as $\epsonetwo{s} - \epsilon_{j_{\max}}(s)\leq 0$ for all $s \in \mc S$, $j\in \{1,2\}$.
This shows that the choice of $P^\star_{A_t|S_{t-1}}(5|s)$ such that $\max \left\{c_{34}^{(1)\star}, c_{34}^{(2)\star} \right\} = c_{13}$ suffices to achieve the criteria \eqref{eq:cuts_criterion1}~-~\eqref{eq:cuts_criterion2}.
\item Case IIb: Choosing the maximal $P^\star_{A_t|S_{t-1}}(5|s)$ $\forall s \in \mc S$ leads to $\max \left\{c_{34}^{(1)\star}, c_{34}^{(2)\star} \right\} < c_{13}$.\\
This immediately satisfies the condition in \eqref{eq:CaseIIcrit1}.
As $P^\star_{A_t|S_{t-1}}(5|s)$ is maximal, we have $P^\star_{A_t|S_{t-1}}(3|s)=0$ $\forall s \in \mc S$. Hence, we have $c_{24}^{(j)\star}=0$, $\forall j \in \{1,2\}$, satisfying the condition \eqref{eq:CaseIIcrit2}.
\end{itemize}
\end{itemize}
This completes the proof.

\section{Proof of Max-Weight Schemes}
\label{sec:proof_prop_maxweight_joint}
This section proves Theorem~\ref{prop:maxweight_new} and Theorem~\ref{prop:maxweight_new_hmm}, showing that the max-weight criteria in \eqref{eq:maxweight_new} and \eqref{eq:maxweight_new_hmm} strongly stabilize all queues in the network, as defined in \eqref{eq:strong_stability}. We use Lyapunov-drift theory to prove the result.

\begin{figure}[ht]
\begin{subfigure}[t]{0.45\textwidth}
 \centering
  \tikzset{>=latex}
\begin{tikzpicture}[scale = 0.5, every node/.style={scale=0.6}]

\node at (2,-5) [fill,circle, inner sep = 1mm] (Q1) {};
\node [above=0.5mm of Q1] () {$\ve Q_{1}$};

\node at (5,-5) [fill,circle, inner sep = 1mm] (Q2) {};
\node [above=0.5mm of Q2] () {$\ve Q_{2}$};

\node at (8,-5) [fill,circle, inner sep = 1mm] (Q3) {};
\node [above=0.5mm of Q3] () {$\ve Q_{3}$};

\node at (11,-5) [fill,circle, inner sep = 1mm] (Q4) {};
\node [above=0.5mm of Q4] () {$\ve Q_{4}$};
\node [right=0.5mm of Q4] () {$\ldots$};

\node at (3.5,-4) [fill,circle, inner sep = 1mm] (E1) {};
\node [right=0.5mm of E1] () {$\ve{\activRV{}{}}_{1}$};

\node at (6.5,-4) [fill,circle, inner sep = 1mm] (E2) {};
\node [right=0.5mm of E2] () {$\ve{\activRV{}{}}_{2}$};

\node at (9.5,-4) [fill,circle, inner sep = 1mm] (E3) {};
\node [right=0.5mm of E3] () {$\ve{\activRV{}{}}_{3}$};

\node at (2.5,-7) [fill,circle, inner sep = 1mm] (A1) {};
\node [below=0.5mm of A1] () {$A_{1}$};

\node at (5.5,-7) [fill,circle, inner sep = 1mm] (A2) {};
\node [below=0.5mm of A2] () {$A_{2}$};

\node at (8.5,-7) [fill,circle, inner sep = 1mm] (A3) {};
\node [below=0.5mm of A3] () {$A_{3}$};

\node at (11.5,-7) [fill,circle, inner sep = 1mm] (A4) {};
\node [below=0.5mm of A4] () {$A_{4}$};
\node [right=0.5mm of A4] () {$\ldots$};

\node at (3.5,-7) [fill,circle, inner sep = 1mm] (Z1) {};
\node [right=0.5mm of Z1] () {$\ve Z_{1}$};

\node at (6.5,-7) [fill,circle, inner sep = 1mm] (Z2) {};
\node [right=0.5mm of Z2] () {$\ve Z_{2}$};

\node at (9.5,-7) [fill,circle, inner sep = 1mm] (Z3) {};
\node [right=0.5mm of Z3] () {$\ve Z_{3}$};

\node at (-1,-9) [fill,circle, inner sep = 1mm] (S0) {};
\node [below=0.5mm of S0] () {$S_{0}$};

\node at (2,-9) [fill,circle, inner sep = 1mm] (S1) {};
\node [below=0.5mm of S1] () {$S_{1}$};

\node at (5,-9) [fill,circle, inner sep = 1mm] (S2) {};
\node [below=0.5mm of S2] () {$S_{2}$};

\node at (8,-9) [fill,circle, inner sep = 1mm] (S3) {};
\node [below=0.5mm of S3] () {$S_{3}$};
\node [right=0.5mm of S3] () {$\ldots$};

\draw[->-=.5] (S0) -- (S1);
\draw[->-=.5] (S1) -- (S2);
\draw[->-=.5] (S2) -- (S3);

\draw[->-=.5] (S0) -- (A1);
\draw[->-=.5] (S1) -- (A2);
\draw[->-=.5] (S2) -- (A3);
\draw[->-=.5] (S3) -- (A4);

\draw[->-=.5] (S1) -- (Z1);
\draw[->-=.5] (S2) -- (Z2);
\draw[->-=.5] (S3) -- (Z3);

\draw[->-=.5] (Q1) -- (Q2);
\draw[->-=.5] (Q2) -- (Q3);
\draw[->-=.5] (Q3) -- (Q4);

\draw[->-=.5] (Q1) -- (A1);
\draw[->-=.5] (Q2) -- (A2);
\draw[->-=.5] (Q3) -- (A3);
\draw[->-=.5] (Q4) -- (A4);

\draw[->-=.5] (A1) -- (Q2);
\draw[->-=.5] (A2) -- (Q3);
\draw[->-=.5] (A3) -- (Q4);

\draw[->-=.5] (Z1) -- (Q2);
\draw[->-=.5] (Z2) -- (Q3);
\draw[->-=.5] (Z3) -- (Q4);

\draw[->-=.5] (Q1) -- (E1);
\draw[->-=.5] (Q2) -- (E2);
\draw[->-=.5] (Q3) -- (E3);

\draw[->-=.5] (E1) -- (Q2);
\draw[->-=.5] (E2) -- (Q3);
\draw[->-=.5] (E3) -- (Q4);

 \end{tikzpicture}
\caption{Hidden State:  $\ensuremath{\Omega}^{t-1} = S_{t-1}$. }
\label{fig:fdg_queues_visible}
\end{subfigure}
\begin{subfigure}[t]{0.45\textwidth}
 \centering
  \tikzset{>=latex}
\begin{tikzpicture}[scale = 0.5, every node/.style={scale=0.6}]

\node at (2,-5) [fill,circle, inner sep = 1mm] (Q1) {};
\node [above=0.5mm of Q1] () {$\ve Q_{1}$};

\node at (5,-5) [fill,circle, inner sep = 1mm] (Q2) {};
\node [above=0.5mm of Q2] () {$\ve Q_{2}$};

\node at (8,-5) [fill,circle, inner sep = 1mm] (Q3) {};
\node [above=0.5mm of Q3] () {$\ve Q_{3}$};

\node at (11,-5) [fill,circle, inner sep = 1mm] (Q4) {};
\node [above=0.5mm of Q4] () {$\ve Q_{4}$};
\node [right=0.5mm of Q4] () {$\ldots$};

\node at (3.5,-4) [fill,circle, inner sep = 1mm] (E1) {};
\node [right=0.5mm of E1] () {$\ve{\activRV{}{}}_{1}$};

\node at (6.5,-4) [fill,circle, inner sep = 1mm] (E2) {};
\node [right=0.5mm of E2] () {$\ve{\activRV{}{}}_{2}$};

\node at (9.5,-4) [fill,circle, inner sep = 1mm] (E3) {};
\node [right=0.5mm of E3] () {$\ve{\activRV{}{}}_{3}$};

\node at (2.5,-6) [fill,circle, inner sep = 1mm] (A1) {};
\node [below=0.5mm of A1] () {$A_{1}$};

\node at (5.5,-6) [fill,circle, inner sep = 1mm] (A2) {};
\node [below=0.5mm of A2] () {$A_{2}$};

\node at (8.5,-6) [fill,circle, inner sep = 1mm] (A3) {};
\node [below=0.5mm of A3] () {$A_{3}$};

\node at (11.5,-6) [fill,circle, inner sep = 1mm] (A4) {};
\node [below=0.5mm of A4] () {$A_{4}$};
\node [right=0.5mm of A4] () {$\ldots$};

\node at (3,-8) [fill,circle, inner sep = 1mm] (Z1) {};
\node [below right=0.5mm of Z1] () {$\ve Z_{1}$};

\node at (6,-8) [fill,circle, inner sep = 1mm] (Z2) {};
\node [below right=0.5mm of Z2] () {$\ve Z_{2}$};

\node at (9,-8) [fill,circle, inner sep = 1mm] (Z3) {};
\node [below right=0.5mm of Z3] () {$\ve Z_{3}$};

\node at (-1,-9) [fill,circle, inner sep = 1mm] (S0) {};
\node [below=0.5mm of S0] () {$S_{0}$};

\node at (2,-9) [fill,circle, inner sep = 1mm] (S1) {};
\node [below=0.5mm of S1] () {$S_{1}$};

\node at (5,-9) [fill,circle, inner sep = 1mm] (S2) {};
\node [below=0.5mm of S2] () {$S_{2}$};

\node at (8,-9) [fill,circle, inner sep = 1mm] (S3) {};
\node [below=0.5mm of S3] () {$S_{3}$};
\node [right=0.5mm of S3] () {$\ldots$};

\draw[->-=.5] (S0) -- (S1);
\draw[->-=.5] (S1) -- (S2);
\draw[->-=.5] (S2) -- (S3);

\draw[->-=.6] (Z1) -- (A2);
\draw[->-=.6] (Z1) -- (A3);
\draw[->-=.6] (Z1) -- (A4);

\draw[->-=.6] (Z2) -- (A3);
\draw[->-=.6] (Z2) -- (A4);

\draw[->-=.6] (Z3) -- (A4);

\draw[->-=.7] (S1) -- (Z1);
\draw[->-=.7] (S2) -- (Z2);
\draw[->-=.7] (S3) -- (Z3);

\draw[->-=.5] (Q1) -- (Q2);
\draw[->-=.5] (Q2) -- (Q3);
\draw[->-=.5] (Q3) -- (Q4);

\draw[->-=.5] (Q1) -- (A1);
\draw[->-=.5] (Q2) -- (A2);
\draw[->-=.5] (Q3) -- (A3);
\draw[->-=.5] (Q4) -- (A4);

\draw[->-=.5] (A1) -- (Q2);
\draw[->-=.5] (A2) -- (Q3);
\draw[->-=.5] (A3) -- (Q4);

\draw[->-=.6] (Z1) -- (Q2);
\draw[->-=.6] (Z2) -- (Q3);
\draw[->-=.6] (Z3) -- (Q4);

\draw[->-=.5] (Q1) -- (E1);
\draw[->-=.5] (Q2) -- (E2);
\draw[->-=.5] (Q3) -- (E3);

\draw[->-=.5] (E1) -- (Q2);
\draw[->-=.5] (E2) -- (Q3);
\draw[->-=.5] (E3) -- (Q4);

 \end{tikzpicture}
\caption{Hidden State:  $\ensuremath{\Omega}^{t-1} = \ve Z^{t-1}$. }
\label{fig:fdg_queues_hidden}
\end{subfigure}
\caption{Bayesian network of the queueing systems. Actions at time $t$ may depend on the previous observations $\ensuremath{\Omega}^{t-1}$ and the current buffer state $\ve Q_t$, where $\ve Q_t$ denotes the buffer state \emph{before} executing action $A_t$.  }
\label{fig:fdg_queues}
\end{figure}
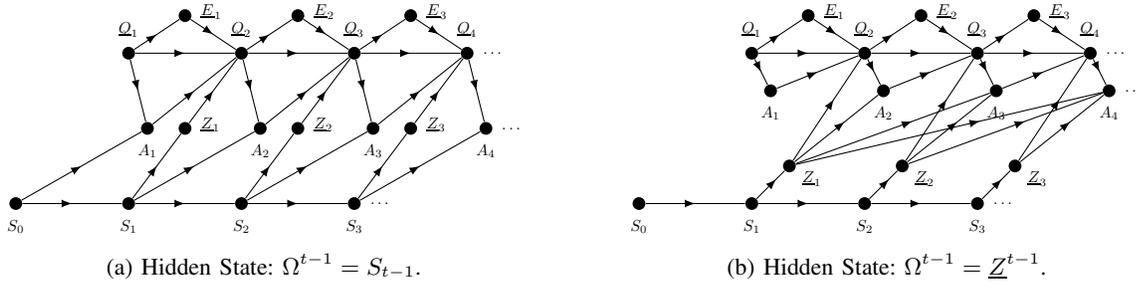

 Recall the dynamics of queues $Q_{1,t}^{(j)}$, $Q_{2,t}^{(j)}$ and $Q_{3,t}^{(j)}$ defined in \eqref{eq:queue_dynamics}:
 \begin{align}
  Q_{1,t+1}^{(j)} &\leq [Q_{1,t}^{(j)} - F_{12,t}^{(j)} - F_{13,t}^{(j)} - F_{14,t}^{(j)}]^+ + F_{01,t}^{(j)}\\
  Q_{2,t+1}^{(j)} &\leq [Q_{2,t}^{(j)} - F_{24,t}^{(j)}]^+ + F_{12,t}^{(j)} + F_{32,t}^{(j)} \label{eq:queue_update2_new}\\
  Q_{3,t+1}^{(j)} &\leq [Q_{3,t}^{(j)} - F_{32,t}^{(j)} - F_{34,t}^{(j)}]^+ + F_{13,t}^{(j)} \label{eq:queue_update3_new}
 \end{align}
The flow variables $F_{lm,t}^{(j)}$ depend on the action $A_t$, the activation variables $\ve{\activRV{}{}}_t$ and the erasures $\ve Z_t$, so the queue state $\ve Q_{t+1}$ is a function of $\ve Q_t$, $A_t$, $\ve{\activRV{}{}}_t$ and $\ve Z_t$.
In the visible case, actions are restricted to depend only on the current queue state $\ve Q_t$ and on the previous channel state $S_{t-1}$, hence are according to a distribution $P_{A_t| \ve Q_t S_{t-1}}$. In the hidden case, it is according to $P_{A_t| \ve Q_t \ve Z^{t-1}}$.
Many steps for the two cases are similar: To harmonize notation, let the actions depend on the \emph{observations}  $\ensuremath{\Omega}^{t-1}$, with the understanding that $\ensuremath{\Omega}^{t-1} = S_{t-1}$ in the visible case, and $\ensuremath{\Omega}^{t-1} = \ve Z^{t-1}$ in the hidden case.
Packet movement decisions may depend on the current queue state, hence are according to $P_{\ve{\activRV{}{}}_t|\ve Q_t}$.
All dependencies are depicted in the Bayesian networks in Fig.~\ref{fig:fdg_queues}.

Define the Lyapunov function $\mathbb L(\ve Q_t)$ as
\begin{align}
 \mathbb L(\ve Q_t) = \sum_{j=1}^2 \sum_{l=1}^3 \left( Q_{l,t}^{(j)} \right)^2
\end{align}
and the $T$-slot conditional Lyapunov drift $\Delta(\ve Q_t)$ as
\begin{align}
 \Delta(\ve Q_t) = \mathbb{E}\left[\mathbb L(\ve Q_{t+T}) - \mathbb L(\ve Q_{t}) \left|\ve Q_t \right. \right]. \label{eq:condLyap}
\end{align}
$\Delta(\ve Q_t)$ measures the expected reduction or increase of the aggregate queue lengths from slot $t$ to slot $t+T$, conditioned on $\ve Q_t$.

Split $\Delta(\ve Q_t)$ into the telescoping sum
\begin{align}
  \Delta(\ve Q_t) =& \left.\mathbb{E}\left[\sum_{\tau=t}^{t+T-1} \mathbb L(\ve Q_{\tau+1}) - \mathbb L(\ve Q_\tau) \right\vert \ve Q_t  \right] \nonumber\\
  =&  \sum_{\tau=t}^{t+T-1} \mathbb{E}\left[ \mathbb L(\ve Q_{\tau+1}) - \mathbb L(\ve Q_\tau) \left| \ve Q_t \right. \right].
  \label{eq:telescopingsum}
  \end{align}
  The individual expectation terms of the sum in \eqref{eq:telescopingsum} depend on the conditioning $\ve Q_t$ only through $\ve Q_\tau$ and $\ensuremath{\Omega}^{\tau-1}$, as $\ve Q_{\tau+1} - \ve Q_\tau \ensuremath{\Omega}^{\tau-1} - \ve Q_t$ forms a Markov chain for $\tau\geq t$.
  Hence, the law of total expectation yields
\begin{align}
 &\Delta(\ve Q_t)= \label{eq:total_exp}
 \sum_{\tau=t}^{t+T-1} \mathbb E \Big[ \mathbb E \left[ \mathbb L(\ve Q_{\tau+1}) - \mathbb L(\ve Q_\tau) \big| \ve Q_\tau \ensuremath{\Omega}^{\tau-1}  \right] \Big| \ve Q_t \Big].
\end{align}
We bound the individual terms inside the inner expectation of \eqref{eq:total_exp} next.
We can use \cite[Lemma 4.3]{georgiadis2006resource}, which states that
for any nonnegative numbers $v,u,\mu,\alpha$ satisfying
$v \leq [u-\mu]^+ + \alpha$, we have
\begin{align}
v^2 &\leq u^2 + \mu^2 + \alpha^2 - 2u(\mu - \alpha).
\end{align}
We apply this lemma and combine it with $(C_{lm,\tau}^{(j)})^2 = C_{lm,\tau}^{(j)} \leq 1$ because $C_{lm,\tau}^{(j)}$ is either $1$ or $0$ and obtain the following bound: %
 \begin{align}
 \mathbb L(\ve Q_{\tau+1})-\mathbb L(\ve Q_\tau)
 &\leq 
 12 - 2\sum_{j=1}^2   Q_{1,\tau}^{(j)} \flowdivRV{1,\tau}{(j)} + Q_{2,\tau}^{(j)} \flowdivRV{2,\tau}{(j)} + Q_{3,\tau}^{(j)} \flowdivRV{3,\tau}{(j)},
 \label{eq:flow_bound}
\end{align}
where $\flowdivRV{l,\tau}{(j)}$ is the flow divergence defined in \eqref{eq:def_flow_div}.
We insert \eqref{eq:flow_bound} into \eqref{eq:total_exp} to obtain 
 \begin{align}
 \Delta(\ve Q_t)  &\stackrel{}{\leq} \sum_{\tau=t}^{t+T-1}12 - 2 \cdot \mathbb E\left[\sum_{j=1}^2 \sum_{l=1}^3 \mathbb E\left[ Q_{l,\tau}^{(j)} \flowdivRV{l,\tau}{(j)} 
\Bigg| \ve Q_\tau \ensuremath{\Omega}^{\tau-1}  \right]  \Bigg| \ve Q_t \right] \label{eq:boundDeltaVis}
\\
&= \sum_{\tau=t}^{t+T-1} 12 -2 \cdot \mathbb E\left[  \sum_{j=1}^2 \sum_{l=1}^3 \sum_{m \in \mc O_l} \mathbb E\left[ F_{lm,\tau}^{(j)} \left( Q_{l,\tau}^{(j)} - Q_{m,\tau}^{(j)} \right) \Bigg| \ve Q_\tau \ensuremath{\Omega}^{\tau-1}  \right] - Q_{1,\tau}^{(j)} R_j  \Bigg| \ve Q_t \right]
\\
&= \sum_{\tau=t}^{t+T-1} 12 -2 \cdot \mathbb E\left[\sum_{j=1}^2 \sum_{l=1}^3 \sum_{m \in \mc O_l} \mathbb E\left[ \activRV{lm,\tau}{(j)} C_{lm,\tau}^{(j)} \left( Q_{l,\tau}^{(j)} - Q_{m,\tau}^{(j)} \right) \Bigg| \ve Q_\tau \ensuremath{\Omega}^{\tau-1}  \right] - Q_{1,\tau}^{(j)} R_j  \Bigg| \ve Q_t \right]
\label{eq:drift_eq},
\end{align}
where $\mc O_l$ contains all queue indices $m$ for which a link from queue $Q_l^{(j)}$ to $Q_m^{(j)}$ exists, with the understanding that $Q_{4,\tau}^{(j)}=0$.

In \eqref{eq:drift_eq}, we have to maximize the inner expectation with respect to $P_{\ve{\activRV{}{}}_\tau|\ve Q_\tau}$ and $P_{A_\tau|\ve Q_\tau \ensuremath{\Omega}^{\tau-1}}$ to find the tightest upper bound on $\Delta(\ve Q_t)$.
Because $C_{lm,\tau}^{(j)}$ is nonnegative, we can first choose $\activRV{lm,\tau}{(j)}=1$ if $( Q_{l,\tau}^{(j)} - Q_{m,\tau}^{(j)} )>0$ and $0$ otherwise, hence we obtain 
 \begin{align}
 \Delta(\ve Q_t) 
&\leq \sum_{\tau=t}^{t+T-1} 12 -2 \cdot \mathbb E\left[\sum_{j=1}^2 \sum_{l=1}^3 \sum_{m \in \mc O_l)} \left[ Q_{l,\tau}^{(j)} - Q_{m,\tau}^{(j)} \right]^+ \mathbb E\left[ C_{lm,\tau}^{(j)}  \bigg| \ve Q_\tau \ensuremath{\Omega}^{\tau-1}  \right] - Q_{1,\tau}^{(j)} R_j  \Bigg| \ve Q_t \right].
\end{align}

The individual terms inside the inner expectation are derived in the following, using the definitions in \eqref{eq:def_F12t1}~-~\eqref{eq:def_F34_new}:
\begin{align}
 \mathbb E\left[ C_{12,\tau}^{(j)}  \big|\ve Q_\tau \ensuremath{\Omega}^{\tau-1}  \right] &= \Pr[A_\tau = j|\ve Q_\tau \ensuremath{\Omega}^{\tau-1}] \big(\epsj{\ensuremath{\Omega}^{\tau-1}} - \epsonetwo{\ensuremath{\Omega}^{\tau-1}}\big)  \\
  \mathbb E\left[ C_{13,\tau}^{(j)}  \big|\ve Q_\tau \ensuremath{\Omega}^{\tau-1}  \right] &= \Pr[A_\tau = 4|\ve Q_\tau \ensuremath{\Omega}^{\tau-1}] \big(1 - \epsonetwo{\ensuremath{\Omega}^{\tau-1}}\big) \\
   \mathbb E\left[ C_{14,\tau}^{(j)}  \big|\ve Q_\tau \ensuremath{\Omega}^{\tau-1}  \right] &= \Pr[A_\tau = j|\ve Q_\tau \ensuremath{\Omega}^{\tau-1}] \big(1-\epsj{\ensuremath{\Omega}^{\tau-1}} \big) \\
  \mathbb E\left[ C_{24,\tau}^{(j)}  \big|\ve Q_\tau \ensuremath{\Omega}^{\tau-1}  \right] &= \Pr[A_\tau = 3|\ve Q_\tau \ensuremath{\Omega}^{\tau-1}] \big(1-\epsj{\ensuremath{\Omega}^{\tau-1}} \big) \\
  \mathbb E\left[ C_{32,\tau}^{(j)}  \big|\ve Q_\tau \ensuremath{\Omega}^{\tau-1}  \right] &= \Pr[A_\tau = 5|\ve Q_\tau \ensuremath{\Omega}^{\tau-1}] \big(\epsj{\ensuremath{\Omega}^{\tau-1}} - \epsonetwo{\ensuremath{\Omega}^{\tau-1}}\big) \\
  \mathbb E\left[ C_{34,\tau}^{(j)}  \big|\ve Q_\tau \ensuremath{\Omega}^{\tau-1}  \right] &= \Pr[A_\tau = 5|\ve Q_\tau \ensuremath{\Omega}^{\tau-1}] \big(1-\epsj{\ensuremath{\Omega}^{\tau-1}}\big),
\end{align}
where the expectation on the LHS is with respect to a distribution $P_{A_\tau|\ve Q_\tau \ensuremath{\Omega}^{\tau-1}}$.
All relations follow because $\ve Z_\tau - \ve Q_\tau \ensuremath{\Omega}^{\tau-1} - A_\tau $ forms a Markov chain.

The action chosen in \eqref{eq:maxweight_new} and \eqref{eq:maxweight_new_hmm} results in the tightest upper bound on $\Delta(\ve Q_t)$ in \eqref{eq:drift_eq}:
The distribution $P_{A_\tau|\ve Q_\tau \ensuremath{\Omega}^{\tau-1}}$ that maximizes the expression inside the conditional expectation for every outcome of $\ve Q_\tau$ and $\ensuremath{\Omega}^{\tau-1}$ also minimizes the upper bound in \eqref{eq:drift_eq}.
The associated optimization problem is a linear program, constrained only by conditions that $P_{A_\tau|\ve Q_\tau \ensuremath{\Omega}^{\tau-1}}$ must be a probability distribution. The optimizer of a linear program lies at the boundary of the constraint set, and thus the optimal conditional distribution is  deterministic. Hence, choosing one action with probability $1$ optimizes the max-weight criterion in \eqref{eq:maxweight_new} and \eqref{eq:maxweight_new_hmm}.

\begin{remark}
 Note that
 \begin{itemize}
  \item the criterion in \eqref{eq:maxweight_new} and \eqref{eq:maxweight_new_hmm} does not depend on $\tau$,
  \item actions are chosen only if the corresponding queues are nonempty, so no transmissions are wasted and $\tilde{F}_{lm,\tau}^{(j)} = C_{lm,\tau}^{(j)}$, unless there is a nonpositive differential backlog $ Q_{l,\tau}^{(j)} - Q_{m,\tau}^{(j)} $.
 \end{itemize}
\end{remark}

The action in \eqref{eq:maxweight_new} and \eqref{eq:maxweight_new_hmm} results in the tightest upper bound in \eqref{eq:drift_eq} under the assumption that action $A_\tau$ can depend on $Q_\tau$ and $\ensuremath{\Omega}^{\tau-1}$. 
Any scheme that bases its decisions for $A_\tau$ on a subset of $Q_\tau \ensuremath{\Omega}^{\tau-1}$ and its decisions for $\ve{\activRV{}{}}_\tau$ according to a distribution $P_{\ve{\activRV{}{}}_\tau}$ will result in a looser bound on $\Delta(\ve Q_t)$.
For the visible case, let the decisions for action $A_\tau$ be drawn randomly from a stationary distribution $P_{A_\tau|S_{\tau-1}}$, as in the probabilistic scheme in Section~\ref{sec:probab_scheme_new_visible}.
For the hidden case, let the decisions be according to $P_{A_\tau|\ve Z^{\tau-1}_{\tau-L}}$, as in the probabilistic scheme in Section~\ref{sec:probab_scheme_new_hmm}.
To harmonize notation, decisions for $A_\tau$ are based on $\ensuremath{{\Omega}}^{\tau-1}_{\tau-L}$, with the understanding that 
$\ensuremath{{\Omega}}^{\tau-1}_{\tau-L} = S_{\tau-1}$ in the visible case and $\ensuremath{{\Omega}}^{\tau-1}_{\tau-L} = \ve Z^{\tau-1}_{\tau-L}$ in the hidden case.

Given that the probabilistic scheme is used, we write the inner expectation in \eqref{eq:boundDeltaVis}
\begin{align}
 \mathbb E\left[ Q_{l,\tau}^{(j)} \flowdivRV{l,\tau}{(j)}\big| \ve Q_\tau \ensuremath{\Omega}^{\tau-1}\right] = 
 Q_{l,\tau}^{(j)} \mathbb E\left[ \flowdivRV{l,\tau}{(j)}\big| \ensuremath{\Omega}^{\tau-1}\right] = Q_{l,\tau}^{(j)} \flowdiv{l}{(j)}{(\ensuremath{{\Omega}}^{\tau-1}_{\tau-L})},
 \label{eq:maxweight_probabilistic}
\end{align}
where $\flowdiv{l}{(j)}{(\ensuremath{{\Omega}}^{\tau-1}_{\tau-L})}$ denotes the flow divergence averaged with respect to some distributions $P_{A_\tau|\ensuremath{{\Omega}}^{\tau-1}_{\tau-L}}$ and $P_{\ve{\activRV{}{}}_\tau}$. 
We use these arguments to further bound $ \Delta(\ve Q_t)$ as follows. The individual steps are explained below.
\allowdisplaybreaks
 \begin{align}
 \Delta(\ve Q_t) 
 &\stackrel{(a)}{\leq} \sum_{\tau=t}^{t+T-1} 12 - 2 \cdot \mathbb E \left[ \sum_{j=1}^2  \sum_{l=1}^3 Q_{l,\tau}^{(j)} \flowdiv{l}{(j)}{(\ensuremath{{\Omega}}^{\tau-1}_{\tau-L})} \bigg|\ve Q_t \right] \label{eq:drift_eq_simpl}\\
 &\stackrel{(b)}{\leq} \sum_{\tau=t}^{t+T-1} 12 + 12 (\tau-t) - 2 \cdot \mathbb E \left[\sum_{j=1}^2  \sum_{l=1}^3 Q_{l,t}^{(j)} \flowdiv{l}{(j)}{(\ensuremath{{\Omega}}^{\tau-1}_{\tau-L})} \bigg|\ve Q_t \right] \label{eq:drift_eq_with_Qbound}\\
 &\stackrel{(c)}{=} 12T + 6T(T-1) - 2 \sum_{j=1}^2 \sum_{l=1}^3 \sum_{\tau=t}^{t+T-1} \sum_{\ensuremath{{\omega}}^L} \Pr[\ensuremath{{\Omega}}^{\tau-1}_{\tau-L}=\ensuremath{{\omega}}^L|\ve Q_t]   Q_{l,t}^{(j)} \flowdiv{l}{(j)}{(\ensuremath{{\omega}}^L)} \\
 &\stackrel{(d)}{=}12T + 6T(T-1)- 2 \sum_{j=1}^2 \sum_{l=1}^3     Q_{l,t}^{(j)} \sum_{\ensuremath{{\omega}}^L} \flowdiv{l}{(j)}{(\ensuremath{{\omega}}^L)}  \sum_{\tau=t}^{t+T-1} \Pr[\ensuremath{{\Omega}}^{\tau-1}_{\tau-L}=\ensuremath{{\omega}}^L|\ve Q_t]\\
 &\stackrel{(e)}{\leq} 12T+6T^2 - 2T\sum_{j=1}^2 \sum_{l=1}^3 Q_{l,t}^{(j)} \left( \left[\sum_{\ensuremath{{\omega}}^L}  P_{\ensuremath{{\Omega}}^L}(\ensuremath{{\omega}}^L) \flowdiv{l}{(j)}{(\ensuremath{{\omega}}^L)} \right] - \varepsilon_{L,T}    \right) \\
 &\stackrel{(f)}{\leq}12T + 6T^2 -2T(\delta-\varepsilon_{L,T}) \sum_{j=1}^2 \sum_{l=1}^3  Q_{l,t}^{(j)} .\label{eq:final_bound_Delta}
\end{align}
Step $(a)$ follows from \eqref{eq:maxweight_probabilistic}.
For step $(b)$ we follow similar steps as in \cite[Sect. 4.9]{neely2010stochastic}: The buffer level $Q_{l,\tau}^{(j)}$ can decrease by at most one packet per time slot:
\begin{align}
 \qquad Q_{l,\tau}^{(j)} &\geq Q_{l,t}^{(j)} - (\tau -t), \quad \text{for }\tau \geq t \label{eq:lb_Ql}.
\end{align}
One obtains \eqref{eq:drift_eq_with_Qbound}, where the expression inside the expectation does not depend on $\ve Q_\tau$ anymore.
Steps $(c)$ writes out the expectation and rearranges terms.
Because the sequence $\ensuremath{{\Omega}}^n$ is stationary and also the probabilistic strategy is stationary, $\flowdiv{l}{(j)}{(\ensuremath{{\omega}}_{\tau-L}^{\tau-1})}$ does not depend on $\tau$ but only on the realization $\ensuremath{{\Omega}}_{\tau-L}^{\tau-1} = \ensuremath{{\omega}}^L$. This is used in step $(d)$.

Step $(e)$ replaces the expression $\sum_{\ensuremath{{\omega}}^L} \flowdiv{l}{(j)}{(\ensuremath{{\omega}}^L)} \sum_{\tau=t}^{t+T-1} \Pr[\ensuremath{{\Omega}}^{\tau-1}_{\tau-L}=\ensuremath{{\omega}}^L|\ve Q_t]$ by the lower bound \linebreak$T \left( \sum_{\ensuremath{{\omega}}^L} \flowdiv{l}{(j)}{(\ensuremath{{\omega}}^L)} P_{\ensuremath{{\Omega}}^L}(\ensuremath{{\omega}}^L) - \varepsilon_{L,T}\right)$ that we derive next:

Define the mixture distribution 
\begin{align}
 \mixdis{T}(\ensuremath{{\omega}}^L|\ve q) = \frac{1}{T} \sum_{\tau=t}^{t+T-1} P_{\ensuremath{{\Omega}}^{\tau-1}_{\tau-L}|\ve Q_t}(\ensuremath{{\omega}}^L|\ve q). \label{eq:def_mixture_distr}
\end{align}
The constant $T$ can be chosen large enough such that
 $\left| \sum_{\ensuremath{{\omega}}^L} \flowdiv{l}{(j)}{(\ensuremath{{\omega}}^L)} \big( P_{\ensuremath{{\Omega}}^L}(\ensuremath{{\omega}}^L) - \mixdis{T}(\ensuremath{{\omega}}^L|\ve q) \right) \big|$
is small, for all $l$, $j$ and $\ve q$.
This follows from 
\begin{align}
 \left| \sum_{\ensuremath{{\omega}}^L} \flowdiv{l}{(j)}{(\ensuremath{{\omega}}^L)} \big( P_{\ensuremath{{\Omega}}^L}(\ensuremath{{\omega}}^L) - \mixdis{T}(\ensuremath{{\omega}}^L|\ve q) \big) \right| &\leq  \sum_{\ensuremath{{\omega}}^L} \left|\flowdiv{l}{(j)}{(\ensuremath{{\omega}}^L)} \right| \left| P_{\ensuremath{{\Omega}}^L}(\ensuremath{{\omega}}^L) - \mixdis{T}(\ensuremath{{\omega}}^L|\ve q)  \right| \nonumber\\
 &\leq \sum_{\ensuremath{{\omega}}^L} \left| P_{\ensuremath{{\Omega}}^L}(\ensuremath{{\omega}}^L) - \mixdis{T}(\ensuremath{{\omega}}^L|\ve q) \right| \leq \varepsilon_{L,T}, \quad \forall  \ve q \label{eq:converge_to_steady}
\end{align}
for some arbitrarily small $\varepsilon_{L,T} >0$. 
The first inequality follows from the triangle inequality, the second one follows from $\left|\flowdiv{l}{(j)}{(\ensuremath{{\omega}}^L)} \right|\leq 1$.
The last step is a bound on the variational distance between the steady-state distribution and the mixture distribution.
We distinguish visible and hidden case: 

In the visible case, \eqref{eq:converge_to_steady} requires the mixture distribution $ \frac{1}{T} \sum_{\tau=t}^{t+T-1} P_{S_{\tau-1}|\ve Q_t}(s|\ve q)$ to converge to the steady-state distribution $\pi_s$ in terms of variational distance for each $\ve q$.
A value of $T$  for an arbitrarily small $\varepsilon_{L,T} >0$ exists if the Markov chain of the channel state process is irreducible and aperiodic, which is a model assumption\footnote{Aperiodicity is not necessarily required due to the Ces\`{a}ro mean in \eqref{eq:converge_to_steady}, but this is beyond the scope of this work. See \cite[Theorem 8.6.1]{horn2012matrix} for details.} in Section~\ref{sec:model}: 
In this case the steady-state distribution $\pi$ is unique (see, e.g., \cite[Theorem 4.3.1]{gallager2013stochastic}) and the distribution $P_{S_{\tau-1}|\ve Q_t}$ converges to $\pi$ for any initial distribution $P_{S_{t-1}|\ve Q_t}$, $\tau > t$. If $P_{S_{\tau-1}|\ve Q_t}$ converges to $\pi$, so does the Ces\`{a}ro mean $ \frac{1}{T} \sum_{\tau=t}^{t+T-1} P_{S_{\tau-1}|\ve Q_t}(s|\ve q)$.
The constant $T$ is thus related to the mixing time of the channel state Markov chain. The decay of $\varepsilon_{L,T}$ with respect to $T$ is $\sim \frac{1}{T}$ and not dependent on $L$.

In the hidden case, \eqref{eq:converge_to_steady} requires the mixture distribution $ \frac{1}{T} \sum_{\tau=t}^{t+T-1} P_{\ve Z^{\tau-1}_{\tau-L}|\ve Q_t}(\ve z^L|\ve q)$ to converge to the stationary distribution $P_{\ve Z^L}(\ve z^L)$ in terms of variational distance for each $\ve q$.
Define the random variable $ \ensuremath{M}_t = (S_{t-L}^{t-1} \ve Z_{t-L}^{t-1})$. Note that the sequence $\ensuremath{M}^n$ is Markov and the corresponding Markov chain is irreducible and aperiodic.
One may verify that $\ensuremath{M}_{\tau+1}$ is independent of $\ve Q_t$ given $\ensuremath{M}_{\tau}$ for $\tau \geq t$ and hence the distribution of $\ensuremath{M}_\tau$ converges to the steady-state distribution of the corresponding Markov chain, for any given initial distribution of $P_{\ensuremath{M}_t|\ve Q_t}$. 
If $\ensuremath{M}_\tau$ is distributed according to the steady-state distribution, so is its component $\ve Z_{\tau-L}^{\tau-1}$, i.e. according to $P_{\ve Z_{\tau-L}^{\tau-1}}(\ve z^L)$. This distribution does not depend on $\tau$ because $\ve Z^n$ is stationary, so we can replace it with $P_{\ve Z^L}(\ve z^L)$.
If $P_{\ve Z_{\tau-L}^{\tau-1}|\ve Q_t}$ converges to $P_{\ve Z^L}$, so does the Ces\`{a}ro mean  $ \frac{1}{T} \sum_{\tau=t}^{t+T-1} P_{\ve Z^{\tau-1}_{\tau-L}|\ve Q_t}(\ve z^L|\ve q)$
One can show that the decay of $\varepsilon_{L,T}$ with respect to $L$ and $T$ is $\sim \frac{ 4^L |\mc S|^L }{T}$, hence 
the constant $T$ has to be chosen significantly larger than $L$.

For step $(f)$, if the rate pair is in the interior of the (approximate) capacity region, i.e. if $(R_1+\bar{\delta}, R_2+\bar{\delta}) \in \bar{\mc C}_{\text{fb}+s}^\text{mem}$ for the visible case and if $(R_1+\bar{\delta}, R_2+\bar{\delta}) \in \bar{\mc C}_{\text{ fb}}^\text{mem}(L)$ for the hidden case, then there exists a constant $\delta >0$ that goes to zero when $\bar{\delta} \rightarrow 0$ such that
\begin{align}
\sum_{\ensuremath{{\omega}}^L} P_{\ensuremath{{\Omega}}^L}(\ensuremath{{\omega}}^L) \flowdiv{l}{(j)}{(\ensuremath{{\omega}}^L)} = \flowdiv{l}{(j)}{}\geq \delta, \quad \forall~l\in \{1,2,3\},~j\in \{1,2\},
\end{align}
where $\delta$ has to be chosen such that $\delta > \varepsilon_{L,T}$.

Using the result in \eqref{eq:final_bound_Delta} and the law of total expectation, we can bound
\begin{align}
 \mathbb E \left[ \mathbb L(\ve Q_{t+T}) - \mathbb L(\ve Q_t) \right] = &\mathbb E \big[ \Delta(\ve Q_t) \big| \ve Q_t \big] \nonumber \\
  \leq  18T^2 - 2T(\delta-\varepsilon_{L,T} ) & \sum_{j=1}^2 \sum_{l=1}^3 \mathbb E\left[  Q_{l,t}^{(j)} \right].
\end{align}
Summing over all time slots $t=1,\ldots,n$ yields
\begin{align}
\sum_{t=1}^{n} &\mathbb E\left[ \mathbb L(\ve Q_{t+T})- \mathbb L(\ve Q_t) \right] = \sum_{t=1}^T \mathbb E\left[\mathbb L(\ve Q_{t+n}) - \mathbb L(\ve Q_t)  \right]  \nonumber \\\leq &18 n T^2 - 2 T (\delta-\varepsilon_{L,T} )   \sum_{t=1}^{n} \sum_{j=1}^2 \sum_{l=1}^3 \mathbb E\left[  Q_{l,t}^{(j)} \right].
 \end{align}
 Rearranging terms gives
 \begin{align}
 \frac{1}{n} \sum_{t=1}^n \sum_{j=1}^2 \sum_{l=1}^3 \mathbb E\left[  Q_{l,t}^{(j)} \right] &\leq \frac{9T}{\delta - \varepsilon_{L,T} } + \frac{\sum_{t=1}^T \mathbb E[\mathbb L(\ve Q_{t}) ]}{2(\delta - \varepsilon_{L,T} )T n},
\end{align}
and taking a $\limsup$ with respect to $n$ on both sides proves strong stability of the queuing network, given that we have $1/T\sum_{t=1}^T \mathbb E[\mathbb L(\ve Q_{t}) ]<\infty$. This is true if the constant $T$ is finite and $\mathbb E[\mathbb L(\ve Q_1)]<\infty$.

\section*{Acknowledgment}
The authors would like to thank Navid Reyhanian, Gianluigi Liva, Gerhard Kramer, Chih-Chun Wang and the reviewers for comments and remarks that led to significant improvements of the paper.

\ifCLASSOPTIONcaptionsoff
  \newpage
\fi

\bibliographystyle{IEEEtran}

\end{document}